\documentclass[11pt, fleqn]{article}
\pdfoutput=1

\usepackage{geometry}
\usepackage{comment}
\usepackage{mathtools}
\usepackage{slashed}
\usepackage[compat=1.0.0]{tikz-feynman}
\usepackage{tabularx,colortbl}
\DeclarePairedDelimiter{\ceil}{\lceil}{\rceil}

\usepackage{jheppub}
\usepackage{amsmath,amssymb,amsthm}
\usepackage{bm}
\usepackage{slashed}
\usepackage{graphicx}
\usepackage[T1]{fontenc}
\usepackage{fix-cm}
\usepackage{makecell}
\usepackage{array}
\usepackage{tabulary}
\usepackage[thinlines]{easytable}
\usepackage[rightcaption]{sidecap}
\usepackage{caption}
\newcolumntype{K}[1]{>{\centering\arraybackslash}p{#1}}

\makeatletter
\newcommand*\bigcdot{\mathpalette\bigcdot@{.5}}
\newcommand*\bigcdot@[2]{\mathbin{\vcenter{\hbox{\scalebox{#2}{$\m@th#1\bullet$}}}}}
\makeatother

\newlength{\dummysp}
\usepackage[font=scriptsize]{caption}
\usepackage{mwe}

\def\p{{\bf p}}

\title{A Convergent Continuum Strong Coupling Expansion For Quantum Mechanics \& Quantum Field Theory}
\author{Erfan Shalchian \footnote{Full Name: M. Erfan Shalchian T. , Email: erfanshalchian@gmail.com, eshalchian@physics.utoronto.ca}}
\affiliation{Department of Physics, University of Toronto, 60 St. George St., Toronto, ON M5S 1A7, Canada}

\abstract{

{\flushleft{W}}e generalize the notion of an asymptotic weak coupling expansion about an exactly solvable model in quantum mechanics and quantum field theory to an all positive value coupling convergent expansion. This is done by rescaling the variables available in the theory by free parameters, then adding and subtracting the exactly solvable model. The rest (initial rescaled theory by free parameters + the subtracted exactly solvable model) is expanded about the added exactly solvable model. Evaluating finite orders of this expansion at its extremum points with respect to the free parameter(s) gives a sequence that converges to the result of the previous asymptotic expansion, with a good convergence rate, at relative strong coupling \footnote{Refer to the first footnote in the Introduction Section for a definition of relative strong coupling.}. We solve for the eigenenergies of the anharmonic, pure anharmonic and double well potential problems using this method by expanding about the symmetrical point of these potentials. Accurate results for the eigenenergies can be obtained for all positive values of the coupling for the anharmonic and pure anharmonic oscillators and at strong coupling for the double well potential. To provide confirmation for the convergent formalism developed for $\phi^4$ theory and QED we improve the electron g-factor calculation at the one loop level using the convergent formalism. Applications of this method are not limited to quantum mechanics or quantum field theory, for example it can also have applications in the context of differential equations.}

\begin{document}

\maketitle
\section{Introduction, outline and summary of results}
\label{sec:1}
It is known that the expansions in the couplings in quantum mechanics and quantum field theory are asymptotic (reference \cite{13} provides an argument for QED). In this work by a simple trick we show that it is possible to modify this asymptotic expansion in a way as to obtain a convergent expansion. To explain the main procedure for obtaining this convergent expansion consider the following one-dimensional integral
\begin{equation}
\label{eq:1.1}
I(\lambda) = \int^{+\infty}_{-\infty} dx \exp \left( - {1 \over 2} x^2 - \lambda x^{4} \right), \ \ \ \ \ \ \ \lambda  > 0 
\end{equation}
it is known that the expansion of the integral \eqref{eq:1.1} in $\lambda$ is an asymptotic series that is Borel summable to the full result of the integral \cite{05}. When summing the terms of this asymptotic series in an ordinary way, the smaller $\lambda$ is the more terms in the expansion become useful and more accurate results are obtained but after a certain order of the expansion the summation starts to diverge. Consider introducing a free parameter $\alpha$ in \eqref{eq:1.1} by rescaling the variable $x \rightarrow x/ \alpha$. After adding and subtracting a quadratic term in the exponent of the integrand we obtain
\begin{flalign}
\label{eq:1.2}
\hspace{-1cm} I(\lambda) & = I^{\alpha}(\lambda) \equiv {1 \over \alpha } \int^{+\infty}_{-\infty} dx \exp \left( - x^2 - \epsilon \left\{ \left( -1 + {1 \over 2 \alpha^2 } \right) x^2 + {\lambda \over \alpha^{4}} x^{4} \right\} \right) = \sum^{\infty}_{n = 0} I^{\alpha}_{n}(\lambda) \epsilon^n \nonumber \\ & = {\sqrt{\pi} \over \alpha } \sum^{\infty}_{n=0} \sum^{n}_{k=0} \bigg{\{} {(-1)^n \over k! (n-k)!} {(4n - 2k -1)!! \over 2^{2n - k}} \left( -1 + {1 \over 2\alpha^2} \right)^k \left( {\lambda \over \alpha^{4} } \right)^{n-k} \bigg{\}}
\end{flalign}
we have also introduced an $\epsilon$ ($ = 1$) parameter to indicate that now the expansion is to be performed with respect to this parameter. In this expansion, similar to as mentioned above, the smaller ${\lambda \over \alpha^4}$ is the more terms in the expansion become useful and therefore more accurate results can be obtained. Since the integral of \eqref{eq:1.2} is independent of $\alpha$ for $\epsilon = 1$ we have the freedom to choose a large value of $\alpha$ to suppress the coupling ${\lambda \over \alpha^4}$, without changing the integral to be evaluated and make more use of the terms in the expansion, therefore obtain more accurate results. This is the main idea behind the convergent expansion method. Note that it is important that we no longer expand in the coupling $\lambda$ as this will result in a cancellation of the introduced $\alpha$ parameter. Although the integral of \eqref{eq:1.2} is independent of $\alpha$, any finite order of its expansion in $\epsilon (= 1)$, defined by $I^{\alpha, n} (\lambda) \equiv \sum^{n}_{k = 0} I^{\alpha}_{k}(\lambda)$, will clearly depend on $\alpha$. Choosing $\alpha$ too large or too small will clearly not give optimum results therefore we expect intermediate values of $\alpha$ to exist that would give optimum results. Since the full integral $I^{\alpha} (\lambda)$ is independent of $\alpha$ we have $d^m I^{\alpha} (\lambda) / d \alpha^m = 0 $ for $m = 1, 2, ...$, therefore it seems that the zero $m$-derivative points of the partial sum $I^{\alpha, n}$, $d^m I^{\alpha, n} (\lambda) / d \alpha^m = 0$, would be good candidates for the optimum values of $\alpha$. For the integral of \eqref{eq:1.2} (and in most other examples), we will be mainly concerned with the zero first derivative points of $I^{\alpha, n}$ but in certain examples we consider later in the paper in Subsection \ref{sec:4.2} the zero points of the higher derivatives will also find applications. Evaluating $I^{\alpha, n} (\lambda)$ at its extremum points, after solving the equation $d I^{\alpha, n} (\lambda) / d \alpha = 0$ for different $n$, we find the results listed in Table \ref{tab:1}.
\begin{table}[h]
\centering
\caption{Numerical values of $I^{\alpha,n} (\lambda) \equiv \sum^{n}_{i=0} I^{\alpha}_{i}(\lambda)$ (for $I^{\alpha} (\lambda) = \sum^{\infty}_{i=0} I^{\alpha}_{i}(\lambda) \epsilon^i$ in relation \eqref{eq:1.2}) at its extremum points with respect to $\alpha$ for different $n$.}
 \begin{tabular}{| p{0.8cm} | c |} 
 \hline
 $\ \ \lambda$ & \begin{tabular}{K{1cm} | K{1cm} | K{1.2cm} | K{2.2cm} | K{4.5cm}}
 $n$ &  1 & 3 & 15 & $I (\lambda )$ of \eqref{eq:1.1} (Exact value)
 \end{tabular} \\
 \Xhline{4\arrayrulewidth}
 $ \ 0.1$ & \begin{tabular}{ K{1cm} | K{1cm} | K{1.2cm} | K{2.2cm} |  K{4.5cm} }
 $\alpha_{\text{ext}}$ & 0.90  & 1.005 & 1.31 & - \\
 \hline
 $I^{\alpha_{\text{ext}}, n} $ & 2.12 & 2.1475 & 2.14970591 & 2.14970593
 \end{tabular} \\
 \Xhline{4\arrayrulewidth}
 $\ \ 1$ & \begin{tabular}{ K{1cm} | K{1cm} | K{1.2cm} | K{2.2cm} |  K{4.5cm} }
 $\alpha_{\text{ext}}$ & 1.36 & 1.59 & 2.20 & - \\
 \hline
 $I^{\alpha_{\text{ext}},n}$ & 1.49 & 1.55 & 1.5548173 & 1.5548178
 \end{tabular} \\
 \Xhline{4\arrayrulewidth}
 $ \  10$ & \begin{tabular}{ K{1cm}  | K{1cm} | K{1.2cm} | K{2.2cm} | K{4.5cm} }
 $\alpha_{\text{ext}}$ & 2.29 & 2.73 & 3.84  & - \\
 \hline
 $I^{\alpha_{\text{ext}},n}$ & 0.91 & 0.959 & 0.9679593 & 0.9679602
 \end{tabular} \\
  \Xhline{4\arrayrulewidth}
 $ \ 10^{10}$ & \begin{tabular}{ K{1cm}  | K{1cm} | K{1.2cm} | K{2.2cm} | K{4.5cm} }
 $\alpha_{\text{ext}}$ & 397 & 476 & 677 & - \\
 \hline
 $I^{\alpha_{\text{ext}},n}$ & .0053 & .00567 & .005732574 & $.005732583$
 \end{tabular} \\
 \hline
\end{tabular}
\label{tab:1}
\end{table}
As can be seen from the results of this Table the expansion of \eqref{eq:1.2} in $\epsilon$ ($ = 1$), evaluated at its extremum points with respect to $\alpha$ at order $n$, produces results that converge to the exact value of the integral with a good convergence rate for all positive values of the coupling $\lambda$. Going to higher orders in the $\epsilon$ expansion with choosing the appropriate extremum points $\alpha_{\text{ext}}$ produces more accurate results for all values of $\lambda > 0$. Also it can be shown that the convergence rate stays level at strong coupling. A discussion on this matter is done in Appendix \ref{sec:A2} .

When $\alpha = 1/\sqrt{2}$ the expansion of \eqref{eq:1.2} reduces to the previous asymptotic expansion in $\lambda$ but with choosing the appropriate extremum points $\alpha_{\text{ext}}$ at each finite order in the expansion we obtain a sequence that converges to the exact result, therefore the $\epsilon (= 1)$ expansion can be considered as a natural generalization of the previously asymptotic weak coupling expansion to an all positive value coupling convergent expansion which has a good convergence rate at relative strong coupling. Furthermore the convergent expansion method can also be applied to the relative strong coupling \footnote{Here we define the notion of relative strong coupling. When performing the expansion of a certain number of terms about an exactly solvable model (e.g. about a Gaussian (path) integral or about the harmonic oscillator Hamiltonian as when solving the time-independent Schrodinger equation perturbatively) there are terms which their expansion is well defined and there might be terms which their expansion is not well defined. The relative strong coupling regime refers to the regime in the coupling space of the theory which the coefficients of the terms which their expansion is well defined is greater than or is of the same order of magnitude of the coefficient of the terms which their expansion is not well defined. As a simple example consider the following one dimensional integral: $\int^{+\infty}_{-\infty} dx \exp \left( - x^2 + 2 \sqrt{\lambda} x^3 - \lambda x^4 \right)$. The term which its expansion about the Gaussian integral is well defined (by itself) is $- \lambda x^4$ and the term which its expansion about the Gaussian integral is not well defined (by itself) is $2 \sqrt{\lambda} x^3$ therefore the relative strong coupling regime would be when $\lambda \gtrsim 2 \sqrt{\lambda}$ (the symbol $\gtrsim$ meaning greater than or of the order of magnitude). Similarly the relative weak coupling regime would be when $\lambda < 2 \sqrt{\lambda}$. In the convergent expansion method the convergence rate is good for the relative strong coupling regime and it becomes worse as we go to the relative weak coupling regime. In the absence of terms which their expansion is not well defined the relative strong coupling regime would correspond to all positive values of the coupling, for example consider the integral: $\int^{+\infty}_{-\infty} dx \exp \left( - {1 \over 2} x^2 - \lambda x^4 \right)$. The convergent expansion of this integral has a good convergence rate for all $\lambda > 0$ as can be seen from Table \ref{tab:1}. As another example consider the following integral: $ \int^{+\infty}_{-\infty} dx \exp \left( a x^2 - \lambda x^4 \right) \rightarrow \int^{+\infty}_{-\infty} dx \exp \left( - x^2 + (1 + a) x^2 - \lambda x^4 \right)$, for $a \geq 0$ the expansion of the $(1+ a)x^2$ term is not well defined since it gives a divergent series, but the expansion of the term $- \lambda x^4$ is well defined therefore the relative strong coupling regime would correspond to when $\lambda \gtrsim a$. Note that for $a = 0$ although the expansion of the term $(1 + 0) x^2$ is not well defined by itself but it can be shown that the convergence rate of the convergent expansion method is good for all $\lambda > 0$ and in particular at weak coupling (this has been verified in numerical calculations ). As a less common example consider the following integral: $\int^{+\infty}_{-\infty} dx \exp \left( - x^2 + \lambda^2 x^3 - \lambda x^4 \right)$, here the relative strong coupling regime is for $\lambda \gtrsim \lambda^2 \rightarrow \lambda \lesssim 1$.}
regime of examples which due to a lack of a quadratic term with the appropriate sign, a straightforward expansion in the coupling is not possible. For example consider the expansion of a double well potential about its symmetrical point or relation \eqref{eq:2.1} for $\kappa = -1$

To provide another simple example, consider the following one dimensional integral
\begin{equation}
\label{eq:1.3}
\bar{I}(\lambda) = \int^{+\infty}_{-\infty} dx \exp \left( - x^2 - \lambda | x^{3} | \right), \ \ \ \ \ \ \ \lambda > 0
\end{equation}
introducing a positive free parameter into the above integral by rescaling $x \rightarrow x /\alpha$, adding and subtracting a quadratic term in the exponent of the integrand we obtain
\begin{flalign}
\label{eq:1.4}
\bar{I}(\lambda) & = \bar{I}^{\alpha} (\lambda) = {1 \over \alpha } \int^{+\infty}_{-\infty} dx \exp \left( - x^2 - \epsilon \bigg{\{} \Big{(} {1 \over \alpha^2} -1\Big{)} x^2 + {\lambda \over \alpha^3} | x^{3} |  \bigg{\}} \right) = \sum^{\infty}_{n = 0} \bar{I}^{\alpha}_{n}(\lambda) \epsilon^n \nonumber \\ & = {1 \over \alpha } \sum^{\infty}_{n=0} \sum^{n}_{k=0} \bigg{\{} {(-1)^n \over k! (n-k)!} \Gamma \Big{(} k + {3 \over 2}(n-k) + {1 \over 2} \Big{)} \left({1 \over \alpha^2} -1 \right)^k \left( {\lambda \over \alpha^{3} } \right)^{n-k} \bigg{\}}
\end{flalign}
Table \ref{tab:2} lists the values of the the partial sum $\bar{I}^{\alpha,n} (\lambda) \equiv \sum^{n}_{i=0} \bar{I}^{\alpha}_{i}$ evaluated at its extremum points with respect to $\alpha$ for different $n$. Again from this Table it can be seen that accurate results can be obtained for all positive values of the coupling $\lambda$.
\begin{table}[h]
\centering
\caption{Numerical values of $\bar{I}^{\alpha,n} (\lambda) \equiv \sum^{n}_{i=0} \bar{I}^{\alpha}_{i}$ (for $\bar{I}^{\alpha} (\lambda) = \sum^{\infty}_{i=0} \bar{I}^{\alpha}_{i} \epsilon^i$ in relation \eqref{eq:1.4}) at its extremum points with respect to $\alpha$.}
 \begin{tabular}{| p{0.8cm} | c |} 
 \hline
 $\ \ \lambda$ & \begin{tabular}{K{1cm} | K{1.5cm} | K{1.7cm} | K{3cm} | K{4.5cm}}
 $n$ &  1 & 3 & 9 & $\bar{I} (\lambda )$ of \eqref{eq:1.3} (Exact value)
 \end{tabular} \\
 \Xhline{4\arrayrulewidth}
 $ \ 0.1$ & \begin{tabular}{ K{1cm} | K{1.5cm} | K{1.7cm} | K{3cm} |  K{4.5cm} }
 $\alpha_{\text{ext}}$ & 1.07  & 1.09 & 1.136 & - \\
 \hline
 $\bar{I}^{\alpha_{\text{ext}}, n} $ & 1.6850 & 1.685964 & 1.6859665147524 & 1.6859665147529
 \end{tabular} \\
 \Xhline{4\arrayrulewidth}
 $\ \ 1$ & \begin{tabular}{ K{1cm} | K{1.5cm} | K{1.7cm} | K{3cm} |  K{4.5cm} }
 $\alpha_{\text{ext}}$ & 1.43 & 1.54 & 1.71 & - \\
 \hline
 $\bar{I}^{\alpha_{\text{ext}},n}$ & 1.317 & 1.3272 & 1.32742113 & 1.32742114
 \end{tabular} \\
 \Xhline{4\arrayrulewidth}
 $ \  10$ & \begin{tabular}{ K{1cm}  | K{1.5cm} | K{1.7cm} | K{3cm} | K{4.5cm} }
 $\alpha_{\text{ext}}$ & 2.60 & 2.88 & 3.31  & - \\
 \hline
 $\bar{I}^{\alpha_{\text{ext}},n}$ & 0.753 & 0.7675 & 0.7682220 & 0.7682222
 \end{tabular} \\
  \Xhline{4\arrayrulewidth}
 $ \ 10^{10}$ & \begin{tabular}{ K{1cm}  | K{1.5cm} | K{1.7cm} | K{3cm} | K{4.5cm} }
 $\alpha_{\text{ext}}$ & 2469 & 2761 & 3210 & - \\
 \hline
 $\bar{I}^{\alpha_{\text{ext}},n}$ & 0.00081 & $0.0008278$ & $0.0008289683$ & $0.0008289687$
 \end{tabular} \\
 \hline
\end{tabular}
\label{tab:2}
\end{table}

The procedure outlined above for modifying an asymptotic expansion in the coupling in a way as to obtain a convergent expansion is pretty general. In this paper we will apply this simple procedure to one dimensional integrals, quantum mechanics and quantum field theory to obtain convergent expansions. The results obtained are pretty impressive. The expansion is convergent with a good convergence rate at relative strong coupling \footnote{At relative weak coupling the convergence rate becomes less efficient. For example refer to relation \eqref{eq:2.1} for $\kappa = -1$ or a double well potential when expanding about its symmetrical point which in these examples the expansion converges to the full result but the convergence rate becomes less efficient at (relative) weak coupling as Tables \ref{tab:7} or \ref{tab:31} show this but the convergence rate at (relative) strong coupling is good. Or consider the discussion related to Table \ref{tab:8} or the double well potential when expanding about one of its vacuums which in these examples the expansion does not show convergence to the full result, as expected, and the convergence rate becomes less efficient at (relative) weak coupling as the results of Tables \ref{tab:8} or \ref{tab:31'} show this.}.

To illustrate how the convergent expansion method works in a simpler context, in Section \ref{sec:2} we apply the procedure described above to one dimensional integrals. A sample of the results obtained in this Section were listed in Table \ref{tab:1}.
In Section \ref{sec:3} we apply the same method to quantum mechanics. In Section \ref{sec:3.1} we develop a convergent formalism for the time-independent Schrodinger equation in one space dimension. In Table \ref{tab:3} we show a sample of the results obtained for the vacuum energy of the anharmonic oscillator.
\begin{table}[h]
\centering
\caption{Applying the convergent expansion in $\epsilon$ to the vacuum energy of the anharmonic oscillator with potential $V_{1}(x) = {1 \over 2} x^2 + \lambda x^4$ gives $E_0 = \sum^{\infty}_{l=0} E^{\alpha}_{0l} \epsilon^l$. Table below lists the values of the partial sum $2E_0^{\alpha, n} = 2\sum^n_{l=0} E^{\alpha}_{0l}$ evaluated at its extremum points with respect to $\alpha$ for different $n$. We consider double the value of the vacuum energy to match the conventions of reference \cite{08}.}
 \begin{tabular}{| c | c |} 
 \hline
 $\lambda$ & \begin{tabular}{ K{1.2cm} | K{3cm} | K{3.2cm} | K{3cm}}
 $n$ & 15 & 25 & $2 E_0$(of \cite{08})
 \end{tabular} \\
 \Xhline{4\arrayrulewidth}
 $1/20$ & \begin{tabular}{ K{1.2cm} | K{3cm} | K{3.2cm} | K{3cm} }
 $\alpha^2_{\text{ext},3}$ & 1.376 & 1.473 & - \\
 \hline
 $2 E_0^{\alpha, n}$ & $1.0652855095439$ & $1.0652855095437$& $1.065 285509 54 (6)$
 \end{tabular} \\
 \Xhline{4\arrayrulewidth}
 $100/2$ & \begin{tabular}{ K{1.2cm} | K{3cm} | K{3.2cm} | K{3cm} }
 $\alpha^2_{\text{ext},3}$ & 10.750 & 12.013 & - \\
 \hline
 $2 E_0^{\alpha, n}$ & $4.9994178$ &  $4.99941755$ & 5.0(1)
 \end{tabular} \\ \hline
 \Xhline{4\arrayrulewidth}
 ${10^{10} / 2}$ & \begin{tabular}{ K{1.2cm} | K{3cm} | K{3.2cm} | K{3cm} }
 $\alpha^2_{\text{ext},3}$  & 4975 & 5563 & - \\
 \hline
 $2 E_0^{\alpha, n}$ & $2284.48119$ & $2284.481040$ & -
 \end{tabular} \\ \hline
\end{tabular}
\label{tab:3}
\end{table}
These results are compared with the values obtained from \cite{08} for $\lambda = 1/20$ and $\lambda = 100/2$. Reference \cite{08} applies the Aitken transformation on 25 terms of the asymptotic series expansion of the vacuum energy $E_0$ in $\lambda$. The digit in parentheses is uncertain. Convergence becomes more difficult at strong coupling using this method (or any other method which is used to improve the convergence rate of an asymptotic series) as can be seen from the results of the above Table listed under reference \cite{08}. However using the convergent expansion method accurate results with a faster convergence rate can be obtained for all positive values of the coupling and the convergence rate stays level at strong coupling (refer to Appendix \ref{sec:A2} for a discussion on this matter). We can clearly see convergence in the above results even for a coupling of order $10^{10}$! We note that the same procedure for obtaining the values of the eigenenergies for $\lambda = 1/20$ and $\lambda = 100/2$ which match the results of reference \cite{08} up to its accurate digits, was applied to a coupling of $\lambda = 10^{10}/2$. The results clearly show convergence therefore are reliable and can be considered as valid \footnote{For the derivation of the numerical results of this work in order to be able to obtain accurate results with arbitrary accuracy we made use of arbitrary precision numbers with a precision up to $100$ digits.}. Also, as mentioned, using this method we obtain a faster convergence rate, for example at $n=15$ for $\lambda =1/20$ the result already matches the result of reference \cite{08} up to its accurate digits. Tables \ref{tab:3'} and \ref{tab:4'} show the values of the eigenenergies obtained for the pure anharmonic potential and the double well potential using the convergent expansion method and compares them with the ones listed in references \cite{12} and \cite{09}, respectively. It is possible to obtain more accurate results for the energy level $\nu = 6$ of the pure anharmonic oscillator that match the results of reference \cite{12} by going to higher orders of the expansion. In Tables \ref{tab:3}, \ref{tab:3'} and \ref{tab:4'} we obtain many sequences of extremum points $\alpha_{\text{ext},i}$ with the higher ones converging faster to the desired energy level. In these Tables the squared value of the extremum point $\alpha^2_{\text{ext},i}$ is shown.
\begin{table}[h]
\centering
\caption{Applying the convergent expansion in $\epsilon$ to the eigenenergies of the pure anharmonic oscillator with potential $V_{0}(x) = {1 \over 2} x^4$ gives $E = \sum^{\infty}_{l=0} E^{\alpha}_{l} \epsilon^l$. Table below lists the values of the partial sum $2E^{\alpha, n} = 2\sum^n_{l=0} E^{\alpha}_{l}$ evaluated at its extremum points with respect to $\alpha$ for different $n$ and different energy levels $\nu = 0, 3 ,6$. We consider double the value of the energy levels to match the conventions of reference \cite{12}.}
 \begin{tabular}{| c | c |} 
 \hline
 $\nu$ & \begin{tabular}{K{1cm} | K{2.8cm} | K{3.4cm} | K{3.8cm}}
 $n$ & 35 & 50 & $2 E$(of \cite{12})
 \end{tabular} \\
 \Xhline{4\arrayrulewidth}
 $3$ & \begin{tabular}{ K{1cm} | K{2.8cm} | K{3.4cm} | K{3.8cm} }
 $\alpha^2_{\text{ext},7}$ & $3.043$ & 3.416 & - \\
 \hline
 $2 E^{\alpha, n}$ & $11.64474551145$ & $11.644745511375$ & $11.644745511378$ \\
 \hline
 $\alpha^2_{\text{ext},8}$ & - & 3.272  & - \\
 \hline
 $2 E^{\alpha, n}$ & - & 11.644745511378 & $11.644745511378$
 \end{tabular} \\
 \Xhline{4\arrayrulewidth}
 $6$ & \begin{tabular}{ K{1cm} | K{2.8cm} | K{3.4cm} | K{3.8cm} }
 $\alpha^2_{\text{ext},11}$ & 3.322 & 3.851 & - \\
 \hline
 $2 E^{\alpha, n}$ & $26.528471183679$  & $26.528471183681$ & $26.528471183682518..$ \\
 \hline
 $\alpha^2_{\text{ext},14}$ & - & 3.507  & - \\
 \hline
 $2 E^{\alpha, n}$ & - & $26.528471183682510$ & $26.528471183682518..$
 \end{tabular} \\
 \Xhline{4\arrayrulewidth}
 $0$ & \begin{tabular}{ K{1cm} | K{2.8cm} | K{3.4cm} | K{3.8cm} }
 $\alpha^2_{\text{ext},3}$ & 2.932 & 3.438 & - \\
 \hline
 $2 E^{\alpha, n}$ & 1.06036209036 & $1.06036209039$ & 1.0604 \\
 \end{tabular} \\ \hline
 \end{tabular}
\label{tab:3'}
\end{table}
\begin{table}[h]
\centering
\caption{Applying the convergent expansion in $\epsilon$ to the eigenenergies of the double well oscillator with potential $V_{\text{-}1} (x) = - {\omega^2} x^2/2 + {\lambda} x^4$ gives $E = \sum^{\infty}_{l=0} E^{\alpha}_{l} \epsilon^l$. Table below lists the values of the partial sum $2E^{\alpha, n} = 2\sum^n_{l=0} E^{\alpha}_{l}$ evaluated at its extremum points with respect to $\alpha$ for different $n$ and different energy levels $\nu = 0, 1$. We consider double the value of the energy levels to match the conventions of reference \cite{09}. }
 \begin{tabular}{| c | c | c | c |} 
 \hline
$\nu$ & $\omega^2$ & $\lambda$ & \begin{tabular}{K{1cm} | K{2.8cm} | K{2.8cm} | K{2.8cm}}
 $n$ & 35 & 65 & $2 E$(of \cite{09})
 \end{tabular} \\
  \Xhline{4\arrayrulewidth}
0 & ${1 \over 2}$ & ${1 / 2}$ & \begin{tabular}{ K{1cm} | K{2.8cm} | K{2.8cm} | K{2.8cm} }
 $\alpha^2_{\text{ext},4}$ & $3.850 \pm .06i$ & 4.780 & - \\
 \hline
 $2 E^{\alpha, n}$ & $0.8700175181$ & $0.870017518375$ & $0.870017518372$ \\
 \hline
 $\alpha^2_{\text{ext},5}$ & - & 4.534 & - \\
 \hline
 $2 E^{\alpha, n}$ & - & $0.870017518372$ & $0.870017518372$
 \end{tabular} \\
 \Xhline{4\arrayrulewidth}
$\nu$ & $\omega^2$ & $\lambda$ & \begin{tabular}{K{1cm} | K{2.8cm} | K{2.8cm} | K{2.8cm}}
 $n$ & 35 & 45 & $2 E$(of \cite{09})
 \end{tabular} \\
  \Xhline{4\arrayrulewidth}
1 & ${1 \over 2}$ & ${1 / 2}$ & \begin{tabular}{ K{1cm} | K{2.8cm} | K{2.8cm} | K{2.8cm} }
 $\alpha^2_{\text{ext},5}$ & $4.385$ & 4.915 & - \\
 \hline
 $2 E^{\alpha, n}$ & $3.3337793286$ & $3.3337793290$ & $3.33377932989$ \\
 \hline
 $\alpha^2_{\text{ext},7}$ & - & 4.282 & - \\
 \hline
 $2 E^{\alpha, n}$ & - & $3.33377932989$ & $3.33377932989$
 \end{tabular} \\ \hline
\end{tabular}
\label{tab:4'}
\end{table}

 In Section \ref{sec:3.2} we apply the same method to formulate a convergent expansion for the path integral formalism.

In Section \ref{sec:4} we apply this method to quantum field theory, in particular to $\phi^4$ theory and QED. As an example to illustrate the validity of the formalism developed in Subsection \ref{sec:4.2.1} we improve the electron g-factor calculation using the convergent expansion method at the one loop level. In Tables \ref{tab:4} and \ref{tab:5} we list a sample of the results obtained. These Tables compare the one loop evaluation of $a_e = (g-2)/2$ using the convergent expansion method given by $a_e (\text{one loop})_{\text{converg. exp.}} = \kappa^{\alpha_{\text{ext}}}_{23} \alpha_{\text{QED}}/ 2\pi $ with the one loop and higher loop evaluations of $a_e$ using the conventional asymptotic expansion method in the coupling $e$ given by $a_e (\text{one loop}) = \alpha_{\text{QED}}/ 2\pi = 0.00116141$ \cite{10} and $a_e (\text{higher loop}) = 0.001159652..$ \cite{06'}, respectively.
\begin{table}[h]
\centering
\caption{Comparison of the one loop evaluation of $a_e = (g-2)/2$ using the convergent expansion method given by $ \kappa^{\alpha_{\text{ext}}}_{23} \alpha_{\text{QED}}/ 2\pi $ with the one loop and higher loop evaluations of $a_e$ using the conventional asymptotic expansion method given by $a_e (\text{one loop}) = \alpha_{\text{QED}}/ 2\pi = 0.00116141$ and $a_e (\text{higher loop}) = 0.001159652..$ , respectively. The evaluation of $ a_e (\text{one loop})_{\text{converg. exp.}} = \kappa^{\alpha_{\text{ext}}}_{23} \alpha_{\text{QED}}/ 2\pi $ at the zero point of its third derivative with respect to $\alpha$ gives significantly more accurate results compared to the conventional one loop evaluation of $a_e$ given by $a_e (\text{one loop}) = \alpha_{\text{QED}}/ 2\pi$. $\beta_1 = 1$, $\beta_2 = \alpha$. }
\begin{tabular}{| K{7cm} | K{2cm} | K{2cm} | K{2cm} | }
 \hline
 $k$ & 1 & 2 & 3 \\
 \Xhline{4\arrayrulewidth}
  $ { d^k \kappa^{\alpha_{\text{ext}}}_{23} / d \alpha^k } = 0$ ; $\alpha_{\text{ext}}$ & 1.0002 & 1.00006 & $ 1.01396 $ \\
 \hline
$a_e (\text{one loop})_{\text{converg. exp.}} = \kappa^{\alpha_{\text{ext}}}_{23} \alpha_{\text{QED}}/ 2\pi $ & .00116141 & .00116141 &  \cellcolor{gray!30} .00115974 \\
 \hline
$ a_e (\text{one loop}) = \alpha_{\text{QED}}/ 2\pi$ & .00116141 & .00116141 & .00116141 \\
 \hline
$ a_e (\text{higher loop}) $ & .00115965 & .00115965 & \cellcolor{gray!30} .00115965 \\
 \hline
\end{tabular}
\label{tab:4}
\end{table}
\begin{table}[h]
\centering
\caption{Comparison of the one loop evaluation of $a_e = (g-2)/2$ using the convergent expansion method given by $ \kappa^{\alpha_{\text{ext}}}_{23} \alpha_{\text{QED}}/ 2\pi $ with the one loop and higher loop evaluations of $a_e$ using the conventional asymptotic expansion method given by $a_e (\text{one loop}) = \alpha_{\text{QED}}/ 2\pi = 0.00116141$ and $a_e (\text{higher loop}) = 0.001159652..$ , respectively. The evaluation of $ a_e (\text{one loop})_{\text{converg. exp.}} = \kappa^{\alpha_{\text{ext}}}_{23} \alpha_{\text{QED}}/ 2\pi $ at the zero point of its third derivative with respect to $\alpha$ gives significantly more accurate results compared to the conventional one loop evaluation of $a_e$ given by $a_e (\text{one loop}) = \alpha_{\text{QED}}/ 2\pi$. $\beta_1 = \alpha$, $\beta_2 = \alpha$.}
\begin{tabular}{| K{7cm} | K{2cm} | K{2cm} | K{2cm} | }
 \hline
 $k$ & 1 & 2 & 3 \\
 \Xhline{4\arrayrulewidth}
  $ { d^k \kappa^{\alpha_{\text{ext}}}_{23} / d \alpha^k } = 0$ ; $\alpha_{\text{ext}}$ & 1.01 & 1.0036 & $ 1.0393 $ \\
 \hline
$a_e (\text{one loop})_{\text{converg. exp.}} = \kappa^{\alpha_{\text{ext}}}_{23} \alpha_{\text{QED}}/ 2\pi $ & .00116152 & .00116145 &  \cellcolor{gray!30} .00115908 \\
 \hline
$ a_e (\text{one loop}) = \alpha_{\text{QED}}/ 2\pi$ & .00116141 & .00116141 & .00116141 \\
 \hline
$ a_e (\text{higher loop}) $ & .00115965 & .00115965 & \cellcolor{gray!30} .00115965 \\
 \hline
\end{tabular}
\label{tab:5}
\end{table}

In applying the convergent $\epsilon$ expansion to QED, when expanding the contribution to $a_e$ ($ = (g-2)/2$) to third order in $\epsilon$ ($= 1$) we obtain the coefficient $ \kappa^{\alpha}_{23}$ which multiplies $\alpha_{\text{QED}}/ 2\pi$, evaluating $ \kappa^{\alpha}_{23} \alpha_{\text{QED}}/ 2\pi$ at the zero point of its third derivative with respect to $\alpha$ produces the results listed in Tables \ref{tab:4} and \ref{tab:5} which clearly show an improvement in the $a_e$ evaluation compared to the conventional one loop evaluation given by $a_e (\text{one loop}) = \alpha_{\text{QED}}/ 2\pi$. To introduce free parameters in the action of QED space-time is rescaled $x^{\mu} \rightarrow x^{\mu} / \alpha $, $A^{\mu} \rightarrow \beta_1 A^{\mu}$ and $\psi \rightarrow \beta_2 \psi$. Tables \ref{tab:4} and \ref{tab:5} show results for different choices of $\beta_1$ and $\beta_2$ in terms of $\alpha$. The results of Tables \ref{tab:4} and \ref{tab:5} are well motivated in Subsection \ref{sec:4.2.2}.
\\
\\
\indent We expect future prospects of this work to be profound. For example currently the main strong coupling formalism known is the lattice formalism which breaks Lorentz invariance however the convergent expansion method developed in this work is an expansion in the continuum theory which preserves the symmetries of the theory and has a good convergence rate at strong coupling.\footnote{ It would be more accurate to say at relative strong coupling. Note that the strong coupling regime of the usual quantum mechanical and quantum field theoretical theories studied is a relative strong coupling regime since usually higher powers of the coupling is assigned to terms with the highest power of the position variable or fields which their expansion is well defined. For example consider the double well potential in quantum mechanics expanded about one of its vacuums: $V(x) = x^2 - 2 \sqrt{\lambda} x^3 + \lambda x^4 $. The term with the highest power of the coupling is $\lambda x^4$ which its expansion is well defined therefore at strong coupling we obtain a better convergence rate when applying the convergent expansion method. Refer to the first footnote in the Introduction Section for a definition of relative strong coupling.}. With noting the applicability of the convergent expansion method to the strong coupling regime of one dimensional perturbed Gaussian integrals and quantum mechanical systems, as studied in this work, we speculate that it will also be possible to study the strong coupling regime of quantum field theories in the continuum using this method. It should also be noted that the applications of this method are not limited to quantum mechanics or quantum field theory. For example in Section \ref{sec:3.1} we apply this method to modify the previously known asymptotic perturbative expansion of the time-independent Schrodinger (differential) equation to a convergent expansion, hence it is clear that this method can also have broader applications in the context of differential equations.

\section{Convergent expansion of one dimensional perturbed Gaussian integrals}
\label{sec:2}
In this Section we develop a convergent expansion for perturbed one dimensional Gaussian integrals. We start from one dimensional integrals to illustrate how the method works in a simpler context. Later we will consider quantum mechanical and quantum field theoretical examples. Consider the following one dimensional integral
\begin{equation}
\label{eq:2.1}
I(\lambda, \kappa) = \int^{+\infty}_{-\infty} dx \exp \left( - { \kappa \over 2} x^2 - \lambda x^4 \right), \ \ \ \ \lambda  > 0, \ \ \kappa \in \{-1, 0 ,+1 \}
\end{equation}
when $\kappa$ is one, \eqref{eq:2.1} can be expanded in the coupling $\lambda$ which results in an asymptotic series in $\lambda$. It is known that this asymptotic series is Borel summable to the full result of the integral \cite{05}. When ordinarily summing the terms of this asymptotic series the smaller $\lambda$ is the more terms in the expansion can be used to obtain a better approximation to the quantity of interest but after a certain order the summation starts to diverge. To obtain a convergent expansion, introduce a free parameter in the integral by rescaling the variable $x \rightarrow {x  / \alpha}$, next add and subtract a quadratic term in the exponent of the integrand. We have
\begin{equation}
\label{eq:2.2}
I(\lambda, \kappa) \! = \! I^{\alpha}(\lambda, \kappa) \! \equiv \! {1 \over \alpha } \! \int^{+\infty \alpha}_{-\infty \alpha} \!\!  dx \exp \! \left( \! - x^2 \! - \epsilon \left\{ \! \left( -1 + \! {\kappa \over 2 \alpha^2 } \right) x^2 + {\lambda \over \alpha^4} x^4 \right\} \! \right) \! = \! \sum^{n}_{i=0} I^{\alpha}_i(\lambda , \kappa) \epsilon^i ,
\end{equation}
in general we allow a complex value for $\alpha = \alpha_1 + i \alpha_2$ with $|\alpha_1| > |\alpha_2|$ in order to have a damping Gaussian integral along the path of integration of \eqref{eq:2.2}. Also we consider $\alpha$ values with a positive real part $\alpha_1 > 0$. We have introduced an $\epsilon$ parameter in the exponent to indicate that the expansion is to be taken with respect to this parameter. Later $\epsilon$ is set to one. In this expansion, similar to as mentioned above, the smaller ${\lambda \over \alpha^4}$ is the more terms in the expansion become useful and more accurate results can be obtained \footnote{When $\kappa = -1$ the expansion of the quadratic term in the $\epsilon$ expansion becomes a divergent expansion since $|-1 + {\kappa \over 2 \alpha^2 } | > 1$ for $\kappa = -1$, in this case other than $\lambda / \alpha^4$ the term ${\kappa \over 2 \alpha^2 }$ should also become smaller in order for more terms in the expansion to become useful, this clearly happens since when choosing a large value of $\alpha$ both $\lambda / \alpha^4$ and ${\kappa \over 2 \alpha^2 }$ are suppressed.}. Since the integral of \eqref{eq:2.2}, for $\epsilon = 1$, is independent of $\alpha$ we have the freedom to choose an arbitrary large value of $\alpha$ and suppress $\lambda /\alpha^4$ (and $\kappa /\alpha^2$ for when $\kappa = -1$), in this case more terms in the $\epsilon$ expansion become useful, therefore more accurate results can be obtained. Note that it is important that we no longer expand in $\lambda$ (e.g. when $\kappa =1$) as this will result in a cancellation of the $\alpha$ parameter and the previous asymptotic expansion is recovered. Although the integral of \eqref{eq:2.2} is independent of $\alpha$, any finite order of its $\epsilon \ ( = 1)$ expansion depends on $\alpha$. Choosing $\alpha$ too large or too small will clearly not give an optimum result at any finite order, therefore we expect intermediate values of $\alpha$ to exist that would give optimum results. Since the full integral $I^{\alpha} (\lambda, \kappa)$ is independent of $\alpha$ we have $d^m I^{\alpha} (\lambda, \kappa) / d \alpha^m = 0 $ for $m = 1, 2, ...$, therefore it seems that the zero $m$-derivative points of the partial sum $I^{\alpha,n} (\lambda , \kappa) = \sum^{n}_{i=0} I^{\alpha}_i(\lambda , \kappa) $ obtained by solving the equation $d^m I^{\alpha, n} (\lambda, \kappa) / d \alpha^m = 0$, would be good candidates for the values of $\alpha$ that would give optimum results for $I^{\alpha,n} (\lambda , \kappa)$ \footnote{For the integral of \eqref{eq:2.2} (and in most other examples) we are mainly concerned with the zero first derivative points of $I^{\alpha, n}$, but in certain examples we consider later in the paper in Subsection \ref{sec:4.2} the zero points of the higher derivatives will also find applications.}. We will see that (a subset of) the extremum values of $I^{\alpha,n} (\lambda , \kappa)$, obtained by evaluating $I^{\alpha,n} (\lambda , \kappa)$ at its first derivative zero points with respect to $\alpha$, form sequences that converge to the exact result. We verify these statements for the integral of \eqref{eq:2.2} in what follows.

Expanding \eqref{eq:2.2} in $\epsilon \ (= 1)$ we obtain
\begin{equation}
\label{eq:2.3}
\begin{split}
 I^{\alpha}(\lambda, \kappa) = \sum^{\infty}_{n=0} I^{\alpha}_n(\lambda, \kappa) = {\sqrt{\pi} \over \alpha } \sum^{\infty}_{n=0} \sum^{n}_{k=0} & \bigg{\{} {(-1)^n \over k! (n-k)!} {(4n - 2k -1)!! \over 2^{2n-k}} \times \\ & \left( -1 + {\kappa \over 2\alpha^2} \right)^k \left( {\lambda \over \alpha^4 } \right)^{n-k} \bigg{\}}
\end{split}
\end{equation}
with $(-1) !! \equiv 1$. Tables \ref{tab:6} and \ref{tab:7} summarize the results of evaluating $I^{\alpha, n}(\lambda, \kappa) \equiv \sum^{n}_{i=0} I^{\alpha}_i (\lambda, \kappa)$ at its extremas with respect to $\alpha$ for different $n$ and for $\kappa = 1,-1$. The sequence of extremum points $\alpha_{\text{ext}}$ in these Tables are obtained by solving the equation $d I^{\alpha, n} / d\alpha = 0$ and using method i) of Appendix \ref{sec:B2}. Here we only obtain one sequence of extremum points based on this method which we have shown in Tables \ref{tab:6} and \ref{tab:7}.
\begin{table}[h]
\centering
\caption{Numerical values of $I^{\alpha,n} (\lambda , 1) = \sum^{n}_{i=0} I^{\alpha}_i(\lambda , 1) $, with $I^{\alpha}_i(\lambda , 1)$ given by relation \eqref{eq:2.3}, at its extremum points with respect to $\alpha$ for different $n$.}
 \begin{tabular}{| p{0.8cm} | c |} 
 \hline
 $\ \ \lambda$ & \begin{tabular}{K{1cm} | K{1cm} | K{2.5cm} | K{1.2cm} | K{2.8cm} | K{1.9cm} | K{2.5cm}}
 $n$ &  1 & 2 & 3 & 4 & 15 & $I (\lambda , 1)$ of \eqref{eq:2.1}
 \end{tabular} \\
 \Xhline{4\arrayrulewidth}
 $ \ 0.1$ & \begin{tabular}{ K{1cm} | K{1cm} | K{2.5cm} | K{1.2cm} | K{2.8cm} | K{1.9cm} |  K{2.5cm} }
 $\alpha_{\text{ext}}$ & 0.90 & 0.96 $\pm$ 0.07 i & 1.005 & 1.045 $\mp$ .052 i & 1.31 & - \\
 \hline
 $I^{\alpha_{\text{ext}}, n}$ & 2.12 & 2.15 $\mp$ .009 i & 2.1475 & 2.1494 $\pm$ 0.00 i & 2.14970591 & 2.14970593
 \end{tabular} \\
 \Xhline{4\arrayrulewidth}
 $\ \ 1$ & \begin{tabular}{ K{1cm} | K{1cm} | K{2.5cm} | K{1.2cm} | K{2.8cm} | K{1.9cm} |  K{2.5cm} }
 $\alpha_{\text{ext}}$ & 1.36 & 1.49 $\pm$ 0.16 i & 1.59 & 1.68 $\pm$ 0.11 i  & 2.20 & - \\
 \hline
 $I^{\alpha_{\text{ext}},n}$ & 1.49 & 1.54 $\mp$ 0.02 i & 1.55 & 1.553 $\mp$ .003 i & 1.5548173 & 1.5548178
 \end{tabular} \\
 \Xhline{4\arrayrulewidth}
 $ \  10$ & \begin{tabular}{ K{1cm}  | K{1cm} | K{2.5cm} | K{1.2cm} | K{2.8cm} | K{1.9cm} | K{2.5cm} }
 $\alpha_{\text{ext}}$ & 2.29 & 2.54 $\mp$ 0.3 i & 2.73 & 2.88 $\mp$ 0.2 i & 3.84  & - \\
 \hline
 $I^{\alpha_{\text{ext}},n}$ & 0.91 & 0.952 $\pm$ 0.02 i & 0.959 & 0.965 $\pm$ .003 i & 0.9679593 & 0.9679602
 \end{tabular} \\
  \Xhline{4\arrayrulewidth}
 $ \ 10^{10}$ & \begin{tabular}{ K{1cm}  | K{1cm} | K{2.5cm} | K{1.2cm} | K{2.8cm} | K{1.9cm} | K{2.5cm} }
 $\alpha_{\text{ext}}$ & 397 & 443 $\mp$ 55 i & 476 & 505 $\mp$ 35 i & 677 & - \\
 \hline
 $I^{\alpha_{\text{ext}},n}$ & .0053 & .0056 $\pm$ .00 i & .00567 & .00571 $\pm$ .00 i & .005732574 & $.005732583$
 \end{tabular} \\
 \hline
\end{tabular}
\label{tab:6}
\end{table}
\begin{table}[h]
\centering
\caption{Numerical values of $I^{\alpha,n} (\lambda , -1) = \sum^{n}_{i=0} I^{\alpha}_i(\lambda , -1) $, with $I^{\alpha}_i(\lambda , -1)$ given by relation \eqref{eq:2.3}, at its extremum points with respect to $\alpha$ for different $n$.}
 \begin{tabular}{| p{0.8cm} | c |} 
 \hline
 $\ \ \lambda$ & \begin{tabular}{K{1cm} | K{1cm} | K{2.5cm} | K{1cm} | K{2.5cm} | K{2cm} | K{2.8cm}}
 $n$ &  1 & 2 & 3 & 4 & 15 & $I (\lambda , -1)$ of \eqref{eq:2.1}
 \end{tabular} \\
 \Xhline{4\arrayrulewidth}
 $ \ 0.1$ & \begin{tabular}{ K{1cm} | K{1cm} | K{2.5cm} | K{1cm} | K{2.5cm} |  K{2cm} |  K{2.8cm} }
 $\alpha_{\text{ext}}$ & 0.56 & 0.65 $\mp$ 0.11 i & 0.71 & 0.77 $\mp$ 0.07 i & 1.106 &- \\
 \hline
 $I^{\alpha_{\text{ext}},n}$ & 4.86 & 5.72 $\pm$ 0.56 i & 6.19 & 6.52 $\pm$ 0.21 i & 6.9502 & 6.9516
 \end{tabular} \\
 \Xhline{4\arrayrulewidth}
 $\ \ 1$ & \begin{tabular}{ K{1cm} | K{1cm} | K{2.5cm} | K{1cm} | K{2.5cm} | K{2cm} | K{2.8cm} }
 $\alpha_{\text{ext}}$ & 1.16 & 1.32 $\mp$ 0.18 i & 1.43 & 1.52 $\mp$ 0.12 i & 2.08 & - \\
 \hline
 $I^{\alpha_{\text{ext}},n}$ & 1.94 & 2.09 $\pm$ 0.08 i & 2.13 & 2.165 $\pm$ 0.02 i & 2.18708 & 2.18710
 \end{tabular} \\
 \Xhline{4\arrayrulewidth}
 $ \ 10$ & \begin{tabular}{ K{1cm}  | K{1cm} | K{2.5cm} | K{1cm} | K{2.5cm} |   K{2cm} | K{2.8cm} }
 $\alpha_{\text{ext}}$ & 2.18 & 2.44 $\mp$ 0.31 i & 2.63 & 2.79 $\mp$ 0.2 i & 3.16 & - \\
 \hline
 $I^{\alpha_{\text{ext}},n}$ & 0.99 & 1.05 $\pm$ 0.03 i & 1.06 & 1.072 $\pm$ .006 i & 1.077255 & 1.077258
 \end{tabular} \\
 \Xhline{4\arrayrulewidth}
 $\ 10^{10}$ & \begin{tabular}{ K{1cm}  | K{1cm} | K{2.5cm} | K{1cm} | K{2.5cm} |  K{2cm} | K{2.8cm} }
 $\alpha_{\text{ext}}$ & 398 & 443 $\mp$ 55 i & 476 & 505 $\mp$ 35 i & 677 & - \\
 \hline
 $I^{\alpha_{\text{ext}},n}$ & .0053 & .0056 $\pm$ .00 i & .00567 & .00571 $\pm$ .00 i & .005732593 & .005732602
 \end{tabular} \\
 \hline
\end{tabular}
\label{tab:7}
\end{table}

A few points are in order:

- The extremum points $\alpha_{\text{ext}}$ form an increasing sequence as a function of $n$. This is expected since in going to higher orders of the expansion we expect to obtain more accurate results and this happens when the coupling is suppressed more \footnote{We will later encounter cases which there are more than one sequence of increasing $\alpha_{\text{ext}}$, with the other ones appearing at higher orders in the expansion. These 2nd, 3rd, etc sequences of $\alpha_{\text{ext}}$, labeled by $\alpha_{\text{ext}, 2}$, $\alpha_{\text{ext}, 3}$, etc have values smaller than the 1st sequence and their $I^{\alpha_{\text{ext}, 2} , n }$, $I^{\alpha_{\text{ext}, 3} , n }$, etc sequences converge faster to the quantity of interest compared to $I^{\alpha_{\text{ext}, 1} , n }$. The discussion in the previous paragraph refers to only one of these sequences. For examples which we obtain many sequences of extremum points refer to Appendix \ref{appen:B} or \ref{appen:C} and for a discussion on two main methods of identifying these sequences of extremum points refer to Appendix \ref{sec:B2}.}. As $n \rightarrow \infty$ we expect this sequence to increase unboundedly since otherwise for any bounded value of $\alpha$ the expansion will diverge after a certain order.

- At even values of $n$ there are no real extremum points for $\alpha$ therefore we have included the complex extremum point with a real part greater than the previous extremum point and smaller than the extremum point after and with the smallest absolute value of the imaginary part possible. In this case we obtain a complex value for $I^{\alpha_{\text{ext}}, n}$ which the real part can be considered, the imaginary part of $I^{\alpha_{\text{ext}},n}$ is not of interest to us and it should eventually go to zero if higher terms in the expansion are summed with the same complex extremum point, since the integral of \eqref{eq:2.2} is real and should be independent of $\alpha$.

- When $\kappa = -1$ from Table \ref{tab:7} it can be seen that the convergence rate becomes worse at (relative) weak coupling. The reason is that in this case the expansion of the quadratic term in the $\epsilon$ expansion is a divergent expansion (note that $| -1 + {\kappa / 2 \alpha^2 }| > 1$ for $\kappa = -1$), therefore it would need the help of the quartic term to form a convergent expansion. When $\lambda $ is small the quartic term cannot provide enough compensation therefore it becomes more difficult to obtain convergence. But Table \ref{tab:7} shows a good convergence rate at (relative) strong coupling and Table \ref{tab:6} shows good convergence rate for all positive values of the coupling $\lambda$.

- When the exponent of the integral of \eqref{eq:2.1} contains higher powers of $x$: $x^5$, $x^6$, ... it is more efficient to associate more powers of $\epsilon$ to these terms. A discussion on this matter is done in Appendix \ref{sec:B1}.

- For a multi-dimensional integral we have the freedom of rescaling each variable with an independent free parameter. In this case it is important to choose these parameters in a way as to obtain more efficient results. We will consider an example of a multi-dimensional integral in Subsection \ref{sec:4.2.2}.
\\
\\
\indent As mentioned the convergence rate in Table \ref{tab:7} becomes worse at weak coupling but the sequence of $I^{\alpha_{\text{ext}},n}(\lambda, -1)$ does converge to the full result of the integral, the reason for this as mentioned was that the expansion of the quadratic term in the $\epsilon$ expansion becomes a divergent expansion, in order to avoid this one might try expanding about one of the minimums of the negative of the exponent in relation \eqref{eq:2.1} for $\kappa = -1$ (e.g. $x_{min} = - 1 / (2\sqrt{\lambda} )$). But in this case although the expansion of the quadratic term in the $\epsilon$ expansion will turn into a convergent expansion as can be seen from \eqref{eq:2.5} but we will also obtain a new term $2 \sqrt{\lambda}x^3$ as can be seen from \eqref{eq:2.4}. As discussed earlier the expansion of the term $2 \sqrt{\lambda}x^3$ in \eqref{eq:2.4} is not a well defined expansion (by itself) but the expansion of the term $- \lambda x^4$ is well defined therefore at the (relative) weak coupling regime $\lambda \lessapprox 2 \sqrt{\lambda} $ the convergence rate will still become worse, furthermore in this case as we will see below the expansion will not show convergence to the full result of the integral. Here we will discuss this and try to find a way to obtain convergence to the full result. From \eqref{eq:2.1} for $\kappa = -1$ and $x \rightarrow x - 1 / (2\sqrt{\lambda} )$ we have
\begin{equation}
\label{eq:2.4}
I(\lambda, -1) = e^{{1 / (16 \lambda)}} \int^{+\infty}_{-\infty} dx \exp \left( -x^2 + 2 \sqrt{\lambda} x^3 - \lambda x^4 \right)
\end{equation}
to obtain a convergent expansion introduce an $\alpha$ parameter in \eqref{eq:2.4} by rescaling $x \rightarrow x/ \alpha$, next add and subtract a Gaussian term in the exponent of the integrand, we obtain
\begin{equation}
\label{eq:2.5}
I(\lambda, -1) \! = \! I^{\alpha}(\lambda, -1) \equiv {1 \over \alpha} e^{{1 / (16 \lambda)}} \! \! \int^{+\infty}_{-\infty} \! \! \! dx \exp \! \left( \! -x^2 \! + \epsilon \bigg{\{} x^2 \Big{(} 1 - {1 \over \alpha^2} \Big{)} \! + 2 {\sqrt{\lambda} \over \alpha^3 } x^3 - {\lambda \over \alpha^4} x^4 \bigg{\}} \! \right)
\end{equation}
after expanding \eqref{eq:2.5} in $\epsilon \ (=1)$ and evaluating the Gaussian integrals we obtain
\begin{equation}
\label{eq:2.6}
\hspace{-0.5cm} \bar{I}^{\alpha}(\lambda, -1) = {\sqrt{\pi}  \over \alpha} e^{{1 / (16 \lambda)}} \sum^{\infty}_{n=0} \underset{j \ \text{even}}{\underset{i+j+k  = n }{\sum^{n}_{i,j,k = 0}}} {1 \over i! j! k!} {(2i + 3j +4k -1)!! \over 2^{i + {3 \over 2}j +2k}}  \Big{(}1 - {1 \over \alpha^2} \Big{)}^i \Big{(} 2 {\sqrt{\lambda} \over \alpha^3}\Big{)}^j \Big{(}-{\lambda \over \alpha^4} \Big{)}^k
\end{equation}
note that the expansion of $I^{\alpha}(\lambda, -1)$ in \eqref{eq:2.5} in $\epsilon \ (=1)$, named $\bar{I}^{\alpha}(\lambda, -1)$, might not be equal to $I^{\alpha}(\lambda, -1)$. The results of evaluating the partial sum $\bar{I}^{\alpha, n}(\lambda, -1) \equiv \sum^{n}_{i=0} \bar{I}^{\alpha}_i(\lambda, -1)$, for $\bar{I}^{\alpha}(\lambda, -1) = \sum^{\infty}_{i=0} \bar{I}^{\alpha}_i(\lambda, -1)$ in \eqref{eq:2.6}, at its extremum points with respect to $\alpha$ are summarized in Table \ref{tab:8} which clearly does not show convergence to the exact result at weak coupling. Also the convergence rate becomes worse at weak coupling, for example consider the results of this Table for $\lambda = 1/100$ which are not showing convergence to a clear number, at least to the order studied.

\begin{table} [h]
\centering
\caption{Numerical values of $\bar{I}^{\alpha, n}(\lambda, -1) \equiv \sum^{n}_{i=0} \bar{I}^{\alpha}_i(\lambda, -1)$ (for $\bar{I}^{\alpha}(\lambda, -1) = \sum^{\infty}_{i=0} \bar{I}^{\alpha}_i(\lambda, -1)$ in \eqref{eq:2.6}) at its extremum points with respect to $\alpha$.}
 \begin{tabular}{| p{1.3cm} | c |} 
 \hline
  & \begin{tabular}{K{1cm} | K{1cm} | K{1.7cm} | K{1.7cm} | K{1cm} | K{4.5cm}}
 $n$ &  9 & 29 & 49 & 69 & $I (\lambda , -1)$ (of \eqref{eq:2.4} or \eqref{eq:2.1})
 \end{tabular} \\
 \Xhline{4\arrayrulewidth}
 $\lambda = {1 \over 100}$ & \begin{tabular}{ K{1cm} | K{1cm} | K{1.7cm} | K{1.7cm} | K{1cm} | K{4.5cm} }
 $\alpha_{\text{ext}}$ & 1.06  & 1.15 & 1.22 & 1.28 & - \\
 \hline
 $\bar{I}^{\alpha_{\text{ext}}, n}$ & 951.3 & 953.9 & 955.2 & 957.3 & 1907.5
 \end{tabular} \\
 \Xhline{4\arrayrulewidth}
 $\lambda = {1 \over 10}$ & \begin{tabular}{ K{1cm} | K{1cm} | K{1.7cm} | K{1.7cm} | K{1cm} | K{4.5cm} }
 $\alpha_{\text{ext}}$ & 1.34 & 1.63 & 1.80 & 1.93 & - \\
 \hline
 $\bar{I}^{\alpha_{\text{ext}},n}$ & 4.33 & 5.93 & 6.70 & 6.91 & 6.95
 \end{tabular} \\
 \Xhline{4\arrayrulewidth}
 $\lambda = 1$ & \begin{tabular}{ K{1cm} | K{1cm} | K{1.7cm} | K{1.7cm} | K{1cm} | K{4.5cm} }
 $\alpha_{\text{ext}}$ & 2.07  & 2.655  & 2.989 & - & - \\
 \hline
 $\bar{I}^{\alpha_{\text{ext}},n}$ & 2.03 & 2.185 & 2.18709 & - & 2.18710
 \end{tabular} \\
 \Xhline{4\arrayrulewidth}
 $\lambda = 10$ & \begin{tabular}{ K{1cm} | K{1cm} | K{1.7cm} | K{1.7cm} | K{1cm} | K{4.5cm} }
 $\alpha_{\text{ext}}$ & 3.50 & 4.57 & 5.18 &- & - \\
 \hline
 $\bar{I}^{\alpha_{\text{ext}},n}$ & 1.07 & 1.07725 & 1.0772581 & - & 1.0772581
 \end{tabular} \\
 \hline
\end{tabular}
\label{tab:8}
\end{table}
The reason for this is that the asymptotic perturbative expansion of \eqref{eq:2.4} in the coupling $\lambda$ does not capture the full result of the integral or in other words it is not Borel summable to the full result of the integral. In general when the perturbative expansion in the coupling does not capture the full result it is clear that we should not expect that its modified convergent expansion to converge to the full result. As mentioned in the literature, e.g. \cite{05}, the presence of other saddles of the exponent of the integrand of \eqref{eq:2.4} in the path of integration along the real axis can result in the non-Borel summability of the asymptotic expansion in the coupling $\lambda$ to the exact result. One solution to this problem is to deform the contour of integration into the complex plane along paths called steepest decent paths such that the exponent along these paths would be a monotonic function on either side of the point of expansion and in this case by a change of variables it is possible to show that the expansion along these steepest decent paths is Borel summable to the exact result. The full integral is then recovered by the sum of the integrals along the steepest decent paths. Another approach called exact perturbation theory (EPT) is to write the integral of \eqref{eq:2.4} in the following way
\begin{equation}
\label{eq:2.7}
I(\lambda, -1) = e^{{1 / (16 \lambda)}} \int^{+\infty}_{-\infty} dx \exp \left( -x^2 - \lambda x^4 \right) \exp \left( 2 {\lambda^{3/2}} x^3 / \lambda_0 \right)
\end{equation}
in this approach no deformation of the contour of the integral is needed and it can be shown that the expansion of \eqref{eq:2.7} in the coupling $\lambda$ is Borel summable to the exact result. To recover the original integral of \eqref{eq:2.4} $\lambda_0$ is set to $\lambda$ later \footnote{For more details regarding these two methods we refer the reader to \cite{05}. A first attempt was made to apply the convergent expansion to the EPT method as to obtain convergence to the exact value of the integral of \eqref{eq:2.4} (or \eqref{eq:2.7} for $\lambda_0 = \lambda$) in the weak coupling regime by associating more powers of $\epsilon$ to the $x^3$ term (e.g. $\epsilon^{3/2} 2 \lambda^{3/2} x^3 / \lambda_0$) which was not very successful and the result showed convergence to the previous value obtained from the convergent expansion of \eqref{eq:2.4}.}.

Here in order to obtain convergence to the full result of the integral of \eqref{eq:2.4} we discuss a different solution to the above problem. The main difficulty in evaluating such integrals is the presence of odd terms in the exponent. When the integral of \eqref{eq:2.5} is expanded in $\epsilon$ (or the integral of \eqref{eq:2.4} is expanded in the coupling $\lambda$) these odd terms give rise to odd integrals which vanish. This can be considered as some loss of information. When evaluating the integral directly we are summing over only positive quantities and hence there is no cancelation happening between the different regions of integration. Therefore if somehow the information of the odd terms is kept in the expansion we might expect that the result of the expansion would converge to the exact result. For this we can rewrite the integral in \eqref{eq:2.4} as follows
\begin{flalign}
\label{eq:2.8}
& I(\lambda, -1) =  I_1(\lambda, -1) + I_2(\lambda, -1) \nonumber  \\ 
& I_1(\lambda, -1) = {1 \over 2} e^{1 \over 16\lambda} \int^{+\infty}_{-\infty} dx \exp \left( -x^2 - 2 \sqrt{\lambda} |x^3| - \lambda x^4 \right)  \\ & I_2(\lambda, -1) = {1 \over 2} e^{1 \over 16\lambda} \int^{+\infty}_{-\infty} dx \exp \left( -x^2 + 2 \sqrt{\lambda} |x^3| - \lambda x^4 \right) \nonumber
\end{flalign}
now there is no odd term in the exponent. Therefore we can expand the integrals $I_1(\lambda, -1)$ and $I_2(\lambda, -1)$ separately and later sum them together to obtain the full result. Introducing an $\alpha$ parameter in the integrals of \eqref{eq:2.8} we obtain
\begin{equation}
\label{eq:2.9}
\begin{split}
& I_1(\lambda, -1) \! = \! I^{\alpha}_1(\lambda, -1) \! \equiv \! {1 \over \alpha} e^{1 \over 16\lambda} \! \int^{\infty}_0 \! \! dx \exp \left( \! -x^2 + \epsilon \left\{ x^2 \left( 1 - {1 \over \alpha^2} \right) - 2 {\sqrt{\lambda} \over \alpha^3 } x^3 - {\lambda \over \alpha^4} x^4 \right\} \! \right) \\ & I_2(\lambda, -1) \! = \! I^{\alpha}_2(\lambda, -1) \! \equiv \! {1 \over \alpha} e^{1 \over 16\lambda} \! \int^{\infty}_0 \! \! dx \exp \left( \! -x^2 + \epsilon \left\{ x^2 \left( 1 - {1 \over \alpha^2} \right) + 2 {\sqrt{\lambda} \over \alpha^3 } x^3 - {\lambda \over \alpha^4} x^4 \right\} \! \right)
\end{split}
\end{equation}
Expanding \eqref{eq:2.9} in $\epsilon \ (=1)$ we have
\begin{equation}
\label{eq:2.10}
\hspace{-0.5cm}\begin{split}
& I^{\alpha}_1(\lambda, -1) = {e^{1 \over 16\lambda} \over \alpha} \Bigg{\{} \sum^{\infty}_{n=0} \! \underset{i+j+k  = n }{\sum^{n}_{i,j,k = 0}} \! {1 \over i!j!k! 2} \Gamma \! \left( i + {{3}j \over 2} +2k + {1 \over 2} \right) \!  \left(1 - {1 \over \alpha^2} \right)^i \! \left(- {2 \sqrt{\lambda} \over \alpha^3} \right)^j \! \left( -{\lambda \over \alpha^4} \right)^k \! \Bigg{\}}  \\ & I^{\alpha}_2(\lambda, -1) = {e^{1 \over 16\lambda} \over \alpha} \Bigg{\{} \sum^{\infty}_{n=0} \underset{i+j+k  = n }{\sum^{n}_{i,j,k = 0}} \! {1 \over i!j!k! 2} \Gamma \left( i + {{3}j \over 2} +2k + {1 \over 2} \right) \!  \left(1 - {1 \over \alpha^2} \right)^i \! \left( {2 \sqrt{\lambda} \over \alpha^3}\right)^j \! \left(-{\lambda \over \alpha^4} \right)^k \! \Bigg{\}}
\end{split}
\end{equation}

The results of Table \ref{tab:8} show that the expansion of \eqref{eq:2.6} does not converge to the exact result at weak coupling whereas from Tables \ref{tab:9}-\ref{tab:12} it can be seen that the expansions of \eqref{eq:2.10} do converge to the exact result. In the next Section we will show how this new method can be applied to the path integral formalism, although the evaluation of the absolute value terms will not be easy in the path integral formalism but possible in principle.
\begin{table} [h]
\centering
\caption{Numerical values of $I^{\alpha, n}_1(\lambda, -1) \equiv \sum^n_{i = 0} I^{\alpha}_{1, i} (\lambda , -1)$ at the extremum points of $\alpha$, with $I^{\alpha}_{1, i}$ from relation \eqref{eq:2.9} given by $I^{\alpha}_1 (\lambda , -1) = \sum^{\infty}_{i = 0} I^{\alpha}_{1, i} (\lambda , -1) \epsilon^i $.}
 \begin{tabular}{| p{1.3cm} | c |} 
 \hline
  & \begin{tabular}{K{1cm} | K{1.5cm} | K{1.5cm} | K{1.5cm} | K{2cm} | K{2cm}}
 $n$ & 1 &  3 & 5 & 7 & Exact
 \end{tabular} \\
 \Xhline{4\arrayrulewidth}
 $\lambda = {1 \over 100}$ & \begin{tabular}{ K{1cm} | K{1.5cm} | K{1.5cm} | K{1.5cm} | K{2cm} | K{2cm} }
 $\alpha_{\text{ext}}$ & 1.13 & 1.18 & 1.21  & 1.24 & - \\
 \hline
 $I^{\alpha_{\text{ext}},n}_{1}$ & 416.7 & 417.45 & 417.455 & 417.455123 & 417.455125
 \end{tabular} \\
 \Xhline{4\arrayrulewidth}
 $\lambda = {1 \over 10}$ & \begin{tabular}{ K{1cm} | K{1.5cm} | K{1.5cm} | K{1.5cm} | K{2cm} | K{2cm} }
 $\alpha_{\text{ext}}$ & 1.356 & 1.466 & 1.543  & 1.604 & - \\
 \hline
 $I^{\alpha_{\text{ext}},n}_{1}$ & 1.298 & 1.308 & 1.30800 & 1.3080155 & 1.3080163
 \end{tabular} \\
 \hline
\end{tabular}
\label{tab:9}
\end{table}
\begin{table}[h]
\centering
\caption{Numerical values of $I^{\alpha, n}_2(\lambda, -1) \equiv \sum^n_{i = 0} I^{\alpha}_{2, i} (\lambda , -1)$ at the extremum points of $\alpha$, with $I^{\alpha}_{2, i}$ from relation \eqref{eq:2.9} given by $I^{\alpha}_2 (\lambda , -1) = \sum^{\infty}_{i = 0} I^{\alpha}_{2, i} (\lambda , -1) \epsilon^i$.}
 \begin{tabular}{| p{1.3cm} | c |} 
 \hline
  & \begin{tabular}{K{1cm} | K{1.5cm} | K{1.5cm} | K{1.5cm} | K{1.5cm} |K{1.5cm}}
 $n$ &  9 & 29 & 49 & 69 & Exact
 \end{tabular} \\
 \Xhline{4\arrayrulewidth}
 $\lambda = {1 \over 100}$ & \begin{tabular}{ K{1cm} | K{1.5cm} | K{1.5cm} | K{1.5cm} | K{1.5cm} | K{1.5cm} }
 $\alpha_{\text{ext}}$ & 0.340 & 0.433 & 0.591 & 0.661 & - \\
 \hline
 $I^{\alpha_{\text{ext}},n}_{2}$ & 861 & 1322 & 1469 & 1488.5 & 1490.0
 \end{tabular} \\
 \Xhline{4\arrayrulewidth}
 $\lambda = {1 \over 10}$ & \begin{tabular}{ K{1cm} | K{1.5cm} | K{1.5cm} | K{1.5cm} | K{1.5cm} | K{1.5cm} }
 $\alpha_{\text{ext}}$ & 0.74  & 1.053 & 1.235 & - & - \\
 \hline
 $I^{\alpha_{\text{ext}},n}_{2}$ & 5.4 & 5.643 & 5.643626 & - & 5.643628
 \end{tabular} \\
 \hline
\end{tabular}
\label{tab:10}
\end{table}
\begin{table}[h]
\centering
\caption{When $I^{\alpha}_{1}(1/100, -1)$ and $I^{\alpha}_{2}(1/100, -1)$ from \eqref{eq:2.10} are evaluated separately then summed we obtain convergence to the exact result $I(1/100, -1)$. }
 \begin{tabular}{| p{1.3cm} | c |}
 \hline
  & \begin{tabular}{K{1.5cm} | K{1.5cm} | K{3cm} | K{2cm}}
 $I^{\alpha_{\text{ext}}, 7}_{1}$ & $I^{\alpha_{\text{ext}}, 69}_{2}$ &  $I^{\alpha_{\text{ext}}, 7}_{1} + I^{\alpha_{\text{ext}},69}_{2}$ & $I(\lambda ,-1)$
 \end{tabular} \\
 \Xhline{4\arrayrulewidth}
 $\lambda = {1 \over 100}$ & \begin{tabular}{ K{1.5cm} | K{1.5cm} | K{3cm}  | K{2cm} }
 417.5 & 1488.5 & 1906 & 1907.5
 \end{tabular}  \\
 \hline
\end{tabular}
\label{tab:11}
\end{table}
\begin{table}[h]
\centering
\caption{When $I^{\alpha}_{1}(1/10, -1)$ and $I^{\alpha}_{2}(1/10, -1)$ from \eqref{eq:2.10} are evaluated separately then summed we obtain convergence to the exact result $I(1/10, -1)$.}
 \begin{tabular}{| p{1.3cm} | c |}
 \hline
  & \begin{tabular}{K{1.7cm} | K{1.5cm} | K{3cm} | K{2cm}}
 $I^{\alpha_{\text{ext}}, 7}_{1}$ & $I^{\alpha_{\text{ext}},49}_{2}$ &  $I^{\alpha_{\text{ext}},7}_{1} + I^{\alpha_{\text{ext}},49}_{2}$ & $I(\lambda ,-1)$
 \end{tabular} \\
 \Xhline{4\arrayrulewidth}
 $\lambda = {1 \over 10}$ & \begin{tabular}{ K{1.7cm} | K{1.5cm} | K{3cm} | K{2cm} }
 1.3080155 & 5.643626 & 6.951642 &  6.951645
 \end{tabular} \\
 \hline
\end{tabular}
\label{tab:12}
\end{table}
\\
At the end of this Section we discuss another way of writing the integral of \eqref{eq:2.2} which would resemble the original expression \eqref{eq:2.1} for $\kappa =1$ and in Section \ref{sec:4} a similar procedure will be applied to quantum field theory. Consider \eqref{eq:2.2} for $\kappa = 1$, after rescaling $x \rightarrow \sqrt{2} x  / \sqrt{ \gamma}$, we obtain
\begin{equation}
\label{eq:2.11}
I^{\alpha}(\lambda, 1) \equiv {\sqrt{2} \over \alpha \sqrt{ \gamma} }\int^{+\infty}_{-\infty} dx \exp \left( - x^2  - 4 \lambda' x^4 \right),  \ \ \ \lambda' \equiv { \epsilon \lambda \over \alpha^4 \gamma^2} , \ \ \gamma \equiv 2 - \epsilon \left( 2 - {1 \over \alpha^2} \right)
\end{equation}
setting $\epsilon = 1$ in the integral of \eqref{eq:2.11} gives $\lambda' \vert_{\epsilon =1} = \lambda$, $\alpha \sqrt{ \gamma} \vert_{\epsilon =1} =1$ and therefore the original expression of \eqref{eq:2.1} for $\kappa=1$ is recovered but now in order to obtain the convergent $\epsilon$ expansion of \eqref{eq:2.2} for $\kappa = 1$ to order $k$ we can first expand the integral of \eqref{eq:2.11} in $\lambda'$ to order $k$ then reorganize everything in the form of an $\epsilon$ expansion to that order, set $\epsilon$ to one and extremize with respect to the free parameter $\alpha$ introduced in the expansion. It is clear that the factor of ${1 \over \alpha \sqrt{ \gamma}} $ in \eqref{eq:2.11} came from rescaling $x$ and it carries information regarding some actual terms of the expansion of \eqref{eq:2.2} in $\epsilon$, therefore it cannot be eliminated or raised to a certain power simply due to the fact that $\alpha \sqrt{ \gamma} \vert_{\epsilon =1} = 1$. If so we will be losing information regarding some terms of the expansion of \eqref{eq:2.2} and will not obtain efficient results as Table \ref{tab:13} clearly shows this. Eliminating the factor of ${1 \over \alpha \sqrt{ \gamma} }$ in \eqref{eq:2.11} and expanding in powers of $\lambda'$ we obtain the following expression
\begin{equation}
\label{eq:2.12}
I^{\alpha}_{\text{triv.}}(\lambda, 1) = \sqrt{2 \pi} \sum^{\infty}_{n=0} {(-1)^n \over n!} {(4n -1)!! \over 2^{2n}} (4 \lambda')^n
\end{equation}
with $\lambda'$ given by \eqref{eq:2.11}. Reorganizing the expansion in \eqref{eq:2.12} in the form of an $\epsilon$ expansion, setting $\epsilon$ to 1 and extremizing with respect to $\alpha$ we obtain the results of Table \ref{tab:13} for $\lambda = 1$. From Table \ref{tab:6} at $n = 4$ the difference between $I^{\alpha_{\text{ext}},n}$ and $I(1, 1)$ is of order $\sim 0.001$ but at order $n = 22$ the difference between $I^{\alpha_{\text{ext}}, n}_{\text{triv.}}$ and $I(1, 1)$ is of order $\sim 0.01$, therefore it is clear that the evaluation becomes less efficient in \eqref{eq:2.12}. We refer to free parameter insertions in perturbative expansions in the coupling such as \eqref{eq:2.12} as trivial insertions of a free parameter since in this insertion no use has been made of the original expression which gave rise to this perturbative series. In other words the free parameter insertion of \eqref{eq:2.12} did not genuinely come from rescaling the variables in the theory as in \eqref{eq:2.11} and it corresponds to merely replacing $\lambda \rightarrow \lambda'$ in the perturbative expansion of \eqref{eq:2.1} in $\lambda$ for $\kappa = 1$ which could have been done for any other perturbative series in $\lambda$.
\begin{table}[h]
\centering
\caption{Numerical values of $I^{\alpha, n}_{\text{triv.}} (\lambda , 1) \equiv \sum^{n}_{i=0} I^{\alpha}_{\text{triv.},i} (\lambda , 1) $ (for $I^{\alpha}_{\text{triv.}} (\lambda , 1) = \sum^{\infty}_{i=0} I^{\alpha}_{\text{triv.},i} (\lambda , 1)$ in relation \eqref{eq:2.12}) at its extremum points with respect to $\alpha$ for $\lambda = 1$.}
 \begin{tabular}{| p{1.3cm} | c |} 
 \hline
 & \begin{tabular}{K{1.2cm} | K{1.2cm} | K{1.2cm} | K{1.2cm} | K{1.2cm}}
 $n$ & 6 & 14 & 22 & $I (\lambda , 1)$
 \end{tabular} \\
 \Xhline{4\arrayrulewidth}
 $\lambda = 1$ & \begin{tabular}{ K{1.2cm} | K{1.2cm}  | K{1.2cm}  | K{1.2cm} |  K{1.2cm} }
 $\alpha_{\text{ext}}$ & 1.71 & 2.02 & 2.22 & - \\
 \hline
 $I^{\alpha_{\text{ext}, n}}_{\text{triv.}}$ & 1.64 & 1.577 & 1.564 & 1.5548
 \end{tabular} \\
 \Xhline{4\arrayrulewidth}
 \hline
 \end{tabular}
 \label{tab:13}
 \end{table}

Therefore given an asymptotic perturbative series in a coupling $\lambda$, it is possible to insert free parameters and the parameter of expansion $\epsilon$ into the perturbative series trivially in many different ways, one simple way is to make a replacement $\lambda \rightarrow \lambda \epsilon / \{ \alpha^2 (1 - \epsilon (1 - 1/\alpha^2)) \}$, then we can rearrange the expansion at each finite order in the form of an expansion in $\epsilon$ to that order, set $\epsilon$ to one and extremize with respect to the free parameter $\alpha$, similar to the discussion related to relation \eqref{eq:2.12}. In this case it is likely that we will obtain a sequence that converges to the result of the previous asymptotic expansion. But with knowing the original expression that gave birth to the asymptotic perturbative series it is possible to insert the free parameters and the parameter of expansion $\epsilon$ into the expansion by rescaling the variables available in the theory, similar to relation \eqref{eq:2.11}, which in this case the convergent expansion in $\epsilon$ obtained would be a lot more efficient compared to when the free parameters and the parameter of expansion $\epsilon$ are inserted in an arbitrary different way into the perturbative series. In the next two Sections we will apply this simple procedure to quantum mechanics and quantum field theory and show how to insert free parameters and the parameter of expansion $\epsilon$ into the theory by rescaling the variables available in the theory.

\section{A convergent formulation for nonrelativistic quantum mechanics}
\label{sec:3}

In this Section we formulate a convergent expansion for nonrelativistic quantum mechanics. In Subsection \ref{sec:3.1} we develop a convergent expansion for the time independent Schrodinger equation and solve for the eigenenergies and eigenfunctions of quantum mechanical systems, in particular we solve for the numerical values of the eigenenergies of the anharmonic, pure anharmonic and the double well oscillator and compare them with the known values in the literature. In Subsection \ref{sec:3.2} we develop a convergent expansion for the path integral formalism.

\subsection{Convergent expansion of the time-independent Schrodinger equation}
\label{sec:3.1}
Consider the time-independent Schrodinger equation for a potential $V(x)$
\begin{equation}
\label{eq:3.1}
\left\{ -{\hbar^2 \over 2 m} {d^2 \over dx^2} + V(x) \right\} \psi = E \psi 
\end{equation}
assume the potential is bounded from below and expandable in powers of $x$, $V(x) = \kappa {1 \over 2} m \omega^2 x^2 + b_3x^3 + b_4x^4 + ... $ for $\kappa \in \{ 0, -1, +1 \}$ and $m, \omega > 0$. To introduce the free parameter $\alpha$ into the equation rescale the variable $x \rightarrow {\sqrt{\hbar} \over \sqrt{ m \omega} } {x \over \alpha }$ (now $x$ is dimensionless) and divide the whole equation by $\alpha^2 \hbar \omega$ \footnote{The general procedure for introducing free parameters into the theory is to first rescale the variables available in the theory then add and subtract a term which we want to perform the expansion about, here being the harmonic oscillator Hamiltonian. In the time independent Schrodinger equation \eqref{eq:3.1} there are two variables to rescale, one is $x$ and the other is the wave function or simply the freedom to divide both sides of the equation by an arbitrary parameter, therefore we can divide the whole equation by $\beta \hbar \omega$ instead of $\alpha^2 \hbar \omega$, after doing this, adding and subtracting a harmonic oscillator Hamiltonian and introducing the parameter of expansion $\epsilon$ we obtain
\begin{equation}
\label{eq:3.2}
\bigg{(} -{1 \over 2} {d^2 \over dx^2} + {\bar{\omega}^2 \over 2} x^2 + \epsilon \Big{\{} {1 \over 2} \Big{(} 1 - { \alpha^2 \over \beta} \Big{)} {d^2 \over dx^2} - {\bar{\omega}^2 \over 2} x^2 \Big{\}} + {1 \over \beta \hbar \omega} V_{\epsilon} \Big{(} {\sqrt{\hbar} \over \sqrt{ m \omega} } {x \over \alpha } \Big{)} \bigg{)} \psi = {E \over \beta \hbar \omega} \psi 
\end{equation}
in order to recast \eqref{eq:3.2} to a more convenient form divide both sides of the equation by $\bar{\gamma} = 1 - \epsilon (1 - \alpha^2 / \beta)$ we obtain
\begin{equation}
\label{eq:3.3}
\bigg{(} -{1 \over 2} {d^2 \over dx^2} + {\widetilde{\omega}^2 \over 2} x^2 + {1 \over \bar{\gamma} \beta \hbar \omega} V_{\epsilon} \Big{(} {\sqrt{\hbar} \over \sqrt{ m \omega} } {x \over \alpha } \Big{)} \bigg{)} \psi = {E \over \bar{\gamma} \beta \hbar \omega} \psi 
\end{equation}
with $\widetilde{\omega}^2 = (\bar{\omega}^2 - \epsilon \bar{\omega}^2 ) / \bar{\gamma}$, now we can solve \eqref{eq:3.3} perturbatively in $\epsilon$ and later reorganize the expansion in terms of an $\epsilon$ expansion to a certain order (note that $\bar{\gamma}$ also carries an $\epsilon$ dependence). In general the different independent free parameters introduced in the theory can be chosen in a way as to obtain more efficient results or to simplify our expressions. Here we have set $\beta = \alpha^2$ in \eqref{eq:3.2} or \eqref{eq:3.3} which leads to \eqref{eq:3.4} and simplifies the expression.}, next add and subtract in the brackets the term $\bar{\omega}^2 x^2/2$, after introducing the parameter of expansion $\epsilon$ we obtain
\begin{equation}
\label{eq:3.4}
\left\{ -{1 \over 2} {d^2 \over dx^2} + {\bar{\omega}^2 \over 2} x^2 + \bar{V}^{\alpha}_{\epsilon} (x) \right\} \psi = {E \over \alpha^2 \hbar \omega} \psi , \ \ \ \bar{V}^{\alpha}_{\epsilon} (x) \equiv - \epsilon {\bar{\omega}^2 \over 2} x^2 + {1 \over \alpha^2 \hbar \omega} V_{\epsilon} \left(  { \sqrt{\hbar} \over \sqrt{m \omega} } {x \over \alpha } \right)
\end{equation}
$\bar{\omega}$ can be chosen as convenient, for the quantum mechanical examples studied in this Subsection we will set it to one but here we will keep it general. $V_{\epsilon}(x)$ corresponds to the $\epsilon$ introduced potential. As a reminder we can no longer expand in the couplings of the theory as this will make the free parameter $\alpha$ to cancel in the expansion and the previous asymptotic expansion is recovered. In general the introduction of the $\epsilon$ parameter should be in a way as to give more efficient results. As noted in the previous Section it is more efficient to include more powers of $\epsilon$ for terms with higher powers of $x$. Based on the discussion in Appendix \ref{sec:B1} we introduce the parameter of expansion $\epsilon$ in the following way
\begin{equation}
\label{eq:3.5}
V_{\epsilon} (x) \equiv \epsilon \{ b_2 x^2 + b_3 x^3 + b_4 x^4 \} + \epsilon^2 \{ b_5 x^5 + ... + b_8 x^8  \} + \epsilon^3 \{ b_9 x^9 + ... + b_{12} x^{12}  \} + ...
\end{equation}
with $b_2 = \kappa {1 \over 2} m \omega^2$, and $\bar{V}^{\alpha}_{\epsilon} (x)$ given by
\begin{equation}
\label{eq:3.6}
\bar{V}^{\alpha}_{\epsilon} (x) \equiv \epsilon \{ \bar{b}^{\alpha}_2 x^2 + \bar{b}^{\alpha}_3 x^3 + \bar{b}^{\alpha}_4 x^4 \} + \epsilon^2 \{ \bar{b}^{\alpha}_5 x^5 + ... + \bar{b}^{\alpha}_8 x^8  \} + \epsilon^3 \{ \bar{b}^{\alpha}_9 x^9 + ... + \bar{b}^{\alpha}_{12} x^{12}  \} + ...
\end{equation}
and the coefficients $\bar{b}^{\alpha}_2, \bar{b}^{\alpha}_3, ... $ are given by
\begin{equation}
\label{eq:3.7}
\bar{b}^{\alpha}_2 = -{\bar{\omega}^2 \over 2} + {\kappa \over 2 \alpha^4}, \ \ \bar{b}^{\alpha}_k = {1 \over \alpha^{k +2} } \left( {\hbar \over m \omega } \right)^{k/2} {b_k \over  \hbar \omega } , \ \ \  k \geq 3
\end{equation}
 after evaluating the expansion to a certain order in $\epsilon$, $\epsilon$ is set to one.
To solve for the eigenenergies and eigenfunctions perturbatively in $\epsilon$ we use the Bender-Wu recursion relation \cite{06} which is an efficient method for solving for the eigenenergies and eigenfunctions of quantum mechanical systems and has been worked out in more detail by Sulejmanpasic and Unsal in \cite{07}. Here we will rederive the recursion relation as to make it suitable for the introduction of a free parameter and the expansion in powers of $\epsilon$. We write the wave function as $\psi(x) =u(x) \exp(- \bar{\omega} x^2/2)$ and insert it in \eqref{eq:3.4} to obtain
\begin{equation}
\label{eq:3.8}
-u''(x) + 2 \bar{\omega} x u'(x) + 2\bar{V}^{\alpha}_{\epsilon} (x) u(x) = 2 \bar{E} u(x) , \ \ \ \bar{E} \equiv { E \over \alpha^2 \hbar \omega} - {\bar{\omega} \over 2}
\end{equation}
to solve for the eigenenergies and eigenfunctions perturbatively in $\epsilon$ expand $u(x) = u_0(x) + \epsilon u_1(x) + ...$ and $\bar{E} = \bar{E}_0 + \bar{E}_1 \epsilon + ... $ in powers of $\epsilon$. After inserting in \eqref{eq:3.8} and equating powers of $\epsilon$ we obtain
\begin{equation}
\label{eq:3.9}
-u''_l(x) + 2\bar{\omega} x u'_l(x) +2 v_{s}(x) u_{l-s}(x) = 2 \bar{E}_{s} u_{l-s}(x) ,\ \ \  l = 0 ,1, ... , \ \ \  s = 0, ..., l
\end{equation}
$v_{n}(x) = 0 $ for $n < 1$ and summation over $s$ is implicit. We have replaced the potential in \eqref{eq:3.8} with the following general form $\bar{V}^{\alpha}_{\epsilon} (x) \rightarrow \widetilde{V}(x) =\sum^{\infty}_{n=1} \epsilon^n v_n(x) $ in \eqref{eq:3.9} for convenience, with $v_n(x)$ a polynomial function in $x$
\begin{equation}
\label{eq:3.10}
v_n (x) = C^n_{m_n} x^{m_n} + ... + C^n_{M_n} x^{M_n}, \ \ \  2 \leq m_n \leq M_n
\end{equation}
relation \eqref{eq:3.9} for $l = 0$ reduces to
\begin{equation}
\label{eq:3.11}
-u''_0(x) + 2 \bar{\omega} x u'_0(x) = 2 \bar{E}_{0} u_0(x)
\end{equation}
which has the solution $u_0(x) = H_{\nu} ( \sqrt{\bar{\omega}} x ) $ and $\bar{E}_{0} = \bar{\omega} \nu$ for $\nu = 0 ,1 , ...$ with $H_{\nu} (x)$ the Hermite polynomial functions \footnote{Here we refer to the Hermite polynomial functions that satisfy the Hermite differential equation: $-H''_{\nu}(x) + 2 x H'_{\nu}(x) = 2 \nu H_{\nu}(x)$ for $\nu = 0, 1, 2, ...$ . }. This gives $E_0 = \alpha^2 \hbar \omega \bar{\omega} ( \nu + 1/2)$. Note that the zero order energy in the $\epsilon$ expansion picks up an $\alpha$ and $\bar{\omega}$ dependence which should not bother us since the expansion term also includes a term proportional to $x^2$ that carries an $\alpha$ and $\bar{\omega}$ dependence.
Next write $u_l(x) = \sum^{K_l}_{k=0} A^k_l x^k$, with $K_l$ (refer to Appendix \ref{sec:A1} for a derivation) given by the following relation
\begin{equation}
\label{eq:3.12}
K_l = \nu + \max \{ a_1M_1 + a_2M_2 +  ... + a_lM_l | a_1 + 2a_2 +  ... + la_l = l, a_i \in \mathbb{N} \cup \{ 0 \}, i = 1, ..., l \}
\end{equation}
and $A^k_l = 0$ for $k < 0$ or $k > K_l$, after inserting $u_l(x)$ in \eqref{eq:3.9} and equating powers of $x$ we obtain
\begin{equation}
\label{eq:3.13}
\hspace{-0.5cm} -A^{k+2}_l (k+2)(k+1) + 2 \bar{\omega} (k - \nu) A^{k}_l = 2 \sum^{l}_{s = 1} \big{\{} \bar{E}_{s} A^{k}_{l-s} - C^{s}_{t_{s}} A^{k - t_{s}}_{l-s} \big{\}} ,  k = 0 , ..., K_l , t_{s} = m_{s} , ..., M_{s}
\end{equation}
summation over $t_s$ is implicit. To solve for $\bar{E}_{l}$ and $A^{k}_{l}$ assume $A^k_{q}$ for $k = 0, ..., K_{q}$ and $\bar{E}_{q}$ are known for $q < l$ and try to solve for $A^k_{l}$ for $k = 0, ..., K_{l}$ and $\bar{E}_{l}$. Similar to \cite{07} for a normalization of the wave functions we take $A^{\nu}_{0} = 1$ and $A^{\nu}_{q} = 0$ for $ q > 0$. First set $k = \nu$ and solve for $\bar{E}_{l}$ for $l > 0$, we obtain
\begin{equation}
\label{eq:3.14}
2\bar{E}_{l} = -A^{\nu+2}_l (\nu+2)(\nu+1) + 2 \sum^{l}_{s = 1} \left\{ C^{s}_{t_{s}} A^{\nu - t_{s}}_{l-s} \right\}
\end{equation}
In order to obtain $\bar{E}_{l}$, $A^{\nu+2}_l$ should be known. For this solve for $A^{k}_l$ for $k >\nu$ from \eqref{eq:3.13}
\begin{equation}
\label{eq:3.15}
 A^{k}_l = {1 \over 2 \bar{\omega} (k - \nu)} \left\{ A^{k+2}_l (k+2)(k+1) + 2 \sum^{l-1}_{s = 1} \bar{E}_{s} A^{k}_{l-s} - 2 \sum^{l}_{s = 1} C^{s}_{t_s} A^{k - t_s}_{l-s}  \right\}
\end{equation}
Note that $A^{k}_0 = 0$ for $k > \nu$ since the Hermite polynomial functions $H_{\nu}$ are of degree $\nu$ and therefore the term $\bar{E}_{l} A^{k}_{0}$ vanishes. From \eqref{eq:3.15} solve for $A^{k}_l$ for $k > \nu$ by starting from $k = K_l$ down to $k = \nu +1$. Note that $A^{k}_l = 0 $ for $k > K_l$ therefore when starting from $k = K_l$ all the quantities on the righthand side of \eqref{eq:3.15} are known by assumption. From here $A^{\nu+2}_l$ is determined and it can be used to solve for $\bar{E}_{l}$ in relation \eqref{eq:3.14}. Once $\bar{E}_{l}$ is known we can use \eqref{eq:3.13} again to solve for the remaining coefficients $A^{k}_l$ for $k < \nu$
\begin{equation}
\label{eq:3.16}
 A^{k}_l = {1 \over 2 \bar{\omega} (k - \nu)} \left\{ A^{k+2}_l (k+2)(k+1) + 2 \sum^{l}_{s = 1} \bar{E}_{s} A^{k}_{l-s} - 2 \sum^{l}_{s = 1} C^{s}_{t_{s}} A^{k - t_{s}}_{l-s} \right\}
\end{equation}

From \eqref{eq:3.8} the partial sum of the eigenenergy $E = \sum^{\infty}_{l=0} E_l \epsilon^l$ for $\epsilon =1$ defined by $E^{\alpha , n} \equiv \sum^{n}_{l=0} E_l $ is given by
\begin{equation}
\label{eq:3.17}
E^{\alpha , n} = \alpha^2 \hbar \omega \sum^{n}_{l=0} \bar{E}_l + {\bar{\omega} \alpha^2 \hbar \omega / 2}
\end{equation}
and the coefficients $A^k$ of the eigenfunction $\psi (x) = \exp (- \bar{\omega} x^2) \sum^{\infty}_{k=0} A^k x^k$ are given by $A^k = \sum^{\infty}_{l=0} A^k_l$. Similar to \eqref{eq:3.17} we can form the partial sum of these coefficients $A^{k,n} \equiv \sum^{n}_{l=0} A^k_l$ and evaluate their extremum values with respect to the free parameter $\alpha$ introduced at order $n$ to obtain a convergent sequence.

Next we apply the above formalism to quantum mechanical systems and solve for their eigenenergies using the convergent expansion. We consider the anharmonic, pure anharmonic and double well potentials $V_{\kappa}(x) = \kappa m \omega^2 x^2/2 + \lambda x^4$ for $\kappa = +1, 0, -1$, $m, \omega > 0$. The corresponding $\alpha$ and $\epsilon$ introduced potential is
\begin{equation}
\label{eq:3.18}
\begin{split}
V_{\kappa}(x) = \kappa m \omega^2 x^2/2 + \lambda x^4 & \longrightarrow \bar{V}^{\alpha}_{\kappa ,\epsilon} (x) = \epsilon \{ \bar{b}^{\alpha}_2 x^2 + \bar{b}^{\alpha}_4 x^4 \}, \\ & \text{with} \ \ \ \bar{b}^{\alpha}_2 = -{1 / 2} + {\kappa / (2 \alpha^4)}, \ \bar{b}^{\alpha}_4 = {\bar{\lambda} / \alpha^6}, \ \bar{\lambda} = {\hbar \lambda / (m^2 \omega^3)}
\end{split}
\end{equation}
and we have set $\bar{\omega} = 1$. From relation \eqref{eq:3.12} it can be verified that $K_l = \nu + 4l$ with $l$ the order of the $\epsilon$ expansion. For this potential relation \eqref{eq:3.14} reduces to
\begin{equation}
\label{eq:3.19}
2\bar{E}_{l} = -A^{\nu+2}_l (\nu+2)(\nu+1) + \Big{(} {\kappa \over \alpha^4 } - 1 \Big{)} A^{\nu - 2}_{l-1} + {2 \bar{\lambda} \over \alpha^6} A^{\nu - 4}_{l-1}
\end{equation}
also for \eqref{eq:3.15} for $k > \nu$ we have 
\begin{equation}
\label{eq:3.20}
 A^{k}_l = {1 \over 2 (k - \nu)} \bigg{\{} A^{k+2}_l (k+2)(k+1) + 2 \sum^{l-1}_{s = 1} \{ \bar{E}_{s} A^{k}_{l-s}\} - \Big{(} {\kappa \over \alpha^4 } - 1 \Big{)} A^{k - 2}_{l-1} - {2 \bar{\lambda} \over \alpha^6} A^{k - 4}_{l-1} \bigg{\}} 
\end{equation}
and \eqref{eq:3.16} to be solved for $k < \nu$ reduces to
\begin{equation}
\label{eq:3.21}
 A^{k}_l = {1 \over 2  (k - \nu)} \bigg{\{} A^{k+2}_l (k+2)(k+1) + 2 \sum^{l}_{s = 1} \{ \bar{E}_{s} A^{k}_{l-s}\} - \Big{(} {\kappa \over \alpha^4 } - 1 \Big{)} A^{k - 2}_{l-1} - {2 \bar{\lambda} \over \alpha^6} A^{k - 4}_{l-1} \bigg{\}}
\end{equation}

The coefficients $A^{k}_0$ for $k = 0, ..., K_0 \ (= \nu)$ correspond to the Hermite polynomial coefficients and are known \footnote{Based on our convention for the normalization of the wave function $A^{\nu}_0 = 1$, $\nu = 0, 1, 2, ...$ therefore the first few Hermite polynomials in our normalization conventions are $H_0 = 1$, $H_1 = x$, $H_2 = x^2 - {1 \over 2}$, $H_3 = x^3 - {3 \over 2} x$, etc. }, also $\bar{E}_0 = \nu$. Therefore starting from $q=1$, with following the procedure outlined in the previous page, we can solve for $A^{k}_q$ for $k = 4q + \nu, ..., 0 $ and $\bar{E}_q$ up to any desired order $q = n$.

Relation \eqref{eq:3.17} can be used to solve for the energy levels of these quantum mechanical systems. Evaluating $E^{\alpha, n}$ at its extremum points with respect to $\alpha$ gives a sequence that converges to the desired energy level. The results of this analysis for the eigenenergies of the anharmonic, pure anharmonic and double well potentials are summarized in Appendix \ref{appen:C} and are compared with the known values in the literature. As can be seen from the Tables of Appendix \ref{appen:C} the results obtained using this method show good convergence for all positive values of the coupling for the anharmonic oscillator, similarly for the pure anharmonic oscillator good convergence rate is obtained for all positive values of the coupling but for the double well potential it becomes more difficult to obtain convergence at weak coupling due to a similar reason mentioned for the one dimensional integrals in the previous Section that when $\kappa = -1$ the expansion of the quadratic term in the $\epsilon$ expansion is not well defined and it becomes a divergent expansion \footnote{$|\bar{b}^{\alpha}_2 | > 1/2$ in relation \eqref{eq:3.18} when $\kappa = -1$ and therefore it becomes greater than the coefficient of $x^2$ in the term ${\bar{\omega}^2 \over 2} x^2$ in \eqref{eq:3.4} for $\bar{\omega} = 1$.} and it would need the help of the quartic term to form a convergent expansion, when $\lambda$ is small the quartic term cannot provide enough compensation therefore it becomes more difficult to obtain convergence\footnote{ Note that for the pure anharmonic oscillator although the expansion of the quadratic term in the $\epsilon$ expansion is not well defined (by itself) but the convergence rate at weak coupling is good (this has been verified in numerical calculations, also refer to the first footnote in the Introduction Section for a similar discussion in the context of one dimensional integrals).}. If we expand about one of the vacuums of the double well potential in order to avoid having a divergent expansion for the quadratic term in the $\epsilon$ expansion, the perturbative expansion, as it is well known, will not converge to the full result but will only converge to the perturbative part of the vacuum energy and will not capture the contribution of the instantons. In this case the potential involves an odd power of $x$ and there will be some loss of information happening in the expansion due to the presence of this odd term similar to the discussion related to Table \ref{tab:8} of Section \ref{sec:2}. But when we expand about the symmetrical point of the double well potential the perturbative expansion converges to the full result, since in this case the potential only involves even terms and therefore there is no loss of information which could have been the case if there were odd terms present in the potential (refer to Section \ref{sec:2} for a discussion on this matter in the simpler context of one dimensional integrals). To see this more explicitly consider the following double well potential $V_{dw}(x) = {\lambda \over 2} (x^2 - {1 \over 4 \lambda})^2 $, expanding this potential about one of its vacuums by replacing $x \rightarrow x - 1 /(2 \sqrt{\lambda})$ we obtain $V_{dw}(x) = {1 \over 2} x^2 - \sqrt{\lambda} x^3 + {\lambda \over 2} x^4 $. For this potential $\kappa$, $m$ and $\omega$ are set to one and from \eqref{eq:3.5}, $b_2 = 1/2$, $b_3 = -\sqrt{\lambda}$ and $b_4 = \lambda /2$, therefore after setting $\hbar$ and $\bar{\omega}$ to one the corresponding $\alpha$ and $\epsilon$ introduced potential of \eqref{eq:3.6} becomes
\begin{equation}
\label{eq:3.21'}
\begin{split}
 V_{dw}(x) = {1 \over 2} x^2 - \sqrt{\lambda} x^3 + {\lambda \over 2} x^4 & \longrightarrow \bar{V}^{\alpha}_{dw, \epsilon} (x) = \epsilon \{ \bar{b}^{\alpha}_2 x^2 + \bar{b}^{\alpha}_3 x^3 + \bar{b}^{\alpha}_4 x^4 \}, \\ & \ \text{with} \ \ \bar{b}^{\alpha}_2 = - {1 \over 2} + {1 \over 2 \alpha^4}, \bar{b}^{\alpha}_3 = - {\sqrt{\lambda} \over \alpha^5}, \bar{b}^{\alpha}_4 =  {\lambda \over 2 \alpha^6}
\end{split}
\end{equation}
now in \eqref{eq:3.21'} $|\bar{b}^{\alpha}_2 | < 1/2$ therefore the expansion of the quadratic term in the $\epsilon$ expansion will be a convergent expansion. Using the potential of \eqref{eq:3.21'} and with following similar steps sketched after relation \eqref{eq:3.18} and using \eqref{eq:3.17} we can solve for the perturbative part of the vacuum energy of the double well potential. The result of this analysis is shown in Table \ref{tab:31'} of Appendix \ref{appen:C} for $\lambda = 3/100$ and compared with the exact value of the vacuum energy from \cite{16}. From this Table it can be seen that the result does not capture the contribution of the instantons and converges only to the perturbative part of the vacuum energy, as expected. Also note that for $\lambda = 3/100$, $\lambda /2 < |-\sqrt{\lambda}|$ therefore we are in the (relative) weak coupling regime and the convergence rate becomes less efficient.

Therefore with applying the convergent expansion method to the time independent Schrodinger equation we can obtain accurate results for the eigenenergies\footnote{By a similar procedure we expect to be able to obtain the coefficients $A^k$ of the wave function by evaluating the extremum values of the partial sum $A^{k,n} \equiv \sum^{n}_{l=0} A^k_l$ and obtaining a sequence that converges to $A^k$.} of the anharmonic oscillator and pure anharomonic oscillator problems for all positive values of the coupling but for the double well potential problem accurate results that capture the full result can only be obtained at strong coupling, since at weak coupling convergence becomes difficult when expanding about the symmetrical point of the double well potential. Another point to consider is that due to the rescaling of $x$ by a free parameter in relation \eqref{eq:3.4} and the modified expansion in the $\epsilon$ parameter the coefficients $A^k$ of the wave function obtained using the convergent expansion method need not be the same as when these coefficients are obtained using the conventional asymptotic expansion in $\lambda$, but we do expect the full wave functions to be equivalent and related in a nontrivial way in terms of their $x$ dependence. We illustrate this point by the following example
\begin{equation}
f(x) = \exp(-x^2) \sum^{\infty}_{j=0} c_j x^j \xrightarrow{\, x \rightarrow x'/\alpha \,} g(x') \equiv f(x'/\alpha) = \exp \left(- x'^2 + \epsilon(x'^2 -x'^2/\alpha^2) \right) \sum^{\infty}_{j=0} c_j x'^j/\alpha^j  \nonumber
\end{equation}
expanding the above exponent in the $\epsilon$ $(=1)$ parameter we obtain the function below
\begin{equation}
g(x') = \exp(-x'^2) \sum^{\infty}_{j=0} c'_j x'^j \nonumber
\end{equation}
it is clear that $c_j$ and $c'_j$ are different coefficients but $g(x')$ and $f(x)$ are related in a trivial way by $g(x') = f(x' / \alpha)$. Similarly we expect the wave functions obtained using the convergent expansion method and the conventional asymptotic expansion method to be equivalent but related in a nontrivial way in terms of their functionality on $x$.

\subsection{Convergent expansion of the path integral formalism}
\label{sec:3.2}
In this Subsection we apply the convergent expansion method to the path integral formalism. Consider the path integral representation of a transition amplitude in quantum mechanics in Euclidean space:
\begin{equation}
\label{eq:3.22}
\langle q_f;t_f| q_i;t_i \rangle =  \langle q_f| \; e^{ - {1 \over \hbar} \hat{H} \Delta t} \; | q_i \rangle \; = N \int \mathcal{D} [q] \; \exp \left( {- {1 \over \hbar} } \int_{t_i}^{t_f} L \; dt \right) \; , \; \; \Delta t = t_f -t_i
\end{equation}
$\hat{H}$ is the Hamiltonian, $L = {1 \over 2} m \dot{q}^2 + V(q)$, $q(t_i) = q_i$ and $q(t_f) = q_f$. Similar to the previous Subsection we take the potential $V(q) = \kappa {1 \over 2} m \omega^2 q^2 + b_3q^3 + b_4q^4 + ...$ with $\kappa = 0, \pm 1$ and $m, \omega > 0$. The formalism can equivalently be applied to Minkowski space.

For a harmonic oscillator Lagrangian $L_{\text{ho}} = {1 \over 2} m \dot{q}^2 + {1 \over 2} m \omega^2 \beta^2_1 q^2$ with zero boundary conditions we have \cite{11} :
\begin{equation}
\label{eq:3.23}
N \int \mathcal{D} [q] \; \exp \left( {- {1 \over \hbar} } \int_{t_i}^{t_f} L_{\text{ho}} \; dt \right) = \left( {m \omega \beta_1 \over 2 \pi \hbar \sinh \omega \beta_1 \Delta t} \right)^{1/2}
\end{equation}
Here $\beta_1$ is a free parameter. Dividing and multiplying \eqref{eq:3.22} by $Z_{\text{ho}}$ we have:
\begin{equation}
\label{eq:3.24}
\langle q_f;t_f| q_i;t_i \rangle = \left( {m \omega \beta_1 \over 2 \pi \hbar \sinh \omega \beta_1 \Delta t} \right)^{1/2} {Z \over Z_{\text{ho}}}
\end{equation}
with $Z_{\text{ho}} \equiv \int \mathcal{D} [q] \; \exp \left( {- {1 \over \hbar} } \int_{t_i}^{t_f} L_{\text{ho}} \; dt \right)$ and $Z = \int \mathcal{D} [q] \; \exp \left( {- {1 \over \hbar} } \int_{t_i}^{t_f} L \; dt \right)$. The boundary conditions for $Z_{\text{ho}}$ are $q(t_i) = q(t_f) = 0$. Expanding the action $S = \int_{t_i}^{t_f} L \; dt$ about the saddle point configuration, $q \rightarrow q_{\text{cl}} + q $, we have $S = S_{\text{cl}} + \int^{t_f}_{t_i} dt \big{\{} {m \over 2} \dot{q}^2 + {1 \over 2}m \omega^2 \beta^2_1 q^2  + {1 \over 2} \big{(} V'' (q_{\text{cl}})$ $- m \omega^2 \beta^2_1 \big{)} q^2 + {1 \over 3!} V''' (q_{\text{cl}}) q^3 + ... \big{\}} $. To introduce free parameters into the path integral rescale $t \rightarrow t / \beta_2$, $q \rightarrow q / \alpha$ in $Z$ and $Z_{\text{ho}}$, add and subtract $(m \dot{q}^2/2 + m \omega^2 q^2/2 )$ in $L$ and $L_{\text{ho}}$ \footnote{In general one might choose to add and subtract a term $(m\dot{q}^2/2 + m\omega^2 \beta^2_3 q^2/2 )$ with $\beta_3$ an arbitrary free parameter. For simplicity we have set $\beta_3 = 1$. In general the introduction of those free parameters will be more useful which lead to the suppression of the couplings.}, we obtain 
\begin{equation}
\label{eq:3.25}
S^{\alpha} = \int_{\beta_2 t_i}^{\beta_2 t_f} \bigg{\{} {m \over 2} \dot{q}^2 + {m \over 2} \omega^2 q^2 + \widetilde{V}^{\alpha}_{\epsilon}(q; q_{\text{cl}}) \bigg{\}} dt
\end{equation}
\begin{equation}
\label{eq:3.26}
S^{\alpha}_{\text{ho}} = \int_{\beta_2 t_i}^{\beta_2 t_f} \bigg{\{} {m \over 2} \dot{q}^2 + {m \over 2} \omega^2 q^2 + \epsilon \Big{\{} \Big{(} {m \beta_2 \over 2 \alpha^2} - {m \over 2} \Big{)} \dot{q}^2 + \Big{(} {m \omega^2 \beta^2_1 \over 2 \beta_2 \alpha^2} - {m \over 2} \omega^2 \Big{)} q^2 \Big{\}} \bigg{\}} dt
\end{equation}
with $\widetilde{V}^{\alpha}_{\epsilon}(q; q_{\text{cl}})$ given by
\begin{flalign}
\label{eq:3.27}
\widetilde{V}^{\alpha}_{\epsilon}(q; q_{\text{cl}}) = & \epsilon { m \over 2} \Big{(} {\beta_2 \over \alpha^2} - 1 \Big{)} \dot{q}^2 + \epsilon {m \omega^2 \over 2 } \Big{(} { \beta^2_1 \over \beta_2 \alpha^2} - 1 \Big{)} q^2 + {\epsilon \over 2 \beta_2 \alpha^2} \left( V'' (q_{\text{cl}}(t / \beta_2) ) - m \omega^2 \beta^2_1 \right) q^2 \nonumber \\ & + {\epsilon \over 3! \beta_2 \alpha^3} V''' (q_{\text{cl}}(t/\beta_2) ) q^3 + ...
\end{flalign}
where we have also introduced the parameter of expansion $\epsilon$ \footnote{This has to be done in a way as to give more efficient results. One way is based on the prescription of Appendix \ref{sec:B1} which the $\epsilon$ parameter is introduced in the following way: $\epsilon^n q^p$ for $4(n-1) < p \leq 4n$.} into the expression for the potential in \eqref{eq:3.27} and in \eqref{eq:3.26}. To rewrite the expression in a more convenient form rescale $q \rightarrow q/ \eta $, with $\eta = (1 - \epsilon(1 - \beta_2 /\alpha^2) )^{1/2}$ we have 
\begin{equation}
\label{eq:3.28}
S'  = \int_{\beta_2 t_i}^{\beta_2 t_f} \bigg{\{} {m \over 2} \dot{q}^2 + {m \over 2} \omega'^2 q^2 + \bar{V}^{\alpha}_{\epsilon}(q ; q_{\text{cl}}) \bigg{\}} dt
\end{equation}
\begin{equation}
\label{eq:3.29}
S'_{\text{ho}} = \int_{\beta_2 t_i}^{\beta_2 t_f} \bigg{\{} {m \over 2} \dot{q}^2 + {m \over 2} \omega'^2 q^2 \bigg{\}} dt
\end{equation}
with $\omega'^2 = \omega^2 \big{(} 1 - \epsilon ( 1 - {\beta^2_1 \over \beta_2 \alpha^2}) \big{)} / \eta^2 $ and $\bar{V}^{\alpha}_{\epsilon}(q ; q_{\text{cl}})$ given by 
\begin{equation}
\label{eq:3.30}
\bar{V}^{\alpha}_{\epsilon}(q ; q_{\text{cl}}) = \epsilon {1/\eta^2 \over 2 \beta_2 \alpha^2} \left( V'' (q_{\text{cl}}(t / \beta_2) ) - m \omega^2 \beta^2_1 \right) q^2 + \epsilon {1/\eta^3 \over 3! \beta_2 \alpha^3} V''' (q_{\text{cl}}(t/\beta_2) ) q^3 \! + ... 
\end{equation}
the boundary conditions for \eqref{eq:3.28} and \eqref{eq:3.29} are $q(\beta_2 t_f ) = q(\beta_2 t_i ) = 0$. Hence the convergent $\epsilon$ expansion of the transition amplitude \eqref{eq:3.24} takes the following form
\begin{equation}
\label{eq:3.31}
\langle q_f;t_f| q_i;t_i \rangle = \langle q_f;t_f| q_i;t_i \rangle^{\alpha} \equiv \left( {m \omega \beta_1 \over 2 \pi \hbar \sinh \omega \beta_1 \Delta t} \right)^{1/2} {Z' \over Z'_{\text{ho}}}
\end{equation}
with $Z'_{\text{ho}} \equiv \int \mathcal{D} [q] \; \exp \left( - {S'_{\text{ho}} / \hbar} \right)$ and $Z' = \int \mathcal{D} [q] \; \exp \left( {- {S' / \hbar} } \right)$, $S'$ and $S'_{\text{ho}}$ given by \eqref{eq:3.28} and \eqref{eq:3.29}, respectively. Therefore \eqref{eq:3.31} can be expanded in $\epsilon$ to a certain order and reorganized in terms of an $\epsilon$ expansion to that order (note that $\omega'$ also carries an $\epsilon$ dependence) and extremized with respect to the free parameter(s) introduced in the expansion after setting $\epsilon$ to one. In general the independent free parameters introduced in the theory can be chosen in a way as to obtain more efficient results or to simplify our expressions. One convenient choice that simplifies the expansion considerably is to set $\beta_1 = \beta_2 = \alpha^2$, in this case $\eta = 1$ and $\omega' = \omega$.

Next we illustrate how we can turn the expansion of \eqref{eq:3.31} into an expansion that we expect to converge to the full result, similar to the method which was discussed for the one dimensional integral in the previous Section. In expanding \eqref{eq:3.31} in $\epsilon$ any term in the expansion that has an odd power of $q$ would vanish due to an odd integral. In performing a direct evaluation of the path integral, e.g. performing a lattice calculation, there is no cancellation happening for the odd terms between the different regions of the integration space since the exponential of a real number is always positive. When expanding these odd terms in a perturbative series this cancellation happens, this can be considered as some loss of information and results in the fact that the perturbative expansion does not converge to the exact result. With retaining the information in these odd terms we expect to obtain an expansion that would converge to the exact result. Although the evaluation of these odd terms, which involve an absolute value, will not be easy in general, but this method involves an interesting message and that is the fact that it is possible to have perturbative expansions in the couplings or the $\epsilon$ parameter (here being the two perturbative $\epsilon$ expansions of $Z'_1$ and $Z'_2$ in \eqref{eq:3.33}) that would capture the full result, in analogy with the discussion of the one dimensional integrals in the previous Section. Consider a general partition function $Z = \int \mathcal{D} [q] \; \exp \left( - S[q]  \right)$. $S[q]$ can be written in terms of an odd part plus an even part, $S[q] = S_e[q] + S_o[q]$, $S_e[q] = (S[q] + S[-q])/2 $ and $S_o[q] = (S[q] - S[-q])/2 $. There are regions of the path integration space (denoted by $ \mathcal{D} [q]_+$) which the quantity $S_o[q] > 0$ and there are regions (denoted by $ \mathcal{D} [q]_-$) which $S_o[q] < 0$. These regions have the same measure and are related to each other by $q(t) \rightarrow - q(t)$, on the other hand the even part of the action is unchanged under $q(t) \rightarrow - q(t)$.

Therefore we have
\begin{flalign}
\label{eq:3.32}
& \int \mathcal{D} [q] \; \exp \left( - S[q]  \right) = \int \mathcal{D} [q]_+ \; \exp \left( - S[q]  \right) + \int \mathcal{D} [q]_- \; \exp \left( - S[q]  \right) \nonumber \\ & = \int \mathcal{D} [q]_+ \; \exp \left( - S_e[q] - \left| S_o[q] \right| \right) + \int \mathcal{D} [q]_- \; \exp \left( - S_e[q] + \left| S_o[q] \right| \right) \nonumber \\ & = {1 \over 2} \int \mathcal{D} [q] \; \exp \left( - S_e[q] - \left| S_o[q] \right| \right) + {1 \over 2} \int \mathcal{D} [q] \; \exp \left( - S_e[q] + \left| S_o[q] \right| \right)
\end{flalign}
using \eqref{eq:3.32}, \eqref{eq:3.31} can be written in the following form
\begin{equation}
\label{eq:3.33}
\langle q_f;t_f| q_i;t_i \rangle^{\alpha} = \left( {m \omega \beta_1 \over 2 \pi \hbar \sinh \omega \beta_1 \Delta t} \right)^{1/2} \left( {Z'_1 \over Z'_{\text{ho}}} + {Z'_2 \over Z'_{\text{ho}}} \right)
\end{equation}
with $Z'_1$ and $Z'_2$ given by
\begin{equation}
\label{eq:3.34}
\begin{split}
& Z'_1 \! = \! {1 \over 2} \! \int \! \mathcal{D} [q] \exp \left( \! - {1 \over \hbar} \int_{\beta_2 t_i}^{\beta_2 t_f} \! \! \bigg{\{} {m \over 2} \dot{q}^2 \! + \! {m \over 2} \omega'^2 q^2 + \bar{V}^{\alpha}_{\epsilon, e}(q ; q_{\text{cl}}) \bigg{\}} dt - { \epsilon \over \hbar} \left| \int_{\beta_2 t_i}^{\beta_2 t_f} \! {1 \over \epsilon} \bar{V}^{\alpha}_{\epsilon, o}(q ; q_{\text{cl}}) dt \right| \right)
 \\ & Z'_2 \! = \! {1 \over 2} \! \int \! \mathcal{D} [q] \exp \left( \! - {1 \over \hbar} \int_{\beta_2 t_i}^{\beta_2 t_f} \! \! \bigg{\{} {m \over 2} \dot{q}^2 \! + \! {m \over 2} \omega'^2 q^2 + \bar{V}^{\alpha}_{\epsilon, e}(q ; q_{\text{cl}}) \bigg{\}} dt + { \epsilon \over \hbar } \left| \int_{\beta_2 t_i}^{\beta_2 t_f} \! {1 \over \epsilon} \bar{V}^{\alpha}_{\epsilon, o}(q ; q_{\text{cl}}) dt \right| \right)
\end{split}
\end{equation}
$\bar{V}^{\alpha}_{\epsilon, e}(q ; q_{\text{cl}}) = \{ \bar{V}^{\alpha}_{\epsilon}(q ; q_{\text{cl}}) + \bar{V}^{\alpha}_{\epsilon}(-q ; q_{\text{cl}}) \} /2$ and $\bar{V}^{\alpha}_{\epsilon, o}(q ; q_{\text{cl}}) = \{ \bar{V}^{\alpha}_{\epsilon}(q ; q_{\text{cl}}) - \bar{V}^{\alpha}_{\epsilon}(-q ; q_{\text{cl}}) \} /2$. In general the evaluation of the absolute value terms in \eqref{eq:3.34} are not easy (but possible in principle) but \eqref{eq:3.33} when expanded in $\epsilon$, in analogy with the one dimensional integrals of \eqref{eq:2.8} and \eqref{eq:2.9} and for when $Z'_1$ and $Z'_2$ from \eqref{eq:3.34} are evaluated separately, is an expansion that is convergent, it is for all positive values of the couplings\footnote{Note that at the relative weak coupling regime the convergence rate can become less efficient.} and it is expected to capture the full result in perturbation theory.

One way to evaluate the absolute value terms is to write them in the form of a square root squared and expand the square root, in the following way
\begin{equation}
\label{eq:3.35}
|S_o[q]| = \sqrt{(S_o[q])^2} = \beta \sqrt{1 - \bar{\epsilon} \{ 1 - (S_o[q])^2 / \beta^2 \} } = \beta \Big{\{} 1 - {\bar{\epsilon} \over 2} \left( 1 - (S_o[q])^2 / \beta^2 \right) +  ...  \Big{\}}
\end{equation}
an $\bar{\epsilon} (= 1)$ parameter is introduced to indicate that the square root is to be expanded in this parameter. $\beta$ is a free parameter similar to $\beta_1$, $\beta_2$ and $\alpha$. Therefore using \eqref{eq:3.35}, \eqref{eq:3.34} can be evaluated to a certain order in the $\bar{\epsilon} (= 1)$ and $\epsilon (= 1)$ expansion and extremized with respect to the free parameter(s) introduced in the theory. 

\section{A convergent formulation for quantum field theory}
\label{sec:4}

In this Section, with following a similar approach as the previous Sections, we develop a convergent formulation for quantum field theory. We consider $\phi^4$ theory and quantum electrodynamics. We will be mostly concerned with the general formalism but in order to provide confirmation for this formalism we apply the convergent expansion method to improve the electron g-factor calculation at the one loop level.

\subsection{A convergent formulation for $\phi^4$ theory}
\label{sec:4.1}
Consider the action of $\phi^4$ theory in renormalized perturbation theory
\begin{equation}
\label{eq:4.1}
S = \int \left\{ {1 \over 2} ( \partial_{\mu} \phi )^2 - {1 \over 2} m^2 \phi^2 - \lambda \phi^4 +\delta_z {1 \over 2} (\partial_{\mu} \phi )^2 - {1 \over 2} \delta_m \phi^2 - \delta_{\lambda} \phi^4 \right\} d^4 x
\end{equation}
to introduce free parameters into the theory rescale $x^{\mu} \rightarrow x^{\mu} / \alpha$, $\phi \rightarrow \beta \phi $, add and subtract the term ${1 \over 2} (\partial_{\mu} \phi )^2 - {1 \over 2} m^2 \phi^2 $ \footnote{Here we have the freedom to add and subtract ${1 \over 2} ( \partial_{\mu} \phi )^2 - {1 \over 2} \bar{m}^2 \phi^2$ with $\bar{m}$ an arbitrary free parameter. For simplicity we have chosen $\bar{m} = m$. In general the introduction of those free parameters will be more useful which can be used to suppress the couplings of the theory.} in the Lagrangian, we obtain
\begin{equation}
\label{eq:4.2}
\begin{split}
S_{\epsilon} \equiv \int d^4 x & \left\{ {1 \over 2} (\partial_{\mu} \phi )^2 - {1 \over 2} m^2 \phi^2 + \epsilon \left\{ \left( {\beta^2 \over 2 \alpha^2}  - {1 \over 2} \right) (\partial_{\mu} \phi )^2 + \left(- {\beta^2 \over 2 \alpha^4} +{1 \over 2} \right)m^2 \phi^2 \right. \right. \\ & \left. \left. - {\beta^4 \over \alpha^4} \lambda \phi^4 +\delta_z {\beta^2 \over 2 \alpha^2} (\partial_{\mu} \phi )^2 - {\beta^2 \over 2 \alpha^4} \delta_m \phi^2 - {\beta^4 \over \alpha^4} \delta_{\lambda} \phi^4 \right\} \right\}
\end{split}
\end{equation}
we keep $\beta$ and $\alpha$ different. If we choose $\beta$ equal to $\alpha$ they would cancel in the term $- \beta^4 / \alpha^4 \lambda \phi^4$ and there will be no suppression in the coupling. Now we have an expansion term with the introduced free parameters and a Gaussian term independent of the free parameters in \eqref{eq:4.2}. Similar to before we want to expand in $\epsilon$ $(=1)$. In this $\epsilon$ expansion we keep the same renormalization conditions as the original theory \eqref{eq:4.1} and later set $\epsilon$ to one and extremize with respect to the free parameter $\alpha$ (after setting $\beta$ equal to a certain power of $\alpha$, e.g. $\beta = \sqrt{\alpha}$ )\footnote{In general there are two free parameters introduced, $\alpha$ and $\beta$ corresponding to the rescaling of space-time and the $\phi$ field, respectively. It might be tempting to extremize with respect to both parameters and search for the extremum point of a two-dimensional surface. However it is not clear whether this will give a significant advantage over setting $\beta$ equal to a certain power of $\alpha$ (e.g. $\beta = \sqrt{\alpha}$) and extremizing with respect to only one parameter, namely $\alpha$. In this work we only consider extremization with respect to one parameter. When setting $\beta$ in terms of $\alpha$ it is important to do this in a way as to obtain more efficient results. Refer to the comment below relation \eqref{eq:4.42} for a discussion on this matter in a simpler context.} to obtain the optimum result. We can recast \eqref{eq:4.2} to a more familiar form by rescaling the field $\phi \rightarrow \{ 1 - \epsilon \left( 1 - {\beta^2 / \alpha^2}  \right) \}^{-1/2} \phi' $, we have
\begin{equation}
\label{eq:4.3}
\begin{split}
S' \equiv \int d^4 x \left\{ {1 \over 2} (\partial_{\mu} \phi' )^2 - {1 \over 2} m'^2 \phi'^2 -  \lambda' \phi'^4 + {\delta'_z \over 2} (\partial_{\mu} \phi' )^2 - {\delta'_m \over 2} \phi'^2 -  \delta'_{\lambda} \phi'^4 \right\}
\end{split}
\end{equation}
with $m'^2$ and $\lambda'$ given by the following relations
\begin{equation}
\label{eq:4.4}
\begin{split}
& m'^2 = m^2 \{ 1 - \epsilon \left( 1 - {\beta^2 / \alpha^4} \right) \} / \{ 1 - \epsilon \left( 1 - {\beta^2 / \alpha^2} \right) \} \\ & \lambda' = {\beta^4 \epsilon \lambda \over \alpha^4} / \{ 1 - \epsilon \left( 1 - {\beta^2 / \alpha^2} \right) \}^2
\end{split}
\end{equation}
and $\delta'_z$, $\delta'_m$ and $\delta'_{\lambda}$ for $\epsilon = 1$ are given by
\begin{equation}
\label{eq:4.5}
\begin{split}
& \delta'_z \vert_{\epsilon = 1} =\left.  {\beta^2 \over \alpha^2} {\epsilon \over 1 - \epsilon(1 - {\beta^2 / \alpha^2} ) } \right\vert_{\epsilon = 1} \delta_z = \delta_z \\ & \delta'_m \vert_{\epsilon = 1} =\left.  {\beta^2 \over \alpha^4} {\epsilon \over 1 - \epsilon(1 - {\beta^2 / \alpha^2} ) } \right\vert_{\epsilon = 1} \delta_m = \delta_m / \alpha^2 \\ & \delta'_{\lambda} \vert_{\epsilon = 1} =\left.  {\beta^4 \over \alpha^4} {\epsilon \over (1 - \epsilon(1 - {\beta^2 / \alpha^2} ) )^2 } \right\vert_{\epsilon = 1} \delta_{\lambda} = \delta_{\lambda}
\end{split}
\end{equation}
in general for an $\epsilon$ value smaller than one clearly $\delta'_z$, $\delta'_m$ and $\delta'_{\lambda}$ will be different from $\delta_z$, $\delta_m$ and $\delta_{\lambda}$ therefore we have only shown their relationship for when $\epsilon$ is set to one. We call \eqref{eq:4.3} the prime theory.
To see which renormalization conditions we will obtain for the prime theory lets apply the transformations above from \eqref{eq:4.1} to \eqref{eq:4.3} to the two point function of  $\phi^4$ theory
\begin{equation}
\label{eq:4.6}
\langle T \phi(x_1) \phi(x_2) \rangle \rightarrow \eta  \langle T \phi'(x'_1) \phi'(x'_2) \rangle , \ \ \ \eta \equiv {\beta^2 \over 1 - \epsilon \left( 1 - {\beta^2 / \alpha^2}  \right)} , \ x'_i = \alpha x_i , \ i =1, 2
\end{equation}
note that the integration variable in \eqref{eq:4.2} is rescaled to $x^{\mu} \rightarrow x^{\mu} / \alpha$ therefore the external points will be rescaled to $x_i \rightarrow \alpha x_i$ for $i =1, 2$ hence $x'_i = \alpha x_i$. Taking the Fourier transform of both sides of \eqref{eq:4.6} with respect to $x_i$ we have
\begin{flalign}
\label{eq:4.7}
& \iint d^4x_1 d^4x_2 e^{i p_1 . x_1} e^{i p_2 . x_2} \langle T \phi(x_1)\phi(x_2) \rangle \sim {i \delta^4(p_1 - p_2) \over p^2_1 - m^2 - M^2(p^2_1) } \\ \label{eq:4.8} & \left( {\eta \over \alpha^8 } \right) \iint d^4x'_1 d^4x'_2 e^{i p'_1 . x'_1} e^{i p'_2 . x'_2} \langle T \phi'(x'_1) \phi'(x'_2) \rangle \sim \left( {\eta \over \alpha^8} \right) {i \delta^4(p'_1 - p'_2) \over p'^2_1 - m'^2 - M'^2(p'^2_1)} \nonumber \\ & = \left( {\eta \over \alpha^8} \right) {i \delta^4(p'_1 - p'_2) \over p'^2_1 - m^2/\alpha^2 - \bar{M}'^2(p'^2_1)}, \ \ \  \bar{M}'^2(p'^2_1) \equiv -m^2/\alpha^2 + m'^2 + M'^2(p'^2_1) &&
\end{flalign}
with $p'_i = p_i/\alpha$ for $i =1, 2$. The symbol $\sim$ means proportional to or equal up to a constant. The renormalization conditions for \eqref{eq:4.8} will be
\begin{equation}
\label{eq:4.9}
\begin{split}
& \left. \bar{M}'^2(p'^2_1)\right\vert_{p'^2_1 = m^2/\alpha^2} = 0 \\ & \left. {d \over d p'^2_1 }  \bar{M}'^2(p'^2_1)\right\vert_{p'^2_1 = m^2 / \alpha^2} = 1 - {\eta \over \alpha^2}
\end{split}
\end{equation}
as previously noted in expanding \eqref{eq:4.2} in $\epsilon$ we want to keep the same renormalization conditions of the original theory \eqref{eq:4.1} therefore the renormalization conditions of \eqref{eq:4.9} for the prime theory are set in a way as to give the same pole and residue of the propagator of the original theory \eqref{eq:4.7} at any order of the $\epsilon$ expansion when we replace $p'_i \rightarrow p_i / \alpha$ in \eqref{eq:4.8}.

Next lets see which renormalization condition we will obtain for the $2$-particle scattering amplitude in the prime theory. The relation between correlation functions and S-matrix elements is given by the LSZ reduction formula \cite{10}:
\begin{equation}
\label{eq:4.10}
\bigg{\{} \prod^4_{j=1} \lim_{p^0_j \rightarrow E_{\p_j} } \int d^4x_j e^{(-1)^j ip_j . x_j} \bigg{\}} \bigg{\langle} T \prod^4_{j=1} \phi (x_j) \bigg{\rangle} \sim \prod^4_{j=1} {i \over p^2_j - m^2 + i \epsilon}  \langle p_2 , p_4 | S | p_3 , p_1 \rangle
\end{equation}
$S = 1 + iT$, $\langle p_2 , p_4 | i T | p_3 , p_1 \rangle = (2 \pi)^4 \delta^4(p_1+p_3 - p_4 - p_2) i \mathcal{M}(p_1,p_3 \rightarrow p_2, p_4)$ and $i \mathcal{M}(p_1,p_3 \rightarrow p_2, p_4) = -i \lambda + ...$ . In the limit which the external 3-momenta go to zero, $\p_j \rightarrow 0$, $i \mathcal{M}(p_1,p_3 \rightarrow p_2, p_4)$ is renormalized to $-i\lambda$. Now lets see how this works out for when the rescalings of \eqref{eq:4.1} to \eqref{eq:4.3} are applied to \eqref{eq:4.10} and in particular for the prime theory \footnote{Note that when applying the rescalings of \eqref{eq:4.1} to \eqref{eq:4.3}, in particular when rescaling the dummy integration variables $x^{\mu} \rightarrow x^{\mu} /\alpha$ the external points of the four point function in $\phi^4$ theory will be rescaled to $x_j \rightarrow \alpha x_j$ but this rescaling should not be applied to the $x_j$ in the Fourier transform integral. A simpler way to see this is that we first apply the rescalings of \eqref{eq:4.1} to \eqref{eq:4.3} to the four point function of $\phi^4$ theory which in particular will result in the rescalings of external points $x_j \rightarrow \alpha x_j$, then we apply the Fourier transform with respect to $x_j$ similar to how it was done for the two point function in relations \eqref{eq:4.6} to \eqref{eq:4.8}.}
\begin{flalign}
\label{eq:4.11}
\left( { \eta^2 \over \alpha^{16}} \right) \prod^4_{j=1} \bigg{\{} \lim_{p'_j \rightarrow {1 \over \alpha} E_{\p_j} } & \int d^4x'_j e^{(-1)^j i p'_j . x'_j} \bigg{\}} \bigg{\langle} T \prod^4_{j=1} \phi' ( x'_j) \bigg{\rangle} \sim \left( { \eta^2 \over \alpha^{16}} \right) \prod^4_{j=1} {i \sqrt{\alpha^2/\eta} \over p'^2_j - m^2 /\alpha^2 + i \epsilon} \nonumber \\ & \langle p'_2 , p'_4 | S' | p'_3 , p'_1 \rangle = \left( { \eta^2 \over \alpha^{16}} \right) { \alpha^{4} \over \eta^2} \prod^4_{i=1} {i \over p'^2_i - m^2/\alpha^2 +i\epsilon } \langle p'_2 , p'_4 | S' | p'_3 , p'_1 \rangle &&
\end{flalign}
with $S' = 1' + iT'$ and $\langle p'_2 , p'_4 | iT' | p'_3 , p'_1 \rangle = (2 \pi)^4 \delta^4(p'_1+p'_3 - p'_4 - p'_2) i \mathcal{M}'(p'_1,p'_3 \rightarrow p'_2, p'_4)$, with $i \mathcal{M}'(p'_1,p'_3 \rightarrow p'_2, p'_4) = {\alpha^4 \over \eta^2} ( -i \lambda' + ... )$ \footnote{As a reminder we note that in the language of Feynman diagrams the square root of the residue of the external propagators have to be absorbed into the scattering amplitude in order to reproduce the LSZ reduction formula hence the factor of ${\alpha^4 \over \eta^2}$ in $i \mathcal{M}'(p'_1,p'_3 \rightarrow p'_2, p'_4) = {\alpha^4 \over \eta^2} (-i\lambda' + ... ) $.} . Therefore the scattering amplitude in the prime theory should be set to $-i \epsilon \lambda$ when $\p'_j \rightarrow 0$ since $\lambda' {\alpha^4 / \eta^2} = \epsilon \lambda$
\begin{equation}
\label{eq:4.12}
i \mathcal{M}'(p'_1,p'_3 \rightarrow p'_2, p'_4) = -i \epsilon \lambda , \ \ \  \p'_j \rightarrow 0
\end{equation}
this completes the discussion of the renormalization conditions of the prime theory. It is clear that when $\epsilon$ is set to one \eqref{eq:4.8} and \eqref{eq:4.11} should reduce to \eqref{eq:4.7} and \eqref{eq:4.10} respectively since in this case we have merely performed a rescaling of space-time $x^{\mu} \rightarrow x^{\mu} / \alpha$ and $\phi \rightarrow \alpha \phi'$ in a closed expression. In order for \eqref{eq:4.8} to reduce to \eqref{eq:4.7} for $\epsilon = 1$ we should have $\bar{M}'^2(p'^2_1) \vert_{\epsilon = 1} = M'^2(p'^2_1) \vert_{\epsilon = 1} = M^2(p^2_1) / \alpha^2$ and for \eqref{eq:4.11} to reduce to \eqref{eq:4.10} when $\epsilon = 1$ we should have ${ 1 \over \alpha^4} \langle p'_2 , p'_4 | S' | p'_3 , p'_1 \rangle \vert_{\epsilon = 1} = \langle p_2 , p_4 | S | p_3 , p_1 \rangle$. This means that if we are using dimensional regularization to regularize the theory and our primed quantities have a left over free parameter dependence when $\epsilon$ is set to one, this left over free parameter dependence should be taken out of the dimensional regularization procedure when we replace $4 \rightarrow d$. As an example to illustrate this point consider $M^2(p^2_1)$ evaluated to one loop order and regularized with dimensional regularization \cite{10}:
\begin{equation}
\label{eq:4.13}
-i M^2(p^2_1) = - {i  \lambda \over 2 } {1 \over (4 \pi)^{d/2}} {\Gamma(1- d/2) \over (m^2)^{1-d/2}} + i(p_1^2 \delta_z - \delta_m)
\end{equation}
in the prime theory this is given by
\begin{equation}
\label{eq:4.14}
-i M'^2(p'^2_1) \rightarrow - {i  \lambda' \over 2 } {1 \over (4 \pi)^{d/2}} {\Gamma(1- d/2) \over (m'^2)^{1-d/2}} + i(p'^2_1 \delta'_z - \delta'_m)
\end{equation}
however it is clear that the quantities evaluated in the prime theory have to reduce to the original theory with considering the appropriate prefactor for correlation functions and scattering amplitudes (e.g. the factor of $( \eta )$ in \eqref{eq:4.6} or the factor of $(\eta^2 / \alpha^{16} )$ in \eqref{eq:4.11}) when $\epsilon$ is set to one and the primed quantities are replaced by their original expressions. Since $m'^2 \vert_{\epsilon =1} = m^2/ \alpha^2$ we have to extract this left over $1/ \alpha^2$ factor from the dimensional regularization procedure as follows
\begin{equation}
\label{eq:4.15}
-i M'^2(p'^2_1) = - {i  \lambda' \over 2 \alpha^2 } {1 \over (4 \pi)^{d/2}} {\Gamma(1- d/2) \over (\alpha^2 m'^2)^{1-d/2}} + i(p'^2_1 \delta'_z - \delta'_m)
\end{equation}
now when $\epsilon$ is set to one, to one loop order we will obtain $-i M'^2(p'^2_1) \vert_{\epsilon =1} = -i M^2(p^2_1) / \alpha^2 $ since we have $\lambda' \vert_{\epsilon =1} = \lambda$, $\alpha^2 m'^2 \vert_{\epsilon =1} = m^2$, $p'^2_1 = p^2_1 / \alpha^2 $ and with applying the renormalization conditions of \eqref{eq:4.9} (or using the general relations of \eqref{eq:4.5}) to one loop order we obtain $\delta'_z \vert_{\epsilon =1} = \delta_z$, $\delta'_m \vert_{\epsilon =1} = \delta_m / \alpha^2$.

To summarize for a convergent expansion of $\phi^4$ theory in renormalized perturbation theory all the calculations can be done in the prime theory with renormalization conditions given by \eqref{eq:4.9} and \eqref{eq:4.12}. If we are evaluating an n-point correlation function: $\langle T \phi'(x'_1) ... \phi'(x'_n) \rangle $, we multiply this by $\eta^{n/2}$, the factor resulting from rescaling each field by $\phi \rightarrow \eta^{1/2} \phi'$ from \eqref{eq:4.1} to \eqref{eq:4.3}, we obtain $\eta^{n/2} \langle T \phi'(x'_1) ... \phi'(x'_n) \rangle $ and if we are evaluating an S-matrix element with $n$ external momenta: $\langle p'_{r+1}, .., p'_n | S' | p'_1, ..., p'_r \rangle $, we multiply this by $( \eta^{n/2} /\alpha^{4n} ) (\alpha^n /\eta^{n/2}) (\alpha^{2n}) = \alpha^{-n} $, we obtain $\alpha^{-n} \langle p'_{r+1}, .., p'_n | S' | p'_1, ..., p'_r \rangle $ \footnote{As stated in the footnote of the previous page when evaluating $i \mathcal{M}'(p'_1, ..., p'_r \rightarrow p'_{r+1}, ..., p'_n) $ the extra factor coming from the square root of the residue of each propagator which is $( \alpha^2 / \eta)^{n/2}$ should also be taken into account.}. Next in order to retrieve the original theory we replace $m'$ and $\lambda'$ by their expressions given by \eqref{eq:4.4} and make the replacements $x'_j \rightarrow \alpha x_j $ for external points when evaluating correlation functions and $p'_j \rightarrow p_j / \alpha$ for external momenta when evaluating S-matrix elements. If we want to evaluate the quantity of interest to order $k$ in the $\epsilon$ expansion we can first evaluate it to order $k$ in the $\lambda'$ expansion then reorganize everything in terms of powers of $\epsilon$ up to order $k$ and finally set $\epsilon$ to one and extremize with respect to the free parameter(s) introduced in the theory \footnote{Here we remind the reader that the terms that we obtain from rearranging the expansion in $\lambda'$ to order $k$ in terms of an $\epsilon$ expansion to order $k$ correspond to the terms in the expansion of \eqref{eq:4.2} in $\epsilon$, in analogy with the one dimensional integral of \eqref{eq:2.11}. When expanding the integral of \eqref{eq:2.11} to order $k$ in $\lambda'$, then rearranging the expansion in terms of an $\epsilon$ expansion to order $k$, the terms obtained from this rearrangement correspond to the terms in the expansion of \eqref{eq:2.2} in $\epsilon$ for $\kappa = 1$ to order $k$. This is to be compared with the expansion of \eqref{eq:2.12}. In \eqref{eq:2.12} when the expansion to order $k$ in $\lambda'$ is rearranged in terms of an expansion to order $k$ in $\epsilon$, the terms obtained from this rearrangement do not correspond to terms coming from an expansion of the integral of a perturbed Gaussian integral and they are just simply a rearrangement of the expansion to order $k$ in $\lambda'$ in terms of an expansion to order $k$ in $\epsilon$.}. As a reminder we note that the $\epsilon (= 1)$ expansion contains the previous asymptotic expansion in the coupling $\lambda$ as a particular case since when the free parameters are set to one the $\epsilon (= 1)$ expansion reduces to the previous asymptotic expansion in the coupling $\lambda$, but now in order to obtain a convergent sequence we have to evaluate every finite order of the $\epsilon (= 1)$ expansion at its extremum points with respect to the free parameter(s) introduced.

\subsection{A convergent formulation for QED}
\label{sec:4.2}
In this Subsection we develop a convergent formulation for quantum electrodynamics. To provide confirmation for this formalism we improve the electron g-factor calculation at the one loop level using the convergent expansion. Consider the action of quantum electrodynamics in Feynman gauge in renormalized perturbation theory \footnote{For convenience we have first written the version of the theory that is suitable for a perturbative expansion then we introduce the free parameters and the parameter of expansion $\epsilon$ into the theory.}
\begin{equation}
\label{eq:4.16}
S \! = \!\! \int \! d^4 x \left\{ \! -{1 \over 2} ( \partial_{\mu} A_{\nu} )^2 \! + \! \bar{\psi} (i \slashed{\partial} \! - \! m) \psi \! - \! e \bar{\psi} \gamma^{\mu} \psi A_{\mu} -{1 \over 4} \delta_3 ( F_{\mu\nu} )^2 \! + \! \bar{\psi} (i \delta_2 \slashed{\partial} \! - \! \delta_m ) \psi \! - \! e \delta_1 \bar{\psi} \gamma^{\mu} \psi A_{\mu} \right\}
\end{equation}
to introduce free parameters in the action rescale space-time $x^{\mu} \rightarrow x^{\mu} / \alpha$ and the fields $A_{\mu} \rightarrow \beta_1 A_{\mu} $, $\psi \rightarrow \beta_2 \psi $, add and subtract $-{1 \over 2} ( \partial_{\mu} A_{\nu} )^2 + \bar{\psi} (i \slashed{\partial} - m ) \psi $ in the Lagrangian, we obtain
\begin{flalign}
\label{eq:4.17}
S \! = & \!\! \int \! d^4 x \bigg{\{} \! -{1 \over 2} ( \partial_{\mu} A_{\nu} )^2 \! + \! \bar{\psi} (i \slashed{\partial} \! - \! m) \psi \! + \epsilon \Big{\{} \Big{(} {1 \over 2} - {\beta^2_1 \over 2 \alpha^2} \Big{)} ( \partial_{\mu} A_{\nu} )^2 \! - \Big{(} 1 - {\beta^2_2 \over \alpha^3} \Big{)} \bar{\psi} i \slashed{\partial} \psi + \Big{(} 1 - {\beta^2_2 \over \alpha^4} \Big{)} m \bar{\psi} \psi \nonumber \\ &  - \! e {\beta_1 \beta^2_2 \over \alpha^4} \bar{\psi} \gamma^{\mu} \psi A_{\mu} -{\beta^2_1 \over 4 \alpha^2} \delta_3 ( F_{\mu\nu} )^2 \! + \delta_2 {\beta^2_2 \over \alpha^3} \bar{\psi} i \slashed{\partial} \psi - \delta_m {\beta^2_2 \over \alpha^4} \bar{\psi} \psi \! - \! e \delta_1 { \beta^2_2 \beta_1 \over \alpha^4} \bar{\psi} \gamma^{\mu} \psi A_{\mu} \Big{\}} \bigg{\}} &&
\end{flalign}
now we want to expand \eqref{eq:4.17} in $\epsilon$ with keeping the same renormalization conditions of the original theory \eqref{eq:4.16}. To transform this to a more familiar form rescale the fields by $A_{\mu} \rightarrow A'_{\mu} / \{ 1 - \epsilon (1 - {\beta^2_1 / \alpha^2}) \}^{1/2} $ and $\psi \rightarrow \psi' / \{ 1 - \epsilon \left( 1 - {\beta^2_2 / \alpha^3} \right) \}^{1/2} $, we obtain
\begin{equation}
\label{eq:4.18}
\begin{split}
S' \! = \!\! \int \! d^4 x \bigg{\{} \! -{1 \over 2} ( \partial_{\mu} A'_{\nu} )^2 \! & + \! \bar{\psi}' (i \slashed{\partial} \! - \! m') \psi' \! - \! e' \bar{\psi}' \gamma^{\mu} \psi' A'_{\mu} -{1 \over 4} \delta'_3 ( F'_{\mu\nu} )^2 \! \\ & + \! \bar{\psi}' (i \delta'_2 \slashed{\partial} \! - \! \delta'_m ) \psi' \! - \! e' \delta'_1 \bar{\psi}' \gamma^{\mu} \psi' A'_{\mu} \bigg{\}}
\end{split}
\end{equation}
with $e'$ and $m'$ given by the following relations
\begin{equation}
\label{eq:4.19}
\begin{split}
& e' = {\epsilon e \over \alpha^4} {\beta_1 \over \{ 1 - \epsilon (1 - {\beta^2_1 / \alpha^2}) \}^{1/2} } {\beta^2_2 \over  1 - \epsilon \left( 1 - {\beta^2_2 / \alpha^3} \right) } \\ & m' = m { 1 - \epsilon (1 - \beta^2_2 / \alpha^4) \over 1 - \epsilon \left( 1 - {\beta^2_2 / \alpha^3}\right) }
\end{split}
\end{equation}
And the counter-terms for $\epsilon = 1$ are related in the following way: $\delta'_3 \vert_{\epsilon =1} = \delta_3$, $\delta'_2 \vert_{\epsilon =1} = \delta_2$, $\delta'_m \vert_{\epsilon =1} = \delta_m /\alpha$ and $\delta'_1 \vert_{\epsilon =1} = \delta_1$. To discuss the renormalization conditions of the prime theory \eqref{eq:4.18} in more detail consider the Fourier transform of the photon two point function in Feynman gauge
\begin{flalign}
\label{eq:4.20}
\!\! \iint \! d^4x_1 d^4x_2 e^{i q . x_1} e^{i \bar{q} . x_2} \langle T A_{\mu}(x_1) A_{\nu}(x_2) \rangle \! \sim \delta^4(q \! - \bar{q}) \left\{  { -i \over q^2 (1 - \Pi(q^2)) } \! \left( g_{\mu\nu} \!  - \! {q_{\mu} q_{\nu} \over q^2} \right) \! + \! {-i \over q^2} \! \left({q_{\mu} q_{\nu} \over q^2} \right) \right\} &&
\end{flalign}
applying the rescalings of \eqref{eq:4.17} and \eqref{eq:4.18} to the two point photon correlation function of \eqref{eq:4.20} and taking its Fourier transform with respect to $x_1$ and $x_2$ we have
\begin{equation}
\label{eq:4.21}
\begin{split}
\left( {\eta_1 \over \alpha^8 } \right) \iint d^4x'_1 d^4x'_2 & e^{i q' . x'_1} e^{i \bar{q}' . x'_2} \langle T A'_{\mu}( x'_1) A'_{\nu}( x'_2) \rangle \sim \\ & \left( {\eta_1 \over \alpha^8 } \right) \delta^4(q' - \bar{q}') \left\{  {-i \over q'^2 (1 - \Pi'(q'^2)) } \left( g_{\mu\nu}  -  {q'_{\mu} q'_{\nu} \over q'^2} \right) + {-i \over q'^2} \left({q'_{\mu} q'_{\nu} \over q'^2} \right) \right\}
\end{split}
\end{equation}
with $q' = q /\alpha$, $\bar{q}' = \bar{q} /\alpha$ and $\eta_1 = {\beta^2_1 / (1 - \epsilon (1 - {\beta^2_1 / \alpha^2}) )}$. Therefore when expanding \eqref{eq:4.17} in $\epsilon$ $(=1)$ in order to have a residue of one for the photon propagator at any order in the $\epsilon$ expansion when we replace $q' \rightarrow q / \alpha$ we should set
\begin{equation}
\label{eq:4.22} 
 \left. \Pi' (q'^2) \right\vert_{q'^2 = 0} = 1 - \eta_1/\alpha^2
\end{equation}
similarly for the fermion two point function we have
\begin{flalign}
\label{eq:4.23}
\iint d^4x_1 d^4x_2 e^{i p . x_1} e^{i \bar{p} . x_2} \langle T \psi(x_1) \bar{\psi}(x_2) \rangle \sim \delta^4(p - \bar{p}) {i \over \slashed{p} - m - \Sigma(\slashed{p}) }
\end{flalign}
applying the rescalings of \eqref{eq:4.17} and \eqref{eq:4.18} we have
\begin{flalign}
\label{eq:4.24}
{\eta_2 \over \alpha^8} \iint d^4x'_1 d^4x'_2 e^{i p' . x'_1} e^{i \bar{p}' . x'_2} \left\langle T \psi'( x'_1) \bar{\psi}'( x'_2) \right\rangle & \sim {\eta_2 \over \alpha^8} \delta^4(p' - \bar{p}') {i \over \slashed{p'} - m' - \Sigma'(\slashed{p}') } \nonumber \\ & = {\eta_2 \over \alpha^8} \delta^4(p' - \bar{p}' ) {i \over \slashed{p}' - m/\alpha - \bar{\Sigma}'(\slashed{p}') }
\end{flalign}
with $\bar{\Sigma}'(\slashed{p}') =  - m/\alpha  + m' + \Sigma'(\slashed{p}')$ and $\eta_2 = {\beta^2_2 / \left( 1 - \epsilon ( 1 - {\beta^2_2 / \alpha^3} ) \right) }$. In order for \eqref{eq:4.24} to have the same renormalization conditions as \eqref{eq:4.23} at any order in the $\epsilon$ expansion when $p' \rightarrow p /\alpha$ we should have
\begin{equation}
\label{eq:4.25}
\begin{split}
& \left. \bar{\Sigma}'(\slashed{p}') \right\vert_{\slashed{p}' = m/\alpha} = 0 \\ 
& \left. {d \over d \slashed{p}'} \bar{\Sigma}'(\slashed{p}') \right\vert_{\slashed{p}' = m/\alpha} = 1 - {\eta_2 \over \alpha^3}
\end{split}
\end{equation}
next we discuss the renormalization condition of the electron vertex function in the prime theory. Consider the Fourier transform of the fermion four point function \footnote{For the discussion of the electron vertex function it is usually assumed that the electron scatters off a heavy target, here in order to simplify the discussion and have two symmetrical electron vertices we consider a fermion four-point function.}. In taking the on shell limit of the external momenta for two incoming fermions with momenta $p_1$ and $p_3$ and two outgoing fermions with momenta $p_2$ and $p_4$ we have
\begin{flalign}
\label{eq:4.26}
& \prod^4_{j=1} \left\{ \lim_{p^0_j \rightarrow E_{\p_j} } \int d^4x_j e^{(-1)^j i p_j . x_j} \right\} \left\langle T \bar{\psi}_{r_1} (x_1) \psi_{r_2} (x_2) \bar{\psi}_{r_3} (x_3) \psi_{r_4} (x_4) \right\rangle \\ & \sim \bigg{\{} \prod_{j=1,3}  { i \bar{u}^{s_j}_{r_j}(p_j) \over p^2_j - m^2 + i\epsilon} \bigg{\}} \bigg{\{} \prod_{j=2,4} {i u^{s_j}_{r_j}(p_j) \over p^2_j - m^2 + i\epsilon} \bigg{\}} \{ \langle p_2, s_2 , p_4, s_4 | S | p_3, s_3 , p_1, s_1 \rangle_{1 \rightarrow 2} + ... \}  \nonumber &&
\end{flalign}
with $S = 1 + iT$ and $\langle p_2, s_2 , p_4, s_4 | iT | p_3, s_3 , p_1, s_1 \rangle_{1 \rightarrow 2} = (2 \pi)^4 \delta^4 (p_1 + p_3 - p_2 - p_4 ) i \mathcal{M}(p_1,p_3 \rightarrow p_2, p_4)_{1 \rightarrow 2} $  with $i \mathcal{M}(p_1 , p_3 \rightarrow p_2 , p_4)_{1 \rightarrow 2} = \bar{u}^{s_2} (p_2) (- i e \gamma^{\mu}) u^{s_1} (p_1) {-i g_{\mu \nu } \over q^2 } \bar{u}^{s_4} (p_4) (- i e \gamma^{\mu}) u^{s_3} (p_3) + ... $ . Summation over $s_j = 1, 2$ for $j =1, ..., 4$ is implicit. The subscript $1 \rightarrow 2$ indicates that we are only considering the process which the fermion $p_1$ scatters into the fermion $p_2$ (and accordingly $p_3$ into $p_4$). The dots indicate higher order corrections. The dots in \eqref{eq:4.26} represent other processes, such as annihilation of $p_1$ and $p_3$ into $p_2$ and $p_4$ and scattering of $p_1$ into $p_4$. We note that here our main concern is to see how the rescalings made in \eqref{eq:4.17} and \eqref{eq:4.18} will effect the renormalization condition for the electron vertex function in the prime theory and we are less concerned with the details of the relation between correlation functions and scattering amplitudes which is standard textbook material, however for completeness of the discussion we include these details here. Now lets see how this will work out when we apply the rescalings of \eqref{eq:4.17} and \eqref{eq:4.18} and in particular for the prime theory
\begin{flalign}
\label{eq:4.27}
& \left( {\eta^2_2 \over \alpha^{16}} \right) \prod^4_{j=1} \left\{ \lim_{p'^0_j \rightarrow {1 \over \alpha} E_{\p_j} } \int d^4x'_j e^{(-1)^j i p'_j . x'_j} \right\} \left\langle T \bar{\psi}'_{r_1} ( x'_1) \psi'_{r_2} ( x'_2) \bar{\psi}'_{r_3} ( x'_3) \psi'_{r_4} (x'_4) \right\rangle \nonumber \\ & \sim \left( {\eta^2_2 \over \alpha^{16}} \right) \bigg{\{} \prod_{j=1,3} {i \sqrt{\alpha^3 / \eta_2} \bar{u}^{s_j}_{r_j}( p'_j) \over p'^2_j - m^2/ \alpha^2 + i \epsilon} \bigg{\}} \bigg{\{} \prod_{j=2,4}  {i \sqrt{\alpha^3 / \eta_2} u^{s_j}_{r_j}(p'_j) \over p'^2_j - m^2 /\alpha^2 + i \epsilon} \bigg{\}} \\ & \hspace{3cm} \left\{ \langle p'_2, s_2 , p'_4, s_4 | S' | p'_3, s_3 , p'_1, s_1 \rangle_{1 \rightarrow 2} + ... \right\} \nonumber \\ & = \left( {\eta^2_2 \over \alpha^{16}} \right) { \alpha^6 \over \eta^2_2} \bigg{\{} \prod_{j=1,3} {i \bar{u}^{s_j}_{r_j}( p'_j) \over p'^2_j - m^2/ \alpha^2 + i \epsilon} \bigg{\}} \bigg{\{} \prod_{j=2,4} { i u^{s_j}_{r_j}(p'_j) \over p'^2_j - m^2/ \alpha^2 + i \epsilon } \bigg{\}} \nonumber \\ & \hspace{3cm} \{  \langle p'_2, s_2 , p'_4, s_4 | S' | p'_3, s_3 , p'_1, s_1 \rangle_{1 \rightarrow 2} + ... \} \nonumber &&
\end{flalign}
with $S' = 1' + iT'$ and $\langle p'_2, s_2 , p'_4, s_4 | iT' | p'_3, s_3 , p'_1, s_1 \rangle_{1 \rightarrow 2} = (2 \pi)^4 \delta^4 (p'_1 + p'_3 - p'_2 - p'_4 ) i \mathcal{M}'(p'_1,p'_3 \rightarrow p'_2, p'_4)_{1 \rightarrow 2} $  with $i \mathcal{M}'(p'_1 , p'_3 \rightarrow p'_2 , p'_4)_{1 \rightarrow 2} = {\alpha^6 \over \eta^2_2} \{ \bar{u}^{s_2} (p'_2) (- i e' \gamma^{\mu}) u^{s_1} (p'_1) {-i g_{\mu \nu } \over q'^2 } \bar{u}^{s_4} (p'_4) (- i e' \gamma^{\mu}) u^{s_3} (p'_3)$ $+ ... \} $. There is also the square root of the residue of the photon propagator in the middle that contributes to the charge renormalization of each vertex in the prime theory, taking that into account and distributing the extra factors symmetrically on both vertices we obtain
\begin{flalign}
\label{eq:4.28}
i \mathcal{M}'(p'_1 , p'_3 & \rightarrow p'_2 , p'_4)_{1 \rightarrow 2} = \\ & \left\{ \bar{u}^{s_2} (p'_2) {\alpha^4 \over \eta_2 \sqrt{\eta_1}} (- i e' \gamma^{\mu}) u^{s_1} (p'_1) {-i \eta_1 g_{\mu \nu } \over \alpha^2 q'^2  } \bar{u}^{s_4} (p'_4) {\alpha^4 \over \eta_2 \sqrt{\eta_1}} (- i e' \gamma^{\mu}) u^{s_3} (p'_3) + ...  \right\} \nonumber &&
\end{flalign}
as noted the factor of $\eta_1/ \alpha^2$ in the middle cancels with the residue of the full photon propagator in the prime theory according to \eqref{eq:4.22}. From the above relation the electron vertex function in the prime theory is $-i {\alpha^4 \over \eta_2 \sqrt{\eta_1}} e' \Gamma'^{\mu}(p'_2, p'_1) =   -i {\alpha^4 \over \eta_2 \sqrt{\eta_1}} e' \{ \gamma^{\mu} + ... \} $, therefore the vertex renormalization condition should be set to
\begin{flalign}
\label{eq:4.29}
-i {\alpha^4 \over \eta_2 \sqrt{\eta_1}} e' \Gamma'^{\mu}(p'_2, p'_1) = -i \epsilon e \gamma^{\mu}, \ \ \ \ \  \text{as} \ \ \   q'^2 = (p'_2 - p'_1)^2 \rightarrow 0
\end{flalign}
since ${\alpha^4 \over \eta_2 \sqrt{\eta_1}} e' = \epsilon e$. Evaluating the extra factors of $\alpha$ and $\eta_2$ in the last line of \eqref{eq:4.27} when replacing $p'_j \rightarrow p_j / \alpha$ we find that there is an extra factor of $ \left( {\eta^2_2 / \alpha^{16}} \right) \left( {\alpha^{6} / \eta^2_2} \right) (\alpha^8) ( 1/ \alpha^2) = 1/ \alpha^4 $ multiplying $\langle p'_2, s_2 , p'_4, s_4 | S' | p'_3, s_3 , p'_1, s_1 \rangle_{1 \rightarrow 2}$ therefore we should have ${1 \over \alpha^4} \langle p'_2, s_2 , p'_4, s_4 | S'$  $| p'_3, s_3 , p'_1, s_1 \rangle_{1 \rightarrow 2} \vert_{\epsilon =1} = \langle p_2, s_2 , p_4, s_4 | S | p_3, s_3 , p_1, s_1 \rangle_{1 \rightarrow 2}$ since when $\epsilon$ is set to one we have merely performed a rescaling of space-time $x^{\mu} \rightarrow x^{\mu} / \alpha$ and the field variables $\psi \rightarrow \alpha^{3/2} \psi' $ and $A_{\mu} \rightarrow \alpha A'_{\mu}$ in a closed expression.

When replacing the primed momenta in terms of their unprimed expressions the ${1 \over \alpha^4}$ factor multiplying $\langle p'_2, s_2 , p'_4, s_4 | S' | p'_3, s_3 , p'_1, s_1 \rangle_{1 \rightarrow 2}$ cancels with an $\alpha^4$ factor coming from the four (primed) momentum conservation delta function, also each vertex should get multiplied by an extra factor of $\alpha$ coming from the replacement of $1 / q'^2$ in the middle photon propagator by $ \alpha^2 / q^2$ and distributing the $\alpha^2$ factor evenly on the two vertices. Note that this $\alpha^2$ factor does not contribute to the photon propagator in the middle since in relation \eqref{eq:4.21} all the $\alpha$ factors cancel when replacing $q' \rightarrow q / \alpha$ and $\bar{q}' \rightarrow \bar{q} / \alpha$ and applying the renormalization condition of \eqref{eq:4.22}. This extra factor of $\alpha$ cancels with a $1/\alpha$ factor coming from replacing the primed momenta of the fermion spin vectors in terms of the unprimed ones, $\bar{u}^{s_j} (p'_j) \rightarrow \bar{u}^{s_j} (p_j) / \sqrt{\alpha}$ for $j = 2, 4$ and $u^{s_j} (p'_j) \rightarrow u^{s_j} (p_j) / \sqrt{\alpha}$ for $j = 1, 3$, therefore we should have $ \Gamma'^{\mu}(p'_2, p'_1)\vert_{\epsilon = 1} = \Gamma^{\mu}(p_2, p_1)$ with $\Gamma^{\mu}(p_2, p_1)$ being the electron vertex function of the original theory \eqref{eq:4.16}.

To summarize, for the evaluation of an S-matrix element in QED in renormalized perturbation theory using the convergent expansion method for n external fermions and m external photons, all the calculations can be done in the prime theory quite similar to the original theory but with renormalization conditions given by \eqref{eq:4.22}, \eqref{eq:4.25} and \eqref{eq:4.29}, then we multiply the S-matrix element in the prime theory by any prefactor as compared to the original theory, this is given by $(\eta^{n/2}_2 \eta^{m/2}_1) ({\alpha^{-4(m+n)}}) (\alpha^{2n})(\alpha^{-n/2}) (\alpha^{2m} ) ({\alpha^3 / \eta_2})^{n/2} ({\alpha^2 / \eta_1})^{m/2} = \alpha^{-n} \alpha^{-m} $, or if we are evaluating an $(n+m)$-point correlation function of $n$ fermion fields and $m$ photon fields the prefactor would be $\eta^{n/2}_2 \eta^{m/2}_1$, if we want to evaluate the correlation function or S-matrix element to order $k$ in the $\epsilon$ expansion we can first evaluate them to order $k$ in the $e'$ expansion, replace $e'$ and $m'$ by their expressions given by \eqref{eq:4.19} and replace the external momenta $p'_j \rightarrow p_j / \alpha$ when evaluating S-matrix elements and the external points $x'_j \rightarrow \alpha x_j$ when evaluating correlation functions, then rearrange the expansion in terms of an $\epsilon$ expansion to order $k$, set $\epsilon$ to one and extremize with respect to the free parameter(s) introduced in the theory \footnote{The discussion here assumes any possible infrared divergences in the theory are appropriately regularized.}.

Therefore given a field theory with identifying the prime theory associated with that field theory and its corresponding renormalization conditions, one can perform the perturbative expansions in the prime theory with considering the appropriate prefactors for correlation functions and S-matrix elements. The previous asymptotic expansion can be obtained by simply setting $\epsilon$ to one and a convergent expansion can be obtained by rearranging the expansion at each finite order in terms of an expansion in $\epsilon$ to that order, setting $\epsilon$ to one and extremizing with respect to the free parameter(s) introduced in the expansion.

\subsubsection{Evaluation of the electron g-factor at one loop using the convergent formalism}
\label{sec:4.2.1}
In order to provide confirmation for the general formalism developed in this Section, in this Subsection we improve the electron g-factor calculation at the one loop level using the convergent expansion method. All the calculations can be done in the prime theory similar to the original theory but with noting that the renormalized mass is $m/\alpha$ and it is different from the propagator mass $m'$. This point will become more clear as we carry out the calculations. Lets consider the one loop contribution to the electron vertex function in the prime theory
\begin{SCfigure}[0.8][h]
\caption{Electron-photon vertex at one loop level in the prime theory of QED}
\includegraphics[scale=0.40]{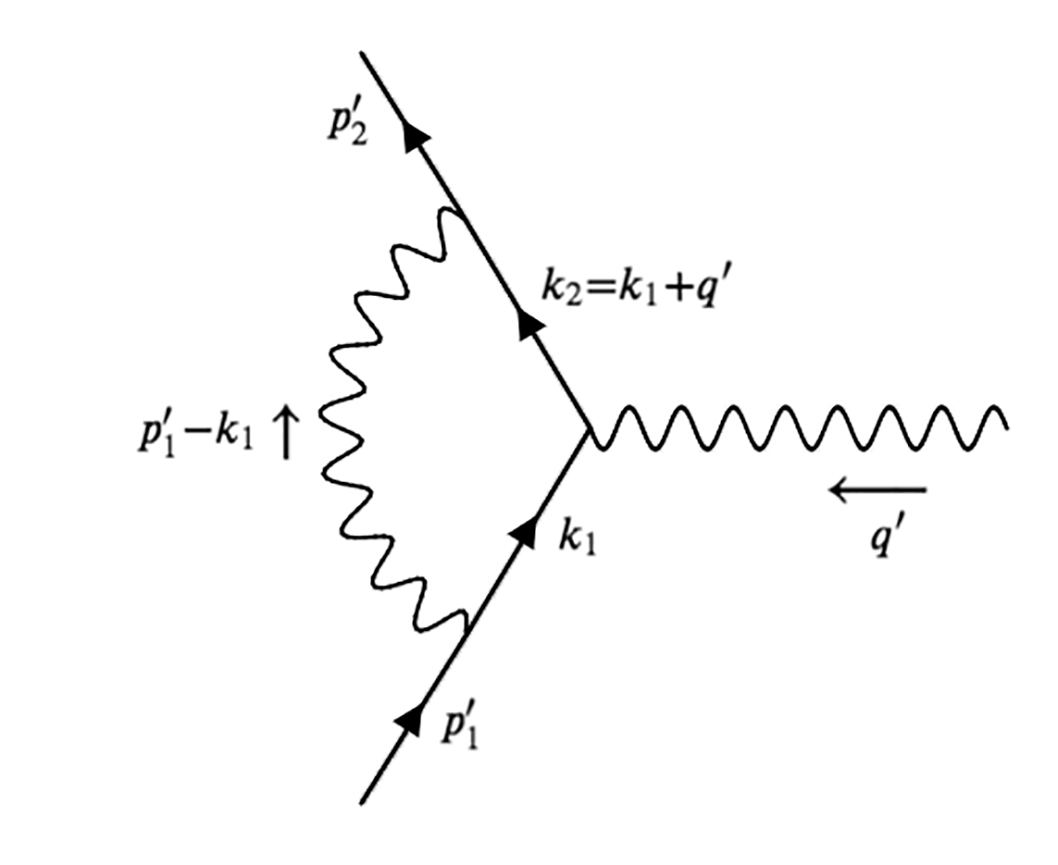}
\end{SCfigure}
\begin{flalign}
\label{eq:4.30}
= 2i e'^2 \int {d^4k_1 \over (2 \pi)^4 } { \bar{u}(p'_2) \left[ \slashed{k}_1 \gamma^{\mu} \slashed{k}_2 + m'^2 \gamma^{\mu} -2m'(k_1 + k_2)^{\mu} \right] u(p'_1) \over ((k_1-p'_1)^2 + i\epsilon) (k^2_2-m'^2 + i\epsilon) (k^2_1-m'^2 + i \epsilon)  }
\end{flalign}
note that in evaluating \eqref{eq:4.30} we have to use the relations $p'^2_1 = m^2/\alpha^2$, $\slashed{p}'_1 u(p'_1) = {m / \alpha} u(p'_1)$, $\bar{u}(p'_2) \slashed{p}'_2 = \bar{u}(p'_2) {m / \alpha}$ as we set the renormalization condition at the physical mass of the original theory. Combining the denominators using the method of Feynman-Schwinger parameters we obtain
\begin{flalign}
\label{eq:4.31}
{1 \over ((k_1-p'_1)^2 + i\epsilon) (k^2_2-m'^2 + i\epsilon) (k^2_1-m'^2 + i \epsilon) } = \int^1_0 dx dy dz \; \delta (x+y+z-1) {2 \over D^3} &&
\end{flalign}
with $D$ given by
\begin{flalign}
\label{eq:4.32}
D = x (k^2_1-m'^2) + y(k^2_2-m'^2) + z(k_1-p'_1)^2 + (x+y+z) i \epsilon
\end{flalign}
after simplifying the above relation using $x+y+z =1$, $p'^2_1= p'^2_2  = m^2/\alpha^2$ and $q'= k_2 - k_1 = p'_2 - p'_1 $ we obtain
\begin{flalign}
\label{eq:4.33}
D = l^2 - \Delta' + i\epsilon
\end{flalign}
with $l = k_1 +y q' -zp'_1$ and $\Delta' = -yx q'^2 - z(1-z) m^2/\alpha^2 + (1-z) m'^2$ . Next we evaluate the numerator
\begin{flalign}
\label{eq:4.34}
N^{\mu} = \bar{u}(p'_2) \left[ \slashed{k}_1 \gamma^{\mu} \slashed{k}_2 + m'^2 \gamma^{\mu} -2m'(k_1 + k_2)^{\mu} \right] u(p'_1)
\end{flalign}
using the relations $k_1 = l - yq' + zp'_1$, $k_2 = l + (1 - y)q' + zp'_1$, $\slashed{p}'_1 u(p'_1) =m/\alpha u(p'_1)$, $\bar{u}(p'_2) \slashed{p}'_2 = \bar{u}(p'_2) m/\alpha $, $\slashed{a} \gamma^{\mu} = 2 a^{\mu} - \gamma^{\mu} \slashed{a}$ and the identities $\int d^4 l \; {l^{\mu} / D^3} = 0 $ and $\int d^4 l \; {l^{\mu} l^{\nu} / D^3} = \int d^4 l \; {{1 \over 4} g^{\mu \nu} / D^3}$ we have \footnote{As a consistency check it can be seen that setting $m /\alpha$ equal to $m'$ we obtain a similar relation as in Section 6.3 of \cite{10}.}
\begin{equation}
\label{eq:4.35}
\begin{split}
N^{\mu} \rightarrow \bar{u}(p'_2) \Big{[} & \gamma^{\mu} \left\{ - l^2 / 2 + m'^2 - z(z+2) ({m / \alpha})^2 + q'^2 (y+z)(1-y) \right\} \\ & + q'^{\mu} (2m' - z {m / \alpha} )(y-x) + p'^{\mu}_+ \{ -2m' z + (z + z^2) {m / \alpha} \} \Big{]} u(p'_1)
\end{split}
\end{equation}
with $p'^{\mu}_+ = p'^{\mu}_1 + p'^{\mu}_2$. After replacing the term proportional to $\bar{u}(p'_2) p'^{\mu}_+ u(p'_1)$ in \eqref{eq:4.35} using the Gordan identity $\bar{u}(p'_2) \gamma^{\mu} {u}(p'_1) = \bar{u}(p'_2) \Big{\{} {p'^{\mu}_+ \over 2 m / \alpha} + {i \sigma^{\mu \nu} q'_{\nu} \over 2 m /\alpha} \Big{\}} {u}(p'_1) $ we obtain
\begin{flalign}
\label{eq:4.36}
& \bar{u}(p'_2) \delta \Gamma'^{\mu}(p'_2, p'_1) {u}(p'_1) = 2ie'^2 \!\! \int \!\! {d^4l \over (2 \pi)^4 } \! \! \int^1_0 \!\! dx dy dz \; \delta (x \! + \! y \! + \! z \! -\! 1) {2 \over D^3} \nonumber \\ & \bar{u}(p'_2) \bigg{[} \gamma^{\mu} \left( - l^2 / 2 + m'^2 - 4 m' {m \over \alpha} z + z^2 ({m / \alpha})^2 + q'^2 (y+z)(1-y) \right)\! \nonumber \\ & + q'^{\mu} (2m' - z {m / \alpha} )(y-x) + \! {i \sigma^{\mu \nu} q'_{\nu} \over 2 m / \alpha} {2m \over \alpha} \Big{\{} 2m' z \! - \! {m \over \alpha}(z + z^2) \Big{\}}  \bigg{]} u(p'_1) &&
\end{flalign}
the term proportional to $q'^{\mu}$ in \eqref{eq:4.36} vanishes when performing the integral over $x$ and $y$, as expected from the Ward identity $q'_{\mu} \Gamma'^{\mu} = 0$. With $\Gamma^{\mu}(p_2, p_1) = \gamma^{\mu} + \delta \Gamma^{\mu}(p_2, p_1) = \gamma^{\mu} F_1(q^2) + {i \over 2 m} {\sigma^{\mu \nu} q_{\nu}} F_2(q^2)$ and in the prime theory $\Gamma'^{\mu}(p'_2, p'_1) = \gamma^{\mu} + \delta \Gamma'^{\mu}(p'_2, p'_1) = \gamma^{\mu} F'_1(q'^2) + {i \over 2 m} { \sigma^{\mu \nu} q_{\nu}} F'_2(q'^2)$ and as previously argued $\Gamma'^{\mu}(p'_2, p'_1)\vert_{\epsilon = 1} = \Gamma^{\mu}(p_2, p_1) $ therefore $F'_2(q'^2)\vert_{\epsilon = 1} = F_2(q^2)$. Hence all we need to do is to calculate $F'_2(q'^2)$ in the prime theory and reorganize the expansion in terms of an $\epsilon$ expansion, set $\epsilon$ to one and extremize with respect to the free parameter(s) introduced in the theory. From \eqref{eq:4.36}, $F'_2(q'^2)$ is given by
\begin{flalign}
\label{eq:4.37}
F'_2(q'^2) = 2ie'^2 \!\! \int \!\! {d^4l \over (2 \pi)^4 } \! \! \int^1_0 \!\! dx dy dz \; \delta (x \! + \! y \! + \! z \! -\! 1) {2 \over D^3} {2m \over \alpha} \left\{ 2m' z \! - \! {m \over \alpha}(z + z^2) \right\}
\end{flalign}
using the identity $\int {d^4l \over (2 \pi)^4 } {1 \over (l^2 - \Delta' + i\epsilon)^3 } = {-i \over 2 (4 \pi)^2}{ 1 \over \Delta' }$ we obtain
\begin{flalign}
\label{eq:4.38}
F'_2(q'^2) = {e'^2 \over (4 \pi)^2}  \! \! \int^1_0 \! dz \int^{1-z}_0 \!\!\! dy {2 \over \Delta'} {2m \over \alpha} \left\{ 2m' z \! - \! {m \over \alpha}(z + z^2) \right\}
\end{flalign}
setting $q'^2 = 0$ we can extract the factor $a_e = (g-2)/2 = F'_2(0) $ evaluated to one loop order, we find
\begin{flalign}
\label{eq:4.39}
F'_2(0) & = {e'^2 \over (4 \pi)^2} \! \! \int^1_0 \! dz {2 \over - z m^2/\alpha^2 + m'^2} {2m \over \alpha} \Big{\{} 2m' z \! - \! {m \over \alpha}(z + z^2) \Big{\}} \nonumber \\ & = {1 \over 2 \pi} {e^2 \over 4 \pi} 2 {e'^2 \over e^2}  \! \! \int^1_0 \! dz {(1 - 2 \alpha {m' / m} ) z + z^2 \over z - \alpha^2 m'^2 /m^2}
\end{flalign}
after simplifying the integrand of \eqref{eq:4.39} using the identity ${(a z + z^2) / (z - b)} = z + a + b + {b (a + b) / (z - b)}$ we have
\begin{equation}
\label{eq:4.40}
\begin{split}
F'_2(0) = {\alpha_{\text{QED} } \over 2 \pi} \kappa (\epsilon) , \ \ \ \ & \kappa(\epsilon) = {e'^2 \over e^2} \Big{\{} 1 + 2 (a + b) + 2b(a+b)( \{ \ln (1-b )^2 \} /2 - \ln b ) \Big{\}} \\ & a \equiv (1 - 2 \alpha {m' / m} ) , \ \ \ \ b \equiv \alpha^2 m'^2 /m^2
\end{split}
\end{equation}
with $m'$ and $e'$ given by \eqref{eq:4.19} and $\alpha_{\text{QED}} = e^2 / 4 \pi$. Setting $\epsilon = 1$ in \eqref{eq:4.40} gives $\kappa (1) = 1$ and therefore the usual one loop calculation of the electron g-factor is recovered but now the goal is to expand $F'_2(0)$ to third order in $\epsilon$ by expanding $\kappa(\epsilon) = \kappa^{\alpha}_2 \epsilon^2 + \kappa^{\alpha}_3 \epsilon^3 + ... $ , set $\epsilon$ to one and extremize with respect to $\alpha$ to obtain the optimum result. The fourth order expansion of $F'_2(0)$ in $\epsilon$ requires the inclusion of the two loop contribution to the electron vertex function. Tables \ref{tab:14} - \ref{tab:17} summarize the results of this analysis. These Tables compare the one loop evaluation of $a_e = (g-2)/2$ using the convergent expansion method given by $ \kappa^{\alpha_{\text{ext}}}_{23} \alpha_{\text{QED}}/ 2\pi $ for $\kappa^{\alpha}_{23} = \kappa^{\alpha}_{2} + \kappa^{\alpha}_{3}$, with the one loop and higher loop evaluations of $a_e$ using the conventional asymptotic expansion method in the coupling $e$ given by $a_e (\text{one loop}) = \alpha_{\text{QED}}/ 2\pi = \kappa^{\alpha = 1}_{23} \alpha_{\text{QED}}/ 2\pi = 0.00116141$ \footnote{Note that when $\alpha = 1$, $\kappa^{\alpha = 1}_{23} = 1$, therefore the convergent expansion evaluation of $a_e$ at one loop reduces to the conventional asymptotic expansion evaluation of $a_e$ at one loop, $\kappa^{\alpha = 1}_{23} \alpha_{\text{QED}}/ 2\pi = \alpha_{\text{QED}}/ 2\pi$} and $a_e (\text{higher loop}) = 0.001159652..$ , respectively, with considering different values for $\beta_1$ and $\beta_2$ in terms of $\alpha$. In these Tables we have included the smallest $\alpha_{\text{ext}}$ value which is greater than one. The zero point of the first derivative of $\kappa^{\alpha}_{23}$ with respect to $\alpha$ does not change the value of $a_e (\text{one loop})_{\text{converg. exp.}}$ considerably as compared to when $\alpha$ is one as is the case for the zero point of the second derivative in Tables \ref{tab:14} - \ref{tab:17}. However the evaluation of $a_e (\text{one loop})_{\text{converg. exp.}} = \kappa^{\alpha}_{23} \alpha_{\text{QED}}/ 2\pi$ at the zero point of its third derivative with respect to $\alpha$ gives a significant improvement to the previous one loop evaluation of $a_e$ given by $a_e (\text{one loop}) = \alpha_{\text{QED}}/ 2\pi $ as can be seen from Tables \ref{tab:14} - \ref{tab:17}. In general for better results and to see an improvement in the value of $a_e$ at the zero point of the first derivative similar to the results of Section \ref{sec:2} or \ref{sec:3} we should have more terms in the expansion that involve the coupling directly, here we only have one term that involves the coupling and the second term comes from the expansion in $\epsilon$ to third order. For this we need to evaluate the two loop contribution to the electron vertex function in the prime theory which is a separate project. Here instead we will try to motivate the results obtained in Tables \ref{tab:14} - \ref{tab:17}. This is discussed in detail in the next Subsection.
\\
\begin{table}[h]
\centering
\caption{$\kappa^{\alpha}_{23} = \kappa^{\alpha}_{2} + \kappa^{\alpha}_{3}$, $\beta_1 = \alpha$, $\beta_2 = \alpha$ }
\begin{tabular}{| K{7cm} | K{2cm} | K{2cm} | K{2cm} | }
 \hline
 $k$ & 1 & 2 & 3 \\
 \Xhline{4\arrayrulewidth}
  $ { d^k \kappa^{\alpha_{\text{ext}}}_{23} / d \alpha^k } = 0$ ; $\alpha_{\text{ext}}$ & 1.01 & 1.0036 & $ 1.0393 $ \\
 \hline
$a_e (\text{one loop})_{\text{converg. exp.}} = \kappa^{\alpha_{\text{ext}}}_{23} \alpha_{\text{QED}}/ 2\pi $ & .00116152 & .00116145 &  \cellcolor{gray!30} .00115908 \\
 \hline
$a_e (\text{one loop}) = \alpha_{\text{QED}}/ 2\pi$ & .00116141 & .00116141 & .00116141 \\
 \hline
$ a_e (\text{higher loop}) $ & .00115965 & .00115965 & \cellcolor{gray!30} .00115965 \\
 \hline
\end{tabular}
\label{tab:14}
\end{table}
\begin{table}
\centering
\caption{$\kappa^{\alpha}_{23} = \kappa^{\alpha}_{2} + \kappa^{\alpha}_{3}$, $\beta_1 = 1$, $\beta_2 = \alpha$ }
\begin{tabular}{| K{7cm} | K{2cm} | K{2cm} | K{2cm} | }
 \hline
 $k$ & 1 & 2 & 3 \\
 \Xhline{4\arrayrulewidth}
  $ { d^k \kappa^{\alpha_{\text{ext}}}_{23} / d \alpha^k } = 0$ ; $\alpha_{\text{ext}}$ & 1.0002 & 1.00006 & $ 1.01396 $ \\
 \hline
$a_e (\text{one loop})_{\text{converg. exp.}} = \kappa^{\alpha_{\text{ext}}}_{23} \alpha_{\text{QED}}/ 2\pi $ & .00116141 & .00116141 &  \cellcolor{gray!30} .00115974 \\
 \hline
$a_e (\text{one loop}) = \alpha_{\text{QED}}/ 2\pi$ & .00116141 & .00116141 & .00116141 \\
 \hline
$ a_e (\text{higher loop}) $ & .00115965 & .00115965 & \cellcolor{gray!30} .00115965 \\
 \hline
\end{tabular}
\label{tab:15}
\end{table}
\begin{table}
\centering
\caption{$\kappa^{\alpha}_{23} = \kappa^{\alpha}_{2} + \kappa^{\alpha}_{3}$, $\beta_1 = \alpha$, $\beta_2 = 1$ }
\begin{tabular}{| K{7cm} | K{2cm} | K{2cm} | K{2cm} | }
 \hline
 $k$ & 1 & 2 & 3 \\
 \Xhline{4\arrayrulewidth}
  $ { d^k \kappa^{\alpha_{\text{ext}}}_{23} / d \alpha^k } = 0$ ; $\alpha_{\text{ext}}$ & 1.0000 & 1.0000 & $ 1.0051 $ \\
 \hline
$a_e (\text{one loop})_{\text{converg. exp.}} = \kappa^{\alpha_{\text{ext}}}_{23} \alpha_{\text{QED}}/ 2\pi $ & .00116141 & .00116141 &  \cellcolor{gray!30} .00116077 \\
 \hline
$a_e (\text{one loop}) = \alpha_{\text{QED}}/ 2\pi$ & .00116141 & .00116141 & .00116141 \\
 \hline
$ a_e (\text{higher loop}) $ & .00115965 & .00115965 & \cellcolor{gray!30} .00115965 \\
 \hline
\end{tabular}
\label{tab:16}
\end{table}
\begin{table}
\centering
\caption{$\kappa^{\alpha}_{23} = \kappa^{\alpha}_{2} + \kappa^{\alpha}_{3}$, $\beta_1 = 1$, $\beta_2 = \alpha^{3/2}$ }
\begin{tabular}{| K{7cm} | K{2cm} | K{2cm} | K{2cm} | }
 \hline
 $k$ & 1 & 2 & 3 \\
 \Xhline{4\arrayrulewidth}
  $ { d^k \kappa^{\alpha_{\text{ext}}}_{23} / d \alpha^k } = 0$ ; $\alpha_{\text{ext}}$ & 1.0063 & 1.0022 & $ 1.0321 $ \\
 \hline
$a_e (\text{one loop})_{\text{converg. exp.}} = \kappa^{\alpha_{\text{ext}}}_{23} \alpha_{\text{QED}}/ 2\pi $ & .00116145 & .00116143 & \cellcolor{gray!30} .00115935 \\
 \hline
$a_e (\text{one loop}) = \alpha_{\text{QED}}/ 2\pi$ & .00116141 & .00116141 & .00116141 \\
 \hline
$ a_e (\text{higher loop}) $ & .00115965 & .00115965 & \cellcolor{gray!30} .00115965 \\
 \hline
\end{tabular}
\label{tab:17}
\end{table}

\subsubsection{A discussion on the results of Tables \ref{tab:14} - \ref{tab:17}}
\label{sec:4.2.2}

In this Subsection we will motivate the results of Tables \ref{tab:14} - \ref{tab:17} and argue that they are genuine, hence confirming the convergent formalism developed in this Section for $\phi^4$ theory and QED.

We first consider a two dimensional integral (a toy model for QED) and try to look for a similar pattern as Tables \ref{tab:14} - \ref{tab:17}. Consider the following two dimensional integral
\begin{flalign}
\label{eq:4.41}
I (\bar{e}) = \int dx dy \exp \left(- (x^2 + y^2) + i \bar{e} y x^2 \right)
\end{flalign}
we can introduce two free parameters into the above expression by rescaling $x \rightarrow x / \alpha$, $y \rightarrow y / \beta$ as follows
\begin{flalign}
\label{eq:4.42}
I (\bar{e}) = I^{\alpha} (\bar{e}) \equiv {1 \over \alpha \beta } \int dx dy \exp \left( - \left( x^2 + y^2 \right) + \epsilon \left\{ (1 - {1 / \alpha^2}) x^2 + (1 - {1 / \beta^2}) y^2 + i { \bar{e} \over \beta \alpha^2} y x^2 \right\} \right) &&
\end{flalign}
whenever we have the possibility of introducing many independent free parameters into our expression it is important to choose them in a way as to obtain more efficient results. For example in \eqref{eq:4.42} if we set $\beta$ equal to one we will obtain more efficient results as compared to setting $\beta$ equal to  $\alpha$ hence we take $\beta = 1$ in \eqref{eq:4.42}. Expanding \eqref{eq:4.42} in $\epsilon$ we have $I^{\alpha}(\bar{e}) = \sum^{\infty}_{n=0} I^{\alpha}_n \epsilon^n$. Defining the partial sum $I^{\alpha, N} \equiv \sum^{N}_{n=0} I^{\alpha}_n$ and extremizing $I^{\alpha, 3}$ with respect to $\alpha$ we obtain the results listed in Tables \ref{tab:18} and \ref{tab:19}. To obtain the partial sum $I^{\alpha, N}$ one convenient way is to rescale $x \rightarrow x / \bar{\eta}$ for $\bar{\eta} = (1 - \epsilon (1 - 1/\alpha^2))^{1/2}$ and expand in $\bar{e}'$, as follows
\begin{flalign}
\label{eq:4.43}
I^{\alpha} (\bar{e}) \equiv {1 \over \alpha \bar{\eta} } \int dx dy \exp \left( - x^2 - y^2 + i \bar{e}' y x^2 \right) = {\pi \over \alpha \bar{\eta}} \underset{k \ \text{even}}{\sum^{\infty}_{k=0}} {i^k \over k!} {(k-1)!! \over 2^{k/2} } {(2k-1)!! \over 2^{k} } \bar{e}'^k && 
\end{flalign}
$\bar{e}' = { \epsilon  \bar{e} / (\alpha^2 \bar{\eta}^2) }$. Therefore to obtain $I^{\alpha, N}$ we can first consider the partial sum of \eqref{eq:4.43} to order $2 \lfloor N/2 \rfloor$ in the $\bar{e}'$ expansion then reorganize the expansion in the form of an $\epsilon$ expansion to order $N$.

\begin{table}[h]
\centering
\captionsetup{width=.75\linewidth}
\caption{ Comparison of the values of $I^{\alpha, 3}$ at the zero points of its 1st, 2nd, and 3rd derivatives with respect to $\alpha$ with $I^{\alpha =1, 3}$ and the exact value $I(\bar{e})$ of \eqref{eq:4.41} for $\bar{e} = 3/10$. $\beta$ is set to one in \eqref{eq:4.42}. }
\begin{tabular}{| K{4cm} | K{1.5cm} | K{1.5cm} | K{1.5cm} | }
 \hline
 $k$ & 1 & 2 & 3 \\
 \Xhline{4\arrayrulewidth}
  $ { d^k I^{\alpha_{\text{ext}},3} / d \alpha^k } = 0$ ; $\alpha_{\text{ext}}$ & 1.169 & 1.076 & $ 1.283 $ \\
 \hline
$ I^{\alpha_{\text{ext}},3} $ & 3.0960 &  \cellcolor{gray!30} 3.0920 & 3.0867 \\
 \hline
$ I^{\alpha = 1, 3} $ & 3.0886 & 3.0886 & 3.0886 \\
 \hline
$ I(\bar{e} = 3/10) $ & 3.0930 & \cellcolor{gray!30} 3.0930 & 3.0930 \\
 \hline
\end{tabular}
\label{tab:18}
\end{table}

\begin{table}[h]
\centering
\captionsetup{width=.75\linewidth}
\caption{ Comparison of the values of $I^{\alpha, 3}$ at zero points of the 1st, 2nd, and 3rd derivatives with respect to $\alpha$ with $I^{\alpha =1, 3}$ and the exact value $I(\bar{e})$ of \eqref{eq:4.41} for $\bar{e} = 1/5$. $\beta$ is set to one in \eqref{eq:4.42}. }
\begin{tabular}{| K{4cm} | K{1.5cm} | K{1.5cm} | K{1.5cm} | }
 \hline
 $k$ & 1 & 2 & 3 \\
 \Xhline{4\arrayrulewidth}
  $ { d^k I^{\alpha_{\text{ext}},3} / d \alpha^k } = 0$ ; $\alpha_{\text{ext}}$ & 1.116 & 1.0566 & $ 1.28 $ \\
 \hline
$ I^{\alpha_{\text{ext}},3} $ & 3.11986 &  \cellcolor{gray!30} 3.11893 & 3.104 \\
 \hline
$ I^{\alpha = 1, 3} $ & 3.11803 & 3.11803 & 3.11803 \\
 \hline
$ I(\bar{e} = 1/5) $ & 3.11899 & \cellcolor{gray!30} 3.11899 & 3.11899 \\
 \hline
\end{tabular}
\label{tab:19}
\end{table}
From these Tables it can be seen that at weak $\bar{e}$ coupling the value of $I^{\alpha,3}$ at the zero point of the second derivative of $I^{\alpha, 3}$ gives significantly more accurate results relative to the value of $I^{\alpha, 3}$ at the zero point of the first derivative of $I^{\alpha, 3}$ and $I^{\alpha = 1, 3}$. The pattern observed here which the zero point of a higher derivative with respect to $\alpha$ in an $\epsilon$ (= 1) expansion to 3rd order gives more accurate results relative to the zero point of the first derivative with respect to $\alpha$ confirms a similar pattern observed in the results of Tables \ref{tab:14} - \ref{tab:17}.

Next we modify the expression $\kappa(\epsilon)$ in \eqref{eq:4.40} in certain reasonable ways as to obtain a new function $\bar{\kappa}(\epsilon)$ and repeat the analysis of Tables \ref{tab:14} - \ref{tab:17}. We will see that in this case the results obtained are not as good as the results of Tables \ref{tab:14} - \ref{tab:17} illustrating how non-trivial it is to obtain the highlighted results of Tables \ref{tab:14} - \ref{tab:17} for $a_e (\text{one loop})_{\text{converg. exp.}}$. Tables \ref{tab:20} and \ref{tab:21} show a similar analysis as Table \ref{tab:15} carried out for $\bar{\kappa}(\epsilon) = \alpha {\kappa}(\epsilon)$ and $\bar{\kappa}(\epsilon) = {\kappa}(\epsilon) / \alpha$. The values obtained in these Tables at the zero point of the third derivative with respect to $\alpha$ are pretty far from the value of $a_e (\text{higher loop})$, showing that the perturbative expansion should be independent of the inserted free parameter in order for its zero $m$-derivative points with respect to the free parameter to produce efficient results, upon multiplying or dividing $\kappa(\epsilon)$ by $\alpha$ clearly the perturbative expansion of $F'_2(0)$ becomes dependent on $\alpha$. In Table \ref{tab:21} there were no zero points for the first derivative of $ \bar{\kappa}^{\alpha}_{23}$.
\begin{table}[h]
\centering
\caption{$ \bar{\kappa}^{\alpha}_{23} \equiv \alpha \kappa^{\alpha}_{23}$, $\beta_1 = 1$, $\beta_2 = \alpha$ }
\begin{tabular}{| K{7cm} | K{1.5cm} | K{1.5cm} | K{1.5cm} | }
 \hline
 $k$ & 1 & 2 & 3 \\
 \Xhline{4\arrayrulewidth}
  $ { d^k \bar{\kappa}^{\alpha_{\text{ext}}}_{23} / d \alpha^k } = 0$ ; $\alpha_{\text{ext}}$ & 1.058 & 1.00006 & $ 1.017 $ \\
 \hline
$ \bar{\kappa}^{\alpha_{\text{ext}}}_{23} \alpha_{\text{QED}}/ 2\pi $ & .0011948 & .0011615 & .0011786 \\
 \hline
$a_e (\text{one loop}) =  \alpha_{\text{QED}}/ 2\pi$ & .0011614 & .0011614 & .0011614 \\
 \hline
$a_e (\text{higher loop}) $ & .0011597 & .0011597 & .0011597 \\
 \hline
\end{tabular}
\label{tab:20}
\end{table}

\begin{table}[h]
\centering
\caption{$ \bar{\kappa}^{\alpha}_{23} \equiv \kappa^{\alpha}_{23} / \alpha$, $\beta_1 = 1$, $\beta_2 = \alpha$ }
\begin{tabular}{| K{7cm} | K{1.5cm} | K{1.5cm} | K{1.5cm} | }
 \hline
 $k$ & 1 & 2 & 3 \\
 \Xhline{4\arrayrulewidth}
  $ { d^k \bar{\kappa}^{\alpha_{\text{ext}}}_{23} / d \alpha^k } = 0$ ; $\alpha_{\text{ext}}$ & - & 1.0001 & $ 1.012 $ \\
 \hline
$ \bar{\kappa}^{\alpha_{\text{ext}}}_{23} \alpha_{\text{QED}}/ 2\pi $ & - & .0011613 & .0011464 \\
 \hline
$a_e (\text{one loop}) =  \alpha_{\text{QED}}/ 2\pi$ & .0011614 & .0011614 & .0011614 \\
 \hline
$a_e (\text{higher loop}) $ & .0011597 & .0011597 & .0011597 \\
 \hline
\end{tabular}
\label{tab:21}
\end{table}
Also lets consider $\bar{\kappa}(\epsilon) = {\kappa}(\epsilon) \eta_2 / \alpha^3$, in this case $\bar{\kappa}(\epsilon =1 ) = 1$. The following Tables summarize the results obtained.
\begin{table}[h]
\centering
\caption{$ \bar{\kappa}^{\alpha}_{23} \equiv \bar{\kappa}^{\alpha}_{2} + \bar{\kappa}^{\alpha}_{3}$ for $ {\kappa}(\epsilon) \eta_2 / \alpha^3 = \bar{\kappa}(\epsilon) = \bar{\kappa}_2 \epsilon^2 + \bar{\kappa}_3 \epsilon^3 + ... $ and $\beta_1 = \alpha$, $\beta_2 = \alpha$ }
\begin{tabular}{| K{7cm} | K{2cm} | K{2cm} | K{2cm} | }
 \hline
 $k$ & 1 & 2 & 3 \\
 \Xhline{4\arrayrulewidth}
  $ { d^k \bar{\kappa}^{\alpha_{\text{ext}}}_{23} / d \alpha^k } = 0$ ; $\alpha_{\text{ext}}$ & 1.0021 & 1.0008 & $ 1.0244 $ \\
 \hline
$ \bar{\kappa}^{\alpha_{\text{ext}}}_{23} \alpha_{\text{QED}}/ 2\pi $ & .00116141 & .00116141 &\cellcolor{gray!15} .00115901 \\
 \hline
$a_e (\text{one loop}) = \alpha_{\text{QED}}/ 2\pi$ & .00116141 & .00116141 & .00116141 \\
 \hline
$a_e (\text{higher loop}) $ & .00115965 & .00115965 & \cellcolor{gray!15} .00115965 \\
 \hline
\end{tabular}
\label{tab:22}
\end{table}
\begin{table}[h]
\centering
\caption{$ \bar{\kappa}^{\alpha}_{23} \equiv \bar{\kappa}^{\alpha}_{2} + \bar{\kappa}^{\alpha}_{3}$ for $ {\kappa}(\epsilon) \eta_2 / \alpha^3 = \bar{\kappa}(\epsilon) = \bar{\kappa}_2 \epsilon^2 + \bar{\kappa}_3 \epsilon^3 + ... $ and $\beta_1 = 1$, $\beta_2 = \alpha$ }
\begin{tabular}{| K{7cm} | K{2cm} | K{2cm} | K{2cm} | }
 \hline
 $k$ & 1 & 2 & 3 \\
 \Xhline{4\arrayrulewidth}
  $ { d^k \bar{\kappa}^{\alpha_{\text{ext}}}_{23} / d \alpha^k } = 0$ ; $\alpha_{\text{ext}}$ & 1.00001 & 1.00001 & $ 1.0094 $ \\
 \hline
$ \bar{\kappa}^{\alpha_{\text{ext}}}_{23} \alpha_{\text{QED}}/ 2\pi $ & .00116141 & .00116141 & \cellcolor{gray!15} .00116023 \\
 \hline
$a_e (\text{one loop}) = \alpha_{\text{QED}}/ 2\pi$ & .00116141 & .00116141 & .00116141 \\
 \hline
$ a_e (\text{higher loop}) $ & .00115965 & .00115965 & \cellcolor{gray!15} .00115965 \\
 \hline
\end{tabular}
\label{tab:23}
\end{table}
\begin{table}[h]
\centering
\caption{$ \bar{\kappa}^{\alpha}_{23} \equiv \bar{\kappa}^{\alpha}_{2} + \bar{\kappa}^{\alpha}_{3}$ for $ {\kappa}(\epsilon) \eta_2 / \alpha^3 = \bar{\kappa}(\epsilon) = \bar{\kappa}_2 \epsilon^2 + \bar{\kappa}_3 \epsilon^3 + ... $ and $\beta_1 = \alpha$, $\beta_2 = 1$ }
\begin{tabular}{| K{7cm} | K{2cm} | K{2cm} | K{2cm} | }
 \hline
 $k$ & 1 & 2 & 3 \\
 \Xhline{4\arrayrulewidth}
  $ { d^k \bar{\kappa}^{\alpha_{\text{ext}}}_{23} / d \alpha^k } = 0$ ; $\alpha_{\text{ext}}$ & - & 1.092 & $ 1.0019 $ \\
 \hline
$ \bar{\kappa}^{\alpha_{\text{ext}}}_{23} \alpha_{\text{QED}}/ 2\pi $ & - & .000886 &\cellcolor{gray!15} .00116121 \\
 \hline
 $ a_e (\text{one loop}) = \alpha_{\text{QED}}/ 2\pi$ & .00116141 & .00116141 & .00116141 \\
 \hline
$ a_e (\text{higher loop}) $ & .00115965 & .00115965 & \cellcolor{gray!15} .00115965 \\
 \hline
\end{tabular}
\label{tab:24}
\end{table}
\begin{table}[h]
\centering
\caption{$ \bar{\kappa}^{\alpha}_{23} \equiv \bar{\kappa}^{\alpha}_{2} + \bar{\kappa}^{\alpha}_{3}$ for $ {\kappa}(\epsilon) \eta_2 / \alpha^3 = \bar{\kappa}(\epsilon) = \bar{\kappa}_2 \epsilon^2 + \bar{\kappa}_3 \epsilon^3 + ... $ and $\beta_1 = 1$, $\beta_2 = \alpha^{3/2}$ }
\begin{tabular}{| K{7cm} | K{2cm} | K{2cm} | K{2cm} | }
 \hline
 $k$ & 1 & 2 & 3 \\
 \Xhline{4\arrayrulewidth}
  $ { d^k \bar{\kappa}^{\alpha_{\text{ext}}}_{23} / d \alpha^k } = 0$ ; $\alpha_{\text{ext}}$ & 1.0063 & 1.0022 & $ 1.0321$ \\
 \hline
$ \bar{\kappa}^{\alpha_{\text{ext}}}_{23} \alpha_{\text{QED}}/ 2\pi $ & .00116145 & .00116143  & \cellcolor{gray!15} .00115935 \\
 \hline
$ a_e (\text{one loop}) = \alpha_{\text{QED}}/ 2\pi$ & .00116141 & .00116141 & .00116141 \\
 \hline
$ a_e (\text{higher loop}) $ & .00115965 & .00115965 & \cellcolor{gray!15} .00115965 \\
 \hline
\end{tabular}
\label{tab:25}
\end{table}
The highlighted results of Tables \ref{tab:22} - \ref{tab:25} for $\bar{\kappa}^{\alpha_{\text{ext}}}_{23} \alpha_{\text{QED}}/ 2\pi$ show an improvement in the evaluation of $a_e$ compared to $ a_e (\text{one loop})$ since we merely multiplied $\kappa(\epsilon)$ by a factor $\eta_2 /\alpha^3$ which is one when $\epsilon = 1$, $\eta_2 /\alpha^3 \vert_{\epsilon =1} = 1$ and therefore independent of the free parameter $\alpha$, but the results are not as good as the results of Tables \ref{tab:14} - \ref{tab:17} since the factor $\eta_2 /\alpha^3$ did not genuinely come from rescaling the variables in the theory and was artificially multiplied. A similar case was discussed in Section \ref{sec:2} when the integral of \eqref{eq:2.11} was multiplied by the factor $\alpha \sqrt{\gamma}$ and after expanding in $\lambda'$ relation \eqref{eq:2.12} was obtained. This factor was equal to one and independent of the free parameter for $\epsilon =1$ but it did not genuinely come from the rescalings of the variables in the theory, therefore the results obtained from relation \eqref{eq:2.12} were not as efficient as the results of relation \eqref{eq:2.11}.
We also considered other possibilities for $\bar{\kappa}(\epsilon)$ such as $\bar{\kappa}(\epsilon) = \kappa(\epsilon) \eta_1 /\alpha^2 $, $\bar{\kappa}(\epsilon) = b \kappa(\epsilon) $ or $\bar{\kappa}(\epsilon) = a \kappa(\epsilon) $ with $a$ and $b$ given by \eqref{eq:4.40} which in all cases $\bar{\kappa}(\epsilon = 1) = 1$ but the results were not as efficient as the results of Tables \ref{tab:14} - \ref{tab:17} at the zero point of the third derivative for a similar reason as mentioned.

\appendix

\section{}
\subsection{Derivation of relation \eqref{eq:3.9} for $K_l$}
\label{sec:A1}
Relation \eqref{eq:3.12}:
\begin{equation}
\label{eq:A1}
K_l = \nu + \max \{ a_1M_1 + a_2M_2 +  ... + a_lM_l \big{|} a_1 + 2a_2 + ... + la_l = l, a_i \in \mathbb{N} \cup \{ 0 \}, i = 1, ..., l \}
\end{equation}
can be easily proven by looking at the potential term $v_{s}(x) u_{l-s}(x)$ in \eqref{eq:3.9}. It is the maximum power of $x$ in this term that determines $K_l$. Lets assume \eqref{eq:A1} is true for $K_{l-1}$ (we know that it is true for when $l - 1 = 0$ since $K_{0} = \nu$, the maximum power of $x$ in the Hermite polynomial functions) and try to prove it for $K_l$. The maximum power that $x$ can have in the term $v_h(x) u_{l-h}(x)$ (with summation over $h = 1, ..., l$ , note that $v_{0}(x) = 0$) is given by
\begin{equation}
\label{eq:A2}
M_{\text{pow} \ \text{of} \ x } \{ v_{h}(x) u_{l-h}(x) \} = \max \{ M_1 + K_{l-1}, M_2 + K_{l-2}, ..., M_{l-1} + K_{1}, M_{l} + K_{0} (= \nu) \}
\end{equation}
it can be shown that \eqref{eq:A2} and \eqref{eq:A1} are the same. For example any of the terms $M_h + K_{l-h}$, $h=1,...,l$ is covered by \eqref{eq:A1}. If $l-h \geq h$
\begin{flalign}
& M_h + K_{l-h} = \nu + M_h + \max \{ a_1M_1 + a_2M_2 +  ... + a_{l-h}M_{l-h} \big{|} a_1 + 2a_2 + ... + (l-h)a_{l-h} = l-h \} \nonumber \\ & = \nu + \max \{ a_1M_1 + ... +  (a_h + 1)M_h +  ... + a_{l-h}M_{l-h} \big{|} a_1 + ... + h(a_h +1)+ ... + (l-h)a_{l-h} = l \} \nonumber &&
\end{flalign}
and if $l-h < h$
\begin{flalign}
M_h + K_{l-h} & = \nu + M_h + \max \{ a_1M_1 + a_2M_2 +  ... + a_{l-h}M_{l-h} \big{|} a_1 + 2a_2 + ... + (l-h)a_{l-h} = l-h \} \nonumber \\ & = \nu + \max \{ a_1M_1 +  ... + a_{l-h}M_{l-h} + M_h \big{|} a_1 + ...+ (l-h)a_{l-h} + h = l \} \nonumber &&
\end{flalign}
for $a_j \in \mathbb{N} \cup \{ 0 \} , j =1, ..., l-h$. Both cases are clearly covered by \eqref{eq:A1}, therefore $\max\{M_h + K_{l-h} \big{|} h = 1, ..., l \} \leq K_l$. It can also be shown that $\max\{M_h + K_{l-h} \big{|} h = 1, ..., l \}$ covers all the possibilities of $K_l$ in relation \eqref{eq:A1} since if any of the coefficients $a_h > 0$ in \eqref{eq:A1} then it would automatically reduce to $M_h + K_{l-h}$, therefore $K_l = \max\{M_h + K_{l-h} \big{|} h = 1, ..., l \}$ and relation \eqref{eq:A1} is proven.

\subsection{Analysis on the convergence rate at strong coupling}
\label{sec:A2}

Here we will show that in the convergent expansion method the convergence rate stays level at strong coupling, at least in the examples discussed here but it is likely that the result is general. First we consider the one dimensional integral of \eqref{eq:2.1} for $\kappa = 1$. For brevity we only show the exact value and the value obtained at order $n = 15$.

\begin{table}[h]
\centering
\caption{Comparison of the numerical values of $I^{\alpha,n} (\lambda , 1) = \sum^{n}_{i=0} I^{\alpha}_i(\lambda , 1) $, with $I^{\alpha}_i(\lambda , 1)$ given by relation \eqref{eq:2.3}, at its extremum point with respect to $\alpha$ for $n = 15$ with the exact value of $I (\lambda, 1 )$ of \eqref{eq:2.1} at strong coupling.}
 \begin{tabular}{| p{0.8cm} | c |} 
 \hline
 $\ \ \lambda$ & \begin{tabular}{K{1cm} | K{4cm} | K{4cm}}
 $n$ & 15 & $I (\lambda, 1 )$ of \eqref{eq:2.1}
 \end{tabular} \\
 \Xhline{4\arrayrulewidth}
 $ \ 10^{10}$ & \begin{tabular}{ K{1cm}  | K{4cm} | K{4cm} }
 $\alpha_{\text{ext}}$ & 677.344 & - \\
 \hline
 $I^{\alpha_{\text{ext}},n}$ & $5.73257385207751 \mathrm{e}\text{-}3$ & $5.73258292087589 \mathrm{e}\text{-}3$
 \end{tabular} \\
 \Xhline{4\arrayrulewidth}
 $ \ 10^{15}$ & \begin{tabular}{ K{1cm}  | K{4cm} | K{4cm} }
 $\alpha_{\text{ext}}$ & 12045.066 & - \\
 \hline
 $I^{\alpha_{\text{ext}},n}$ & $3.22366860723383 \mathrm{e}\text{-}4$ & $3.22367370708450 \mathrm{e}\text{-}4$
 \end{tabular} \\
 \Xhline{4\arrayrulewidth}
 $ \ 10^{20}$ & \begin{tabular}{ K{1cm}  | K{4cm} | K{4cm} }
 $\alpha_{\text{ext}}$ & 214195 & - \\
 \hline
 $I^{\alpha_{\text{ext}},n}$ & $1.81280208622337 \mathrm{e}\text{-}5 $ & $1.81280495408032 \mathrm{e}\text{-}5$
 \end{tabular} \\
 \hline
\end{tabular}
\label{tab:27'}
\end{table}

From Table \ref{tab:27'} we obtain:
\begin{flalign}
\label{eq:A3}
|5.73258292087589 - 5.73257385207751|/5.73258292087589 = 1.58197422 \mathrm{e}\text{-}6 \nonumber \\
| 3.22367370708450 - 3.22366860723383 | / 3.22367370708450 = 1.58199965 \mathrm{e}\text{-}6 \\
| 1.81280495408032 - 1.81280208622337 | / 1.81280495408032 = 1.58199973 \mathrm{e}\text{-}6 \nonumber
\end{flalign}

From \eqref{eq:A3} it is clear that the convergence rate stays level at strong coupling and it is of order $10^{-6}$ at $n = 15$. In the above Table we have shown the numbers up to 15 significant digits which is accurate enough to produce the accuracy of the convergence rates shown. Next we consider the vacuum energy of the anharmonic oscillator. To show that the convergence rate stays level at strong coupling we perform an order of magnitude estimate of the convergence rate although a more accurate analysis is possible by studying higher orders of the expansion. Here we consider the values obtained at order $n=25$ to be accurate compared to the ones obtained at order $n = 15$ and repeat the same analysis as above.

\begin{table}[h]
\centering
\caption{Partial sum of the eigenenergies $2E^{\alpha,n} = 2\sum^n_{l=0} E_{l} $ in relation \eqref{eq:3.17} for the anharmonic oscillator of relation \eqref{eq:3.18} ($\kappa = 1$) evaluated at their extremum point with respect to $\alpha$ for $n = 15$ and $n = 25$ at strong coupling. $\hbar$, $m$ and $\omega$ are set to one.}
 \begin{tabular}{| c | c |} 
 \hline
 $\lambda$ & \begin{tabular}{ K{1.2cm} | K{3cm} | K{3.2cm} }
 $n$ & 15 & 25
 \end{tabular} \\
 \Xhline{4\arrayrulewidth}
 $10^4/2$ & \begin{tabular}{ K{1.2cm} | K{3cm} | K{3.2cm} }
 \hline
 $\alpha^2_{\text{ext},3}$ & 49.76 & 55.64 \\
 \hline
 $2 E^{\alpha, n}$ & $22.861610326$ & $22.861608872$
 \end{tabular} \\ \hline
 \Xhline{4\arrayrulewidth}
 ${10^{10} / 2}$ & \begin{tabular}{ K{1.2cm} | K{3cm} | K{3.2cm} }
 \hline
 $\alpha^2_{\text{ext},3}$ & 4975 & 5563 \\
 \hline
 $2 E^{\alpha, n}$ & $2284.4811863$ & $2284.4810400$
 \end{tabular} \\ \hline
 \Xhline{4\arrayrulewidth}
 ${10^{15} / 2}$ & \begin{tabular}{ K{1.2cm} | K{3cm} | K{3.2cm} }
 \hline
 $\alpha^2_{\text{ext},3}$ & 230926 & 258218 \\
 \hline
 $2 E^{\alpha, n}$ & $106036.21585$ & $106036.20906$
 \end{tabular} \\ \hline
\end{tabular}
\label{tab:28'}
\end{table}

From Table \ref{tab:28'} we obtain:
\begin{flalign}
\label{eq:A4}
|22.861608872 - 22.861610326 |/ 22.861608872 \approx 6 \mathrm{e}\text{-}8 \nonumber \\
| 2284.4810400 - 2284.4811863 | / 2284.4810400 \approx 6 \mathrm{e}\text{-}8 \\
| 106036.20906 - 106036.21585 | / 106036.20906 \approx 6 \mathrm{e}\text{-}8 \nonumber
\end{flalign}
therefore from \eqref{eq:A4} it is clear that the convergence rate stays level at strong coupling.

\section{}
\label{appen:B}

In this Appendix we will discuss two main methods of classifying the sequences of extremum points $\alpha_{\text{ext}}$ and show how to introduce the $\epsilon$ parameter of expansion into the theory as to obtain more efficient results.

\subsection{Introducing the parameter of expansion $\epsilon$}
\label{sec:B1}
The proposed method of introducing the $\epsilon$ parameter is that we associate $\epsilon^{\ceil{p/4}}$ to $x^p$. More explicitly for $4(n-1) < p \leq 4n$, $n \in \mathbb{N}$ we introduce the $\epsilon$ parameter in the following way: $\epsilon^n x^p$.

As an example we consider the following integral
\begin{flalign}
\label{eq:B1}
I & = \int dx \exp \left( -x^2 - x^4 - x^6 - x^{10} \right) \nonumber \\  & = {1 \over \alpha} \int dx \exp \left( - x^2 + \left\{ \epsilon_1 \left( 1 - {1 \over \alpha^2 } \right) x^2 - \epsilon_2 {x^4 \over \alpha^4} - \epsilon_3 {x^6 \over \alpha^6} - \epsilon_4 {x^{10} \over \alpha^{10}} \right\} \right) \equiv I^{\alpha}_{\epsilon_i}
\end{flalign}
$\epsilon_i$ for $i =1, 2, 3 ,4$ are set to one later. We will consider 4 different ways of introducing the $\epsilon$ parameter: (i) $\epsilon_1 = \epsilon_2 = \epsilon_3 = \epsilon_4 =\epsilon$, (ii) $\epsilon_n = \epsilon^n$ for $n =1, 2, 3$ and $\epsilon_4 = \epsilon^5$ quite similar to an $\hbar$ expansion, (iii) $\epsilon_1 = \epsilon_2 = \epsilon$, $\epsilon_3 = \epsilon_4 = \epsilon^2$ and (iv) the suggested way above which is $\epsilon_1 = \epsilon_2 = \epsilon $, $\epsilon_3 = \epsilon^2 $ and $\epsilon_4 = \epsilon^3 $. We will see that method (iv) is more efficient. Expanding \eqref{eq:B1} we obtain the following series
\begin{flalign}
\label{eq:B2}
I^{\alpha}_{\epsilon_i} = & {1  \over \alpha} \sum^{\infty}_{n=0} \underset{ \sum_i p_i = n }{\sum^{n}_{p_1, .., p_4 = 0}} { \sqrt{\pi} \over p_1! p_2! p_3! p_4! } {(2p_1 + 4p_2 +6p_3 + 10p_4  -1)!! \over 2^{p_1 + {2} p_2 + 3p_3 + 5 p_4}} \times \\ &  \epsilon^{p_1}_1  \left( 1 - {1 \over \alpha^2} \right)^{p_1} \left( - { \epsilon_2 \over \alpha^4} \right)^{p_2} \left( - { \epsilon_3 \over \alpha^6} \right)^{p_3} \left(- { \epsilon_4 \over \alpha^{10}} \right)^{p_4} \nonumber
\end{flalign}

Tables \ref{tab:26} - \ref{tab:29} summarize the results of an epsilon expansion using the above ways.
From Table \ref{tab:29} it can be clearly seen that the suggested method produces more efficient results.
\begin{table}[h]
\centering
\caption{Numerical values of $I^{\alpha,n}_{\epsilon_i} \equiv \sum^{n}_{k=0} I^{\alpha}_{\epsilon_i, k} $ evaluated at its extremum points with respect to $\alpha$, for $\epsilon_i = \epsilon$, $i =1, ..., 4$ and $I^{\alpha}_{\epsilon_i} = \sum^{\infty}_{k=0} I^{\alpha}_{\epsilon_i, k} \epsilon^k $ in relation \eqref{eq:B2}.}
\begin{tabular}{| K{1.4cm}  | K{2cm} | K{2cm} | K{2cm} | K{2.2cm} |K{1.5cm} |}
 \hline
  $n$ & 3 & 6 & 15 & 20 & Exact \\
 \Xhline{4\arrayrulewidth}
  $\alpha_{\text{ext},1}$ & 2.56 & $3.23 \pm .15i$ & 4.533 & $5.06 \pm .07i$ & - \\
 \hline
  $I^{\alpha_{\text{ext},1},n}_{\epsilon_i}$ & 1.18 & 1.21 $\mp$ .02i & 1.232 & $1.238 \mp .002i$ & 1.251287 \\
 \Xhline{4\arrayrulewidth}
  $\alpha_{\text{ext},2}$ & $2.54 \pm .44i$ & $3.22 \pm .44i$ & $4.531 \pm.17i$ & $5.05 \pm .21i$ & - \\
 \hline
  $I^{\alpha_{\text{ext},2},n}_{\epsilon_i}$ & $1.20 \mp .08i$ & $1.22 \mp .05i$ & $1.233 \mp .01i$ & $1.239 \mp .01i$ & 1.251287 \\
 \hline
\end{tabular}
\label{tab:26}
\end{table}
\begin{table}[h]
\centering
\caption{Numerical values of $I^{\alpha,n}_{\epsilon_i} \equiv \sum^{n}_{k=0} I^{\alpha}_{\epsilon_i, k} $ evaluated at its extremum points with respect to $\alpha$, for $\epsilon_i = \epsilon^i$, $i =1,2,3$, $\epsilon_4 = \epsilon^5$ and $I^{\alpha}_{\epsilon_i} = \sum^{\infty}_{k=0} I^{\alpha}_{\epsilon_i, k} \epsilon^k $ in relation \eqref{eq:B2}.}
\begin{tabular}{| K{1.4cm} | K{1cm} | K{1cm} | K{1cm} | K{1cm} | K{1.5cm} |}
 \hline
  $n$ & 3 & 6 & 15 & 20 & Exact \\
 \Xhline{4\arrayrulewidth}
  $\alpha_{\text{ext},1}$ & 1.96 & 2.76 & 4.23 & 4.80 & - \\
 \hline
  $I^{\alpha_{\text{ext},1}, n }_{\epsilon_i}$ & 1.38 & 1.30 & 1.26 & 1.256 & 1.251287 \\
 \hline
  $\alpha_{\text{ext},2}$ & - & 1.62 & 3.45 & 4.10 & - \\
 \hline
  $I^{\alpha_{\text{ext},2}, n}_{\epsilon_i}$ & - & 0.64 & 1.23 & 1.243 & 1.251287 \\
 \hline
\end{tabular}
\label{tab:27}
\end{table}
\begin{table}[h]
\centering
\caption{Numerical values of $I^{\alpha,n}_{\epsilon_i} \equiv \sum^{n}_{k=0} I^{\alpha}_{\epsilon_i, k} $ evaluated to its extremum points with respect to $\alpha$, for $\epsilon_1 = \epsilon_2 = \epsilon$, $\epsilon_3 = \epsilon^2$, $\epsilon_4 = \epsilon^2$ and $I^{\alpha}_{\epsilon_i} = \sum^{\infty}_{k=0} I^{\alpha}_{\epsilon_i, k} \epsilon^k $ in relation \eqref{eq:B2}.}
\begin{tabular}{| K{1.4cm} | K{1cm} | K{2.5cm} | K{3cm} | K{3cm} | K{2cm} |}
 \hline
  $n$ & 3 & 6 & 15 & 20 & Exact \\
 \Xhline{4\arrayrulewidth}
  $\alpha_{\text{ext},1}$ & 2.26 & $2.77 \pm 0.19i$ & $ 3.82 \pm .52i $ & $ 4.25 \pm .65 i$ & - \\
 \hline
  $I^{\alpha_{\text{ext},1}, n}_{\epsilon_i}$ & 1.239 & $1.2495 \mp .004$ & $1.2526 \pm 1\mathrm{e}{\text{-}3} i$ & $1.2510 \pm 1\mathrm{e}{\text{-}3} i$  & $1.251287$ \\
 \hline
  $\alpha_{\text{ext},2}$ & 1.80 & 2.54 & $3.77 \pm .27i$ & $4.20 \pm .41i$ & - \\
 \hline
  $I^{\alpha_{\text{ext},2}, n}_{\epsilon_i}$ & 1.14 & 1.257 & $1.251492 \mp 3\mathrm{e}{\text{-}4} i$ & 1.251488 & $1.251287$ \\
 \hline
\end{tabular}
\label{tab:28}
\end{table}
\begin{table}[h]
\centering
\caption{Numerical values of $I^{\alpha,n}_{\epsilon_i} \equiv \sum^{n}_{k=0} I^{\alpha}_{\epsilon_i, k} $ evaluated at its extremum points with respect to $\alpha$, for $\epsilon_1 = \epsilon_2 = \epsilon$, $\epsilon_3 = \epsilon^2$, $\epsilon_4 = \epsilon^3$ and $I^{\alpha}_{\epsilon_i} = \sum^{\infty}_{k=0} I^{\alpha}_{\epsilon_i, k} \epsilon^k $ in relation \eqref{eq:B2}.}
\begin{tabular}{| K{1.4cm} | K{1cm} | K{1cm} | K{2.5cm} | K{3cm} | K{2cm} |}
 \hline
  $n$ & 3 & 6 & 15 & 20 & Exact \\
 \Xhline{4\arrayrulewidth}
  $\alpha_{\text{ext},1}$ & 2.16 & 2.80 & $ 3.86 \mp .07 $ & $ 4.31 \pm .12 i$ & - \\
 \hline
  $I^{\alpha_{\text{ext},1}, n}_{\epsilon_i}$ & 1.264 & 1.2518 & 1.25120 $\pm$ $.00 i$ & $1.251274 \mp .00 i$  & $1.251287$ \\
 \hline
  $\alpha_{\text{ext},2}$ & - & 2.43 & 3.59 & 4.15 & - \\
 \hline
  $I^{\alpha_{\text{ext},2}, n}_{\epsilon_i}$ & - & 1.24 & 1.2515 & 1.251300 & $1.251287$ \\
 \hline
\end{tabular}
\label{tab:29}
\end{table}
\\
\subsection{A classification of the sequences of extremum points $\alpha_{\text{ext}}$}
\label{sec:B2}
For the evaluation of the Gaussian integrals of Section \ref{sec:2} and the eigenenergies of the quantum systems in Section \ref{sec:3}, finding the extremum points $\alpha_{\text{ext}}$ at order $n$ reduces to finding the roots of a polynomial in $\alpha$ of some degree, say $\bar{m}$, with real coefficients. This polynomial has $\bar{m}$ roots. In what follows we discuss two different ways of identifying the sequences of extremum points:

i) One way to classify the different sequences of extremum points is based on a realness criteria. In this approach we only consider the positive real extremum points with the exception that at some orders of $n$ in order to fill in the gap of the elements of the real sequences of extremum points we might also need to consider a complex extremum point with a real part greater than the previous real element of the sequence and smaller than the real element after and the smallest absolute value of the imaginary part possible that satisfies the condition $\text{Re}(\alpha_{\text{ext}}) > | \text{Im}(\alpha_{\text{ext}}) |$. Lets assume at order $n$ of the expansion we obtain $\bar{m}$ roots (or extremum points) with $m$ of them $r_1, ..., r_{m}$ being real and positive and also lets assume in this order we need to consider a complex extremum point $w$ with the properties mentioned above. If these extremum points are organized in terms of decreasing real value $r_1 \geq r_2 \geq... \geq \text{Re}(w) \geq ... \geq r_{m}$, then $r_1$ would be considered as the element of the first sequence of real extremum points, $r_2$ as the element of the 2nd sequence of real extremum points, etc.

All of the sequences of extremum points in the Tables of Sections \ref{sec:1}, \ref{sec:2} and \ref{sec:3} are identified based on method i).

For example in the Tables of Section \ref{sec:1} and Section \ref{sec:2} (with the only exception being Table \ref{tab:13}) at odd orders of $n$ we only obtain one real positive extremum point therefore we only have one sequence of positive real extremum points but at even orders there are no real roots thus in order to fill in the gap of the elements of the positive real sequence of extremum points at even orders of $n$, we consider the extremum point with a real part that is larger than the previous element of the sequence and smaller than the element after and with the smallest absolute value of imaginary part possible as part of this sequence at even orders of $n$. For Table \ref{tab:13} it is the opposite, at even orders of $n$ we obtain one real positive extremum point but at odd orders there are no real extremum points therefore we can consider the complex extremum point with the properties of $w$ described above at odd orders of $n$ as part of this sequence.

The sequences of extremum points in the Tables of Appendix \ref{appen:C} are also specified based on method i) above but in these cases we obtain many sequences of real extremum points.

ii) In the second method we remove the realness constraint. Lets assume at order $n$ of the expansion we obtain $\bar{m}$ roots with $m'$ of them $r_1, ..., r_{m'}$ having a positive real part and satisfying $\text{Re}(r_i) > | \text{Im}(r_i) |$ for $i =1, ..., m'$. In this sequence of $r_1, ..., r_{m'}$ the conjugate pair roots are considered as one element. If these extremum points are organized in terms of decreasing real part $\text{Re}(r_1) \geq \text{Re}(r_2) \geq... \geq \text{Re}(r_{m})$, then $r_1$ would be considered as the element of the first sequence of extremum points, $r_2$ as the element of the 2nd sequence of extremum points, etc.

In the examples of the previous Subsection \ref{sec:B1} we have organized the sequences of extremum points based on method ii). For example in Table \ref{tab:29} at order $n = 20$ the extremum point with the largest real part is $4.31 \pm 0.12i$ therefore it is associated to the first sequence of extremum points and the extremum point with the second largest real part is $4.15$ therefore it is associated to the second sequence of extremum points, etc.

Whether the sequences of extremum points are organized based on method i) or ii) it is the higher sequences that will eventually converge to the quantity of interest faster as we go to higher orders in the expansion.

\section{Numerical value of the eigenenergies obtained using the convergent expansion method}
\label{appen:C}

Here we include the Tables of numerical data obtained for the eigenenergies of the quantum systems discussed in Subsection \ref{sec:3.1}. We consider double the value of the eigenenergies in Tables \ref{tab:30}, \ref{tab:31} and \ref{tab:32} as to match with the conventions of \cite{08}, \cite{09} and \cite{12}.

Table \ref{tab:30} shows the values of the vacuum energy of the anharmonic oscillator and compares it with the values of \cite{08}. The Table shows good convergence rate for all positive values of the coupling.

Table \ref{tab:31} shows the values of the vacuum and the first excited state energy of the double well potential and compares it with the values of \cite{09}. The convergence becomes more difficult at weak coupling due to the reason mentioned in Subsection \ref{sec:3.1}. But the convergence rate is good at strong coupling.

To show that using the convergent expansion method accurate results can be obtained for large positive values of the coupling, in Tables \ref{tab:30} and \ref{tab:31} we have included a few examples of results obtained for the eigenenergies for very large values $\lambda$. The same procedure that was used to obtain the eigenenergies of Tables \ref{tab:30} and \ref{tab:31} for smaller values of $\lambda$ that match the results of references \cite{08} and \cite{09} was applied to the larger values of the coupling. The results clearly show convergence therefore are reliable and can be considered as valid.

Table \ref{tab:31'} shows the evaluation of the perturbative part of the vacuum energy of the double well potential of \eqref{eq:3.21'} and compares it with the exact value stated in \cite{16}. Further details are explained in the caption of this Table.

Table \ref{tab:32} shows the eigenenergies of the pure anharmonic oscillator for levels $\nu = 0, 3 ,6$ and compares it with the values listed in \cite{12}. The pure anharmonic oscillator eigenenergies for levels $\nu = 0, 6$ and $\nu = 3$ were first computed in \cite{14} and \cite{15}, respectively. They were reevaluated using another method by \cite{12}. It is possible to obtain more accurate results for the level number $\nu = 6$ that match the results of reference \cite{12} by going to higher orders in the expansion.

It is quite remarkable that the accuracy required for the $\alpha^2_{\text{ext},i}$ in order to produce the results of the energy levels with the accuracy shown is just a few significant digits. So the accuracy of $\alpha^2_{\text{ext},i}$ shown in the Tables below are enough to produce the accuracy of the numerical values of the energy levels.

The coefficients of the potential in \eqref{eq:3.18} are a function of $\alpha^4$ and $\alpha^6$, therefore the partial sum of the eigenenergies in \eqref{eq:3.17} for the quantum mechanical examples studied would be a function of $\alpha^2$, $\alpha^4$ and $\alpha^6$. The data of Tables \ref{tab:30}, \ref{tab:31} and \ref{tab:32} were taken by the replacement of $\alpha^2 = \alpha'$ and finding the zero first derivative points of $E^{\alpha, n}$ with respect to $\alpha'$, therefore the squared of the extremum points with respect to $\alpha$ are shown in these Tables.

The convergence rate for the energy levels becomes better for the higher sequences. For example the value obtained for $2 E^{\alpha, n}$ for $\lambda = 1/20$ at order $n = 25$ for $\alpha^2_{\text{ext},3} = 1.473$ in Table \ref{tab:30} is a lot closer to the value of reference \cite{08} compared to the value obtained at order $n = 25$ for $\alpha^2_{\text{ext},1} = 1.994$.

The sequences of extremum points in Tables \ref{tab:30} - \ref{tab:32} are identified based on method i) of Appendix \ref{sec:B2} and we have shown the highest sequence available at the largest order studied. For example in Table \ref{tab:32} for $\nu = 6$, $n = 50$ there were not more than a total of $14$ sequences of real extremum points, therefore we have shown these sequences up to the $14$th one. For a discussion on two main methods for identifying the sequences of extremum points refer to Appendix \ref{sec:B2}. In Table \ref{tab:31'} we only obtain one sequence of real extremum points at least to the order shown in this Table.

As mentioned the convergence rate becomes better for higher sequences therefore it is possible to obtain an estimate of the accuracy obtained for the energy levels. For example with taking the value obtained for the vacuum energy in Table \ref{tab:30} at order $n = 25$ for the third sequence of real extremum points to be accurate compared to the ones obtained at order $n = 15$, it is possible to obtain an estimate of the accuracy obtained at order $n = 15$. For example for the coupling of $\lambda = 10^{10}/2$ we have $|2284.481040 - 2284.48119 |/ 2284.481040 \sim 10^{-7} $, meaning that the values obtained for the vacuum energy of the anharmonic oscillator for $\lambda = 10^{10}/2$ at order $n = 15$ for the 3rd sequence have an accuracy of order one out of $10^{7}$.

In these Tables when the extremum point is imaginary we have not included the imaginary part of the eigenenergy as it is very small and it is not of interest to us for a similar reason as mentioned in Section \ref{sec:1} for the one-dimensional integrals.
\begin{table}[h]
\centering
\caption{Partial sum of the eigenenergies $2E^{\alpha,n} = 2\sum^n_{l=0} E_{l} $ in relation \eqref{eq:3.17} for the anharmonic oscillator of relation \eqref{eq:3.18} ($\kappa = 1$) evaluated at their extremum points with respect to $\alpha$ for different $n$. $\hbar$, $m$ and $\omega$ are set to one.}
 \begin{tabular}{| c | c |} 
 \hline
 $\lambda$ & \begin{tabular}{ K{1.2cm} | K{2cm} | K{3cm} | K{3.2cm} | K{3cm}}
 $n$ & 6 & 15 & 25 & $2 E$(of \cite{08})
 \end{tabular} \\
 \Xhline{4\arrayrulewidth}
 $1/20$ & \begin{tabular}{ K{1.2cm} | K{2cm} | K{3cm} | K{3.2cm} | K{3cm} }
 $\alpha^2_{\text{ext},1}$ & 1.307 & 1.669 &1.994 & - \\
 \hline
 $2 E^{\alpha, n}$ & $1.06528547$ & $1.065285503$ & $1.065 285507$ & $1.065 285509 54 (6)$ \\
 \hline
 $\alpha^2_{\text{ext},2}$ & 1.256 & 1.402 & 1.620 & - \\
 \hline
 $2 E^{\alpha, n}$ & $1.06528549$ & $1.0652855095440$ & $1.0652855095438$& $1.065 285509 54 (6)$ \\
 \hline
 $\alpha^2_{\text{ext},3}$ & - & 1.376 & 1.473 & - \\
 \hline
 $2 E^{\alpha, n}$ & - & $1.0652855095439$ & $1.0652855095437$& $1.065 285509 54 (6)$
 \end{tabular} \\
 \Xhline{4\arrayrulewidth}
 $1/2$ & \begin{tabular}{ K{1.2cm} | K{2cm} | K{3cm} | K{3.2cm} | K{3cm} }
 $\alpha^2_{\text{ext},1}$ & 2.257 & 3.207 & 3.985 & - \\
 \hline
 $2 E^{\alpha, n}$ & $1.39234$ & $1.392346$ & $1.392348$ & 1.39235(5) \\
 \hline
 $\alpha^2_{\text{ext},2}$ & 2.106 & 2.523 & 3.087 & - \\
 \hline
 $2 E^{\alpha, n}$ & $1.39235$ & $1.392351646$ & $1.392351644$ & 1.39235(5) \\
 \hline
 $\alpha^2_{\text{ext},3}$ & - & 2.453 & 2.711 & - \\
 \hline
 $2 E^{\alpha, n}$ & - & $1.392351645$ & $1.3923516415$& 1.39235(5)
 \end{tabular} \\
 \Xhline{4\arrayrulewidth}
 $100/2$ & \begin{tabular}{ K{1.2cm} | K{2cm} | K{3cm} | K{3.2cm} | K{3cm} }
 $\alpha^2_{\text{ext},1}$ & 9.774 & 14.411 & 18.117 & - \\
 \hline
 $2 E^{\alpha, n}$ & $4.99921$ & $4.99927$ & $4.99930$ & 5.0(1) \\
 \hline
 $\alpha^2_{\text{ext},2}$ & 9.016 & 11.091 & 13.835 & - \\
 \hline
 $2 E^{\alpha, n}$ & $4.9993$ & $4.9994179$ & $4.99941777$ & 5.0(1) \\
 \hline
 $\alpha^2_{\text{ext},3}$ & - & 10.750 & 12.013 & - \\
 \hline
 $2 E^{\alpha, n}$ & - & $4.9994178$ &  $4.99941755$ & 5.0(1)
 \end{tabular} \\ \hline
 \Xhline{4\arrayrulewidth}
 $10^4/2$ & \begin{tabular}{ K{1.2cm} | K{2cm} | K{3cm} | K{3.2cm} | K{3cm} }
 $\alpha^2_{\text{ext},1}$ & 45.22 & 66.79 & 84.01 & - \\
 \hline
 $2 E^{\alpha, n}$ & $22.8606$ & $22.86086$ & $22.86102$ & - \\
 \hline
 $\alpha^2_{\text{ext},2}$ & 41.68 & 51.34 & $64.11$ & - \\
 \hline
 $2 E^{\alpha, n}$ & $22.861$ & $22.8616106$ & $22.8616101$ & - \\
 \hline
 $\alpha^2_{\text{ext},3}$ & - & 49.76 & 55.64 & - \\
 \hline
 $2 E^{\alpha, n}$ & - & $22.8616103$ & $22.86160887$ & -
 \end{tabular} \\ \hline
 \Xhline{4\arrayrulewidth}
 ${10^{10} / 2}$ & \begin{tabular}{ K{1.2cm} | K{2cm} | K{3cm} | K{3.2cm} | K{3cm} }
 $\alpha^2_{\text{ext},1}$ & 4521 & 6678 & 8401 & - \\
 \hline
 $2 E^{\alpha, n}$ & $2284.38$ & $2284.406$ & $2284.422$ & - \\
 \hline
 $\alpha^2_{\text{ext},2}$ & 4167 & 5134 & $6410$ & - \\
 \hline
 $2 E^{\alpha, n}$ & $2284.419$ & $2284.48122$ & $2284.481160$ & - \\
 \hline
 $\alpha^2_{\text{ext},3}$ & - & 4975 & 5563 & - \\
 \hline
 $2 E^{\alpha, n}$ & - & $2284.48119$ & $2284.481040$ & -
 \end{tabular} \\ \hline
\end{tabular}
\label{tab:30}
\end{table}
\begin{table}[h]
\centering
\caption{Partial sum of the eigenenergies $2E^{\alpha,n} = 2\sum^n_{l=0} E_{l} $ in relation \eqref{eq:3.17} for the double well potential of relation \eqref{eq:3.18} ($\kappa = -1$) evaluated at their extremum points with respect to $\alpha$ for different $n$. From \eqref{eq:3.18} $\bar{\lambda} = \lambda / \omega^3$ with $\hbar$ and $m$ set to one. $\nu$ corresponds to the energy level. $\nu = 0$ is the vacuum energy, $\nu = 1$ is the first excited states eigenenergy. }
 \begin{tabular}{| c | c | c | c | c |} 
 \hline
$\nu$ & $\omega^2$ & $\lambda$ & $\bar{\lambda}$ & \begin{tabular}{K{1cm} | K{2cm} | K{2.8cm} | K{2.8cm} | K{2.8cm}}
 $n$ & 25 & 45 & 65 & $2 E$(of \cite{09})
 \end{tabular} \\
 \Xhline{4\arrayrulewidth}
0 & $4$ & ${1 / 2}$ & ${1 \over 16}$ & \begin{tabular}{ K{1cm} | K{2cm} | K{2.8cm} | K{2.8cm} | K{2.8cm} }
 $\alpha^2_{\text{ext},3}$ & 1.037 & 1.437 &1.766 & - \\
 \hline
 $2 E^{\alpha, n}$ & $-1.731$ & $-1.734$  & $-1.732$ & $-1.71035..$ \\
 \hline
 $\alpha^2_{\text{ext},4}$ & - & $1.261 \pm .02i$ & 1.521 & - \\
 \hline
 $2 E^{\alpha, n}$ & - & $-1.699$ & $-1.698$ & $-1.71035..$ \\
 \hline
 $\alpha^2_{\text{ext},5}$ & - & - & 1.425 & - \\
 \hline
 $2 E^{\alpha, n}$ & - & - & $-1.708$& $-1.71035..$
 \end{tabular} \\
 \Xhline{4\arrayrulewidth}
$\nu$ & $\omega^2$ & $\lambda$ & $\bar{\lambda}$ & \begin{tabular}{K{1cm} | K{2cm} | K{2.8cm} | K{2.8cm} | K{2.8cm}}
 $n$ & 15 & 35 & 65 & $2 E$(of \cite{09})
 \end{tabular} \\
  \Xhline{4\arrayrulewidth}
0 & ${1 \over 2}$ & ${1 / 2}$  & $\sqrt{2}$ & \begin{tabular}{ K{1cm} | K{2cm} | K{2.8cm} | K{2.8cm} | K{2.8cm} }
 $\alpha^2_{\text{ext},2}$ & 3.271 & 4.821 & 6.453 & - \\
 \hline
 $2 E^{\alpha, n}$ & $0.8700179$ & $0.87001774$ & $0.87001768$ & $0.870017518372$ \\
 \hline
 $\alpha^2_{\text{ext},4}$  & - & $3.850 \pm .06i$ & 4.780 & - \\
 \hline
 $2 E^{\alpha, n}$ & - & $0.8700175181$ & $0.870017518375$ & $0.870017518372$ \\
 \hline
 $\alpha^2_{\text{ext},5}$  & - & - & 4.534 & - \\
 \hline
 $2 E^{\alpha, n}$ & - & - & $0.870017518372$ & $0.870017518372$
 \end{tabular} \\
 \Xhline{4\arrayrulewidth}
 $\nu$ & $\omega^2$ & $\lambda$ & $\bar{\lambda}$ & \begin{tabular}{K{1cm} | K{2cm} | K{2.8cm} | K{2.8cm} | K{2.8cm}}
 $n$ & 15 & 35 & 50 & -
 \end{tabular} \\
  \Xhline{4\arrayrulewidth}
0 & $1$ & $10^{10}$ & $10^{10}$ & \begin{tabular}{ K{1cm} | K{2cm} | K{2.8cm} | K{2.8cm} | K{2.8cm} }
 $\alpha^2_{\text{ext},2}$ & 6468 & 9385 & 11047 & - \\
 \hline
 $2 E^{\alpha, n}$  & $2878.26563$ & $2878.26552$ & $2878.26550$ & - \\
 \hline
 $\alpha^2_{\text{ext},3}$  & 6268 & 7957 & 9332 & - \\
 \hline
 $2 E^{\alpha, n}$ & $2878.26559$ & $2878.26540466$ & $2878.26540478$ & - \\
 \hline
 $\alpha^2_{\text{ext},4}$ & - & $7553 \pm 110i$ & $8312$ & - \\
 \hline
 $2 E^{\alpha, n}$ & - & $2878.26540494 $ & $2878.26540501$ & -
 \end{tabular} \\
  \Xhline{4\arrayrulewidth}
$\nu$ & $\omega^2$ & $\lambda$ & $\bar{\lambda}$ & \begin{tabular}{K{1cm} | K{2cm} | K{2.8cm} | K{2.8cm} | K{2.8cm}}
 $n$ & 15 & 35 & 45 & $2 E$(of \cite{09})
 \end{tabular} \\
  \Xhline{4\arrayrulewidth}
1 & ${1 \over 2}$ & ${1 / 2}$  & $\sqrt{2}$ & \begin{tabular}{ K{1cm} | K{2cm} | K{2.8cm} | K{2.8cm} | K{2.8cm} }
 $\alpha^2_{\text{ext},3}$ & 3.73 & $5.463$ & 6.142 & - \\
 \hline
 $2 E^{\alpha, n}$ & $3.333778$ & $3.3337784$ & $3.3337786$ & $3.33377932989$ \\
 \hline
 $\alpha^2_{\text{ext},5}$  & - & $4.385$ & 4.915 & - \\
 \hline
 $2 E^{\alpha, n}$ & - & $3.3337793286$ & $3.3337793290$ & $3.33377932989$ \\
 \hline
 $\alpha^2_{\text{ext},7}$  & - & - & 4.282 & - \\
 \hline
 $2 E^{\alpha, n}$ & - & - & $3.33377932989$ & $3.33377932989$
 \end{tabular} \\ \hline
\end{tabular}
\label{tab:31}
\end{table}
\begin{table}[h]
\centering
\caption{ Evaluating the vacuum energy of the double well potential when expanding about one of its vacuums in relation \eqref{eq:3.21'}, using relation \eqref{eq:3.17} $E^{\alpha, n} = \sum^n_{l=0} E_{l}$, shows convergence to only the perturbative part of the vacuum energy, as expected. This is to be compared with the exact value of the vacuum energy for $\lambda = 3/100$ which is $0.4531$. As stated in \cite{16} the difference between the lowest two energy levels is of order $0.02$, therefore with the exact vacuum energy being at $0.4531$, we expect the first excited state energy to be at $\approx 0.473$. The result in this Table is showing convergence to only the perturbative part of the vacuum energy which is $\approx 0.463$ and does not take into account the contribution of the instantons. The instanton contributions will result in the splitting of the perturbative vacuum energy value of $\approx 0.463$ into a lower vacuum energy of $\approx 0.453$ and a higher first excited state of $\approx 0.473$. In this case since the potential of \eqref{eq:3.21'} involves an odd term there is a possibility that the perturbative expansion of the Schrodinger equation does not capture the full result as can be seen from this Table. This is to be compared with the results of Table \ref{tab:31} which show convergence to the full result of the double well potential eigenenergies since these results have been evaluated using the potential of relation \eqref{eq:3.18} for $\kappa = -1$ which only involves even terms. }
 \begin{tabular}{| c | c |} 
 \hline
 $\lambda$ & \begin{tabular}{K{1cm} | K{2cm} | K{2cm} | K{2cm} | K{2cm} | K{2cm}}
 $n$ & 15 & 35 & 55 & 75 & $E_0$(of \cite{16})
 \end{tabular} \\
 \Xhline{4\arrayrulewidth}
 ${3 / 100}$ & \begin{tabular}{ K{1cm} | K{2cm} | K{2cm} | K{2cm} | K{2cm} | K{2cm} }
 $\alpha_{\text{ext}}$ & 1.069 & 1.117 & 1.155 &1.187 & - \\
 \hline
 $E^{\alpha, n}$ & $0.4648$ & $0.4639$ & $0.4637$  & $0.4636$ & $0.4531$ \\
 \hline
 \end{tabular} \\ \hline
\end{tabular}
\label{tab:31'}
\end{table}
\begin{table}[h]
\centering
\caption{Partial sum of the eigenenergies $2E^{\alpha,n} = 2\sum^n_{l=0} E_{l} $ in relation \eqref{eq:3.17} for the pure anharmonic oscillator of relation \eqref{eq:3.18} ($\kappa = 0$) evaluated at their extremum points with respect to $\alpha$ for different $n$. $\hbar$, $m$ and $\omega$ are set to one. $\nu$ corresponds to the energy level. $\nu = 0$ is the vacuum energy, $\nu = 3$ is the third excited states eigenenergy, etc.}
 \begin{tabular}{| c | c | c |} 
 \hline
 $\nu$ & $\lambda$ & \begin{tabular}{K{1cm} | K{2.2cm} | K{2.8cm} | K{3.4cm} | K{3.5cm}}
 $n$ & 20 & 35 & 50 & $2 E$(of \cite{12})
 \end{tabular} \\
 \Xhline{4\arrayrulewidth}
 $3$ & 1/2 & \begin{tabular}{ K{1cm} | K{2.2cm} | K{2.8cm} | K{3.4cm} | K{3.5cm} }
 $\alpha^2_{\text{ext},5}$ & $2.76 \pm .04 i$ & 3.465 & 4.067 & - \\
 \hline
 $2 E^{\alpha, n}$ & 11.644745500 & $11.644745505$ & $11.644745507$ & $11.644745511378$ \\
 \hline
 $\alpha^2_{\text{ext},7}$ & - & $3.043$ & 3.416 & - \\
 \hline
 $2 E^{\alpha, n}$ & -  & $11.64474551145$ & $11.644745511375$ & $11.644745511378$ \\
 \hline
 $\alpha^2_{\text{ext},8}$ & -  & - & 3.272  & - \\
 \hline
 $2 E^{\alpha, n}$ & - & - & 11.644745511378 & $11.644745511378$
 \end{tabular} \\
 \Xhline{4\arrayrulewidth}
 $6$ & 1/2 & \begin{tabular}{ K{1cm} | K{2.2cm} | K{2.8cm} | K{3.4cm} | K{3.5cm} }
 $\alpha^2_{\text{ext},8}$ & 3.064 & 3.825 & 4.478 & - \\
 \hline
 $2 E^{\alpha, n}$ & 26.528471187 & $26.528471186$ & $26.528471185$ & $26.528471183682518..$ \\
 \hline
 $\alpha^2_{\text{ext},11}$ & -  & 3.322 & 3.851 & - \\
 \hline
 $2 E^{\alpha, n}$ & - & $26.528471183679$  & $26.528471183681$ & $26.528471183682518..$ \\
 \hline
 $\alpha^2_{\text{ext},14}$ & -  & - & 3.507  & - \\
 \hline
 $2 E^{\alpha, n}$ & - & - & $26.528471183682510$ & $26.528471183682518..$
 \end{tabular} \\
 \Xhline{4\arrayrulewidth}
 $0$ & 1/2 & \begin{tabular}{ K{1cm} | K{2.2cm} | K{2.8cm} | K{3.4cm} | K{3.5cm} }
 $\alpha^2_{\text{ext},2}$ & 2.699 & 3.457 & 4.070 & - \\
 \hline
 $2 E^{\alpha, n}$ & 1.06036216  & $1.06036213$ & $1.06036212$ & 1.0604 \\
 \hline
 $\alpha^2_{\text{ext},3}$ & $2.442 \pm 0.05$  & 2.932 & 3.438 & - \\
 \hline
 $2 E^{\alpha, n}$ & 1.0603620938 & 1.06036209036 & $1.06036209039$ & 1.0604 \\
 \hline
 $\alpha^2_{\text{ext},4}$ & - & $2.783 \pm 0.04i $ & 3.062 & - \\
 \hline
 $2 E^{\alpha, n}$ & - & $1.06036209046$ & $1.06036209048$ & 1.0604
 \end{tabular} \\ \hline
 \end{tabular}
\label{tab:32}
\end{table}

\clearpage
\acknowledgments

The author was benefitted from discussions with Bob Holdom, especially related to the Introduction of the paper.

\end{document}